\documentclass[%
twocolumn,
 amsmath,amssymb,
 aps,
prd,
superscriptaddress,showpacs]{revtex4-2}

\usepackage[utf8]{inputenc}
\usepackage{graphicx}
\usepackage{dcolumn}
\usepackage{bm}
\usepackage{lineno}
\usepackage{float}
\usepackage{xspace}
\usepackage{placeins}
\usepackage{gensymb}

\begin{document}

\preprint{v3.0}

\newcommand{\numu}{$\nu_\mu$\xspace}
\newcommand{\nue}{$\nu_e$\xspace}
\newcommand{\nuebar}{$\overline{\nu}_e$\xspace}
\newcommand{\dcp}{$\delta_{CP}$\xspace}
\newcommand{\thtt}{$\theta_{23}$\newcommand}
\newcommand{\dmsq}{$|\Delta m^2_{32}|$\xspace}
\newcommand{\nova}{NOvA\xspace}
\newcommand{\mb}{MiniBooNE\xspace}
\newcommand{\mub}{MicroBooNE\xspace}
\newcommand{\nuance}{\textsc{nuance}}
\newcommand{\minerva}{MINERvA\xspace}
\newcommand{\addcite}{{\color{red} ADD CITATION}~}
\newcommand{\tmu}{\ensuremath{T_{\mu}}\xspace}
\newcommand{\pmu}{\ensuremath{|\textbf{p}_{\mu}|}\xspace}
\newcommand{\cost}{\ensuremath{\cos{\theta_{\mu}}}\xspace}
\newcommand{\enu}{\ensuremath{E_{\nu}}\xspace}
\newcommand{\qq}{\ensuremath{Q^{2}}\xspace}
\newcommand{\enuqe}{\ensuremath{E_{\nu}^{\textrm{QE}}}\xspace}
\newcommand{\qqqe}{\ensuremath{Q^{2}_{\textrm{QE}}}\xspace}
\newcommand{\pf}{\ensuremath{p_{F}}\xspace}
\newcommand{\eb}{\ensuremath{E_{b}}\xspace}
\newcommand{\carbon}{C\ensuremath{^{12}}\xspace}
\newcommand{\oxygen}{O\ensuremath{^{16}}\xspace}
\newcommand{\ma}{\ensuremath{M_{\textrm{A}}}\xspace}
 \newcommand{\maeff}{\ensuremath{M^{\mbox{\scriptsize{eff}}}_{\textrm{A}}}\xspace}
\newcommand{\mares}{\ensuremath{M^{\mbox{\scriptsize{RES}}}_{\textrm{A}}}\xspace}
\newcommand{\maqe}{\ensuremath{M_{\textrm{A}}^{\mbox{\scriptsize{QE}}}}\xspace}
\newcommand{\rpm}{\raisebox{.2ex}{$\scriptstyle\pm$}}


\title{Measurement of Differential Cross Sections for $\nu_\mu$-Ar Charged-Current Interactions with Protons and no Pions in the Final State with the MicroBooNE Detector}

\newcommand{\Bern}{Universit{\"a}t Bern, Bern CH-3012, Switzerland}
\newcommand{\BNL}{Brookhaven National Laboratory (BNL), Upton, NY, 11973, USA}
\newcommand{\UCSB}{University of California, Santa Barbara, CA, 93106, USA}
\newcommand{\Cambridge}{University of Cambridge, Cambridge CB3 0HE, United Kingdom}
\newcommand{\StKates}{St. Catherine University, Saint Paul, MN 55105, USA}
\newcommand{\CIEMAT}{Centro de Investigaciones Energ\'{e}ticas, Medioambientales y Tecnol\'{o}gicas (CIEMAT), Madrid E-28040, Spain}
\newcommand{\Chicago}{University of Chicago, Chicago, IL, 60637, USA}
\newcommand{\Cincinnati}{University of Cincinnati, Cincinnati, OH, 45221, USA}
\newcommand{\CSU}{Colorado State University, Fort Collins, CO, 80523, USA}
\newcommand{\Columbia}{Columbia University, New York, NY, 10027, USA}
\newcommand{\FNAL}{Fermi National Accelerator Laboratory (FNAL), Batavia, IL 60510, USA}
\newcommand{\Granada}{Universidad de Granada, Granada E-18071, Spain}
\newcommand{\Harvard}{Harvard University, Cambridge, MA 02138, USA}
\newcommand{\IIT}{Illinois Institute of Technology (IIT), Chicago, IL 60616, USA}
\newcommand{\KSU}{Kansas State University (KSU), Manhattan, KS, 66506, USA}
\newcommand{\Lancaster}{Lancaster University, Lancaster LA1 4YW, United Kingdom}
\newcommand{\LANL}{Los Alamos National Laboratory (LANL), Los Alamos, NM, 87545, USA}
\newcommand{\Manchester}{The University of Manchester, Manchester M13 9PL, United Kingdom}
\newcommand{\MIT}{Massachusetts Institute of Technology (MIT), Cambridge, MA, 02139, USA}
\newcommand{\Michigan}{University of Michigan, Ann Arbor, MI, 48109, USA}
\newcommand{\Minnesota}{University of Minnesota, Minneapolis, MN, 55455, USA}
\newcommand{\NMSU}{New Mexico State University (NMSU), Las Cruces, NM, 88003, USA}
\newcommand{\Otterbein}{Otterbein University, Westerville, OH, 43081, USA}
\newcommand{\Oxford}{University of Oxford, Oxford OX1 3RH, United Kingdom}
\newcommand{\PNNL}{Pacific Northwest National Laboratory (PNNL), Richland, WA, 99352, USA}
\newcommand{\Pitt}{University of Pittsburgh, Pittsburgh, PA, 15260, USA}
\newcommand{\Rutgers}{Rutgers University, Piscataway, NJ, 08854, USA}
\newcommand{\StMarys}{Saint Mary's University of Minnesota, Winona, MN, 55987, USA}
\newcommand{\SLAC}{SLAC National Accelerator Laboratory, Menlo Park, CA, 94025, USA}
\newcommand{\SDSMT}{South Dakota School of Mines and Technology (SDSMT), Rapid City, SD, 57701, USA}
\newcommand{\Maine}{University of Southern Maine, Portland, ME, 04104, USA}
\newcommand{\Syracuse}{Syracuse University, Syracuse, NY, 13244, USA}
\newcommand{\TelAviv}{Tel Aviv University, Tel Aviv, Israel, 69978}
\newcommand{\Tennessee}{University of Tennessee, Knoxville, TN, 37996, USA}
\newcommand{\UTA}{University of Texas, Arlington, TX, 76019, USA}
\newcommand{\Tufts}{Tufts University, Medford, MA, 02155, USA}
\newcommand{\VTech}{Center for Neutrino Physics, Virginia Tech, Blacksburg, VA, 24061, USA}
\newcommand{\Warwick}{University of Warwick, Coventry CV4 7AL, United Kingdom}
\newcommand{\Yale}{Wright Laboratory, Department of Physics, Yale University, New Haven, CT, 06520, USA}
\newcommand{\listerThanks}{Now at University of Wisconsin--Madison}

\affiliation{\Bern}
\affiliation{\BNL}
\affiliation{\UCSB}
\affiliation{\Cambridge}
\affiliation{\StKates}
\affiliation{\CIEMAT}
\affiliation{\Chicago}
\affiliation{\Cincinnati}
\affiliation{\CSU}
\affiliation{\Columbia}
\affiliation{\FNAL}
\affiliation{\Granada}
\affiliation{\Harvard}
\affiliation{\IIT}
\affiliation{\KSU}
\affiliation{\Lancaster}
\affiliation{\LANL}
\affiliation{\Manchester}
\affiliation{\MIT}
\affiliation{\Michigan}
\affiliation{\Minnesota}
\affiliation{\NMSU}
\affiliation{\Otterbein}
\affiliation{\Oxford}
\affiliation{\PNNL}
\affiliation{\Pitt}
\affiliation{\Rutgers}
\affiliation{\StMarys}
\affiliation{\SLAC}
\affiliation{\SDSMT}
\affiliation{\Maine}
\affiliation{\Syracuse}
\affiliation{\TelAviv}
\affiliation{\Tennessee}
\affiliation{\UTA}
\affiliation{\Tufts}
\affiliation{\VTech}
\affiliation{\Warwick}
\affiliation{\Yale}

\author{P.~Abratenko} \affiliation{\Tufts} 
\author{M.~Alrashed} \affiliation{\KSU}
\author{R.~An} \affiliation{\IIT}
\author{J.~Anthony} \affiliation{\Cambridge}
\author{J.~Asaadi} \affiliation{\UTA}
\author{A.~Ashkenazi} \affiliation{\MIT}
\author{S.~Balasubramanian} \affiliation{\Yale}
\author{B.~Baller} \affiliation{\FNAL}
\author{C.~Barnes} \affiliation{\Michigan}
\author{G.~Barr} \affiliation{\Oxford}
% ...

\collaboration{The MicroBooNE Collaboration}
\thanks{microboone\_info@fnal.gov}\noaffiliation


\date{\today}

\begin{abstract}
We present an analysis of MicroBooNE data with a signature of one muon, no pions, and at least one proton above a momentum threshold of 300~MeV/c (CC0$\pi$Np).
This is the first differential cross section measurement of this topology in neutrino-argon interactions. 
We achieve a significantly lower proton momentum threshold than previous carbon and scintillator-based experiments. 
Using data collected from a total of approximately $1.6 \times 10^{20}$ protons-on-target, we measure the muon neutrino cross section for the CC0$\pi$Np interaction channel in argon at MicroBooNE in the Booster Neutrino Beam which has a mean energy of around 800~MeV.
We present the results from a data sample with estimated efficiency of 29\% and  purity of 76\% as differential cross sections in five reconstructed variables: the muon momentum and polar angle, the leading proton momentum and polar angle, and the muon-proton opening angle. We include smearing matrices that can be used to ``forward-fold'' theoretical predictions for comparison with these data.
We compare the measured differential cross sections to a number of recent theory predictions demonstrating largely good agreement
with this first-ever data set on argon.

\end{abstract}

\maketitle


\section{Introduction}
\label{sec:intro}

A comprehensive understanding of neutrino interactions is one of the core needs of neutrino oscillation experiments~\cite{nustec-review}.  These measurements are 
an important component of systematic uncertainties in both existing neutrino oscillation experiments, such as T2K~\cite{Abe:2018wpn} and NOvA~\cite{Acero:2019ksn}, and future programs and experiments such as SBN~\cite{Antonello:2015lea}, DUNE~\cite{Abi:2020evt}, and Hyper-Kamiokande~\cite{HyperK_design_report}.
In many oscillation analyses, for example ~\cite{Abe:2018wpn,Acero:2019ksn}, a lack of understanding of neutrino interactions is limiting the precision of such measurements.  At this time, the interaction information available is predominantly on light targets such as carbon.
For future experiments, an accurate modeling of neutrino interactions with argon is required; this is a primary goal of the \mub  experiment~\cite{Acciarri:2016smiDET}. We report on the first differential cross-section measurement of CC0$\pi$Np interactions on argon, including measurements of proton kinematics.

The understanding of neutrino interactions comes through cross-section measurements of various channels.
The charged-current quasielastic (CCQE) interaction~\cite{nustec-review} is considered to be very important because it forms a significant contribution in many accelerator-based neutrino oscillation experiments, and because the final state topology is simple with an easily identifiable lepton.
Early experiments on deuterium targets, e.g. \cite{Kitagaki:1983px} were able to identify true CCQE interactions by identifying hadrons in the final state.
These were the first measurements of the axial form factor.
More recent experiments, K2K~\cite{Gran:2006jn}, used nuclear targets and detection of hadrons in the final state with more advanced detectors.
They still focused on the goal of measuring the nucleon axial form factor.
\mb~\cite{miniboone-ccqe} pioneered many of the analysis methods used today.
It is located along the same neutrino beam as \mub, but with a mineral oil (CH$_2$) target.  
The interpretation of these data was complicated because of the presence of other interactions such as multinucleon ($2p2h$) interactions~\cite{Martini:2011wp,Nieves:2011ppTHEORY} where the primary interaction is with two nucleons, and pion production where the pion is absorbed in the residual nucleus.
These data provided evidence for the importance of  the $2p2h$ interaction in neutrino interactions.
 Events from these alternate mechanisms have different proton multiplicities and kinematic distributions compared with CCQE events. When only the muon is detected, the event can be easily mistaken as a CCQE interaction leading to a bias in neutrino energy estimations.

To avoid this problem, a common signal definition used is CC$0\pi$ or ``CCQE-like'' where the final state has one muon and any number of protons but no pions above the detection threshold of the experiment.
Components of $2p2h$ and pion production, followed by pion absorption, are then included.
As a result of using a broader signal definition, backgrounds are easier to handle and the associated model dependence in the result is greatly decreased.

Recent \minerva~\cite{Betancourt:2017uso,Ruterbories:2018gub} and T2K~\cite{Abe:2018pwo} CC$0\pi$ results use this signal definition and include events where protons are required as one component of the signal to better differentiate between models. 
One \minerva measurement~\cite{Betancourt:2017uso} using a range of targets (carbon, iron, and lead) showed growing problems describing the magnitude of the data with increasing atomic number.
Along with the CC$0\pi$ cross section measurement, T2K published proton momenta and multiplicity distributions in their most recent paper~\cite{Abe:2018pwo}.
Each experiment has a characteristic proton detection threshold: 450 MeV/c (kinetic energy of 102.3 MeV) for \minerva~\cite{Betancourt:2017uso,Ruterbories:2018gub}, and 500 MeV/c (kinetic energy of 124.9 MeV) for T2K~\cite{Abe:2018pwo}.
A recent MicroBooNE measurement focused on single proton final states in a region of phase space where CCQE is expected to dominate \cite{uB_CCQE_2020}.  The largest differences between data and predictions were seen at forward muon angles.

Pion production interaction events are also included in the event sample for this measurement.  
Both experimental and theoretical understanding of pion production processes are needed~\cite{nustec-review}.
In addition, the component of these events that satisfy the signal definition of this measurement is not well understood because models of both pion production and pion absorption in the nuclear medium are required.

Theoretical development has benefited from previous work for electron interactions where many of the same reaction mechanisms are used.
The $2p2h$ mechanism was developed for electron interaction modeling~\cite{Megias:2016fjk} and then imported to neutrino models~\cite{Martini:2011wp}. 
Although all event generator Monte Carlo algorithms now include $2p2h$ mechanisms, neutrino data give only indirect evidence for it, in contrast with electron scattering where the evidence is more conclusive. 
Relevant neutrino data were published by ArgoNeuT including kinematics for a two-proton sample~\cite{Argoneut-2p}. 
Because their sample size is small, they could select and analyze events through a combined manual and automated analysis that enabled an impressively low threshold of 21 MeV in proton kinetic energy.
However, there is still a strong need for more detailed information about the protons in the final state of neutrino interactions.

This article presents an analysis of a sample of charged-current events with one muon and at least one proton in the final state in argon.
Measuring the outgoing proton increases the sensitivity to nuclear effects relative to a measurement of inclusive muon kinematics such as~\cite{Adams:2019iqcINCL} while keeping a more inclusive signal definition than the aforementioned analysis of one-proton final states~\cite{uB_CCQE_2020} retains a higher statistics data sample. According to the signal definition adopted, the highest energy (leading) proton must have a momentum between 300~MeV/c and 1200~MeV/c (
see Sec.~\ref{sec:event_sel} for details), the muon must have a momentum greater than 100~MeV/c, and there must be no pions or other mesons in the final state.
Any number of final state neutrons is permitted.
We refer to this signal definition as CC$0\pi Np$ (where N $\ge$ 1) for the remainder of this article.
Events from this CC$0\pi Np$ signal definition are primarily populated by CC quasi-elastic interactions, but with significant components from multinucleon interactions ($2p2h$) and events where pions are produced but then absorbed in the nucleus.
These different components have different signatures in the five kinematic variables we measure, and as such, these data can be used to build and test models in interaction generators.

The cross-sections presented here are measured differentially in the kinematics of the muon and leading proton in each event.  
In addition to the muon momentum and angle, measured distributions of the leading proton momentum and angle, and opening angle between the muon and leading proton are presented.  
By presenting these spectra for CC0$\pi$Np events, a broad picture of muon neutrino interactions in argon is provided and model dependence in these results is decreased.
To best describe these data, comparisons need to include all contributing mechanisms listed in the signal definition (in the preceding paragraph) and should be folded with the smearing matrices provided because the data are not corrected for detector resolution effects.
A breakdown according to the interactions implemented in the Monte Carlo program used in this data analysis and comparisons with various event generator codes is presented in Sec~\ref{sec:results}.

\section{\mub Experiment} 
\label{sec:expt}
\begin{figure}
  \centering
  \includegraphics[width=0.45\textwidth]{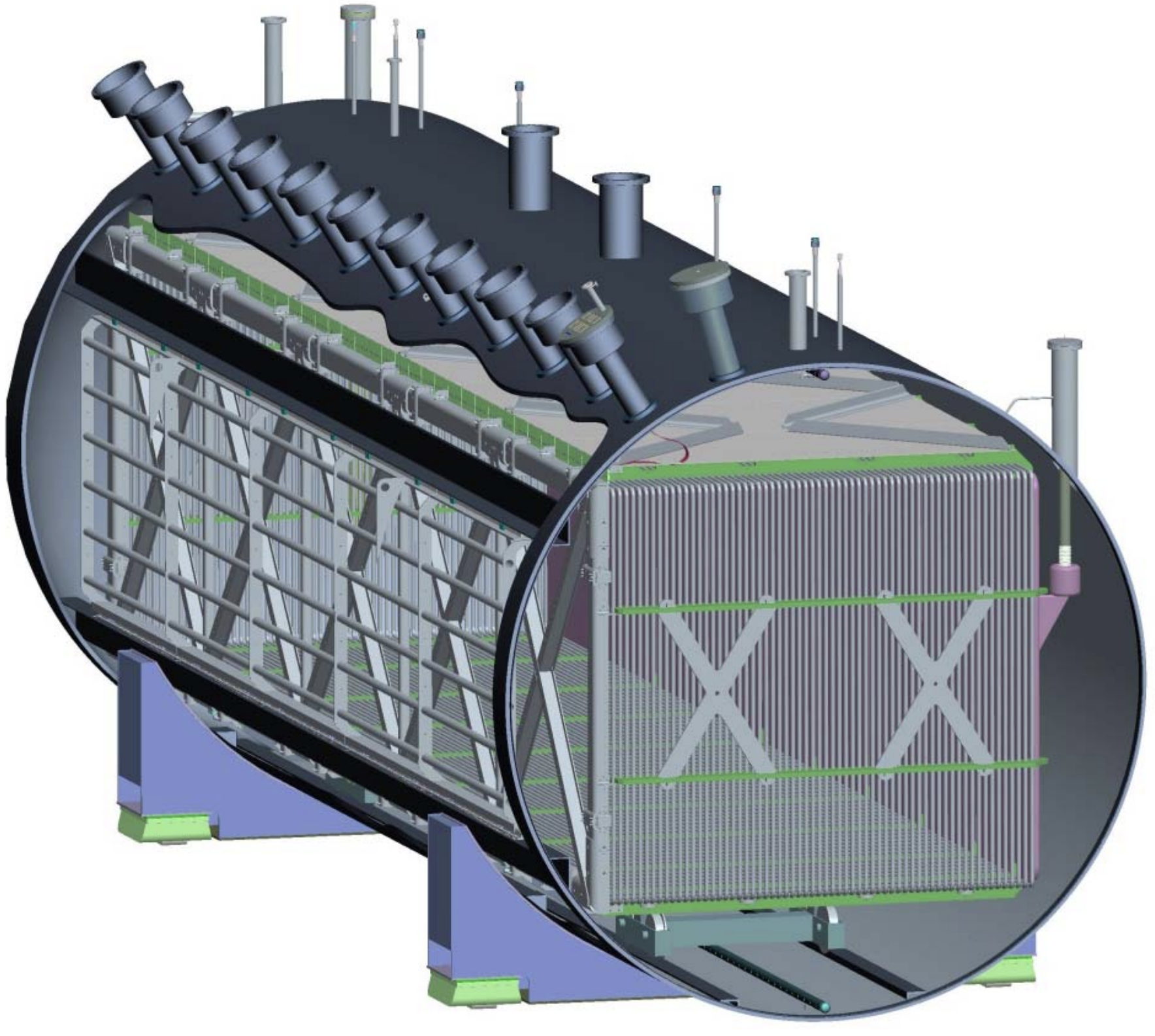}
  \caption{A schematic drawing of the MicroBooNE LArTPC as installed inside cryostat. }
  \label{fig:microboone_TPC}
\end{figure}

The \mub experiment~\cite{Acciarri:2016smiDET} consists of a liquid argon time projection chamber (LArTPC) in the Fermilab Booster Neutrino Beam (BNB). The detector consists of a cylindrical cryostat filled with approxiamately 170 tons of liquid argon.
Inside this cryostat is a 10.36~m (L) $\times$ 2.56~m (W) $\times$ 2.32~m (H) rectangular TPC, shown in Figure \ref{fig:microboone_TPC}, which is sensitive to charge produced in 85 tons of the liquid.
The TPC operates at an electric field of 273~V/cm, provided by a cathode held at $-$70~kV and kept uniform by a field cage around the TPC, though the local electric field is modified by up to 15\% by the presence of positive ions in the detector, known as the space charge effect~\cite{uBUVLaser2019, uBcosmicSCE_2020}.
Ionization electrons drift in this electric field towards three planes of wires forming the anode.
It takes 2.3~ms for an electron to drift from the cathode to the first anode plane.
The innermost two planes of wires are angled at $\rpm$60$^{\circ}$ from the vertical and detect induced signals from electrons as they drift past the wire planes.
The final plane has vertical wires that collect drifting ionization electrons.
In total there are 8192 wires with a separation of 3~mm between any two adjacent wires and between each wire plane. The detector coordinate system is defined with the TPC electron drift direction oriented in the negative $x$-direction, $y$-direction vertically aligned, and the $z$-direction parallel to the neutrino beam. The coordinate system origin is at the upstream edge of the anode wires and equidistant between the top and bottom field cage, and the axes form a right-handed set. We also define the polar angle from the $z$-axis, $\theta$, and the azimuthal angle around the $z$-axis, $\phi$.

Behind the anode plane is an array of 32 8-inch Hamamatsu PMTs.
In front of each PMT is an acrylic disc coated in tetraphenyl butadiene (TPB).
The 128~nm scintillation light produced by argon excited by charged particles is shifted to visible wavelengths by this TPB coating, allowing detection by the PMTs.
The PMTs have a time resolution of a few nanoseconds which is several orders of magnitude smaller than the TPC drift time.
The timing information from the PMTs is used initially to provide a trigger for data collection, and later the PMT signals are associated with TPC activity to further reject cosmic-ray background.

Data collection begins with a hardware trigger, which comes either from the Fermilab accelerator complex or from a function generator in the detector's electronics racks (known as the ``external trigger'' and used for cosmic background estimation).
Accelerator signals veto the external triggers so there is a one-to-one match between accelerator spills and MicroBooNE beam triggers during data taking.
Once a hardware trigger is received, the data acquisition system reads in the PMT data and determines if there is light consistent with the presence of a neutrino interaction during the 1.6~$\mu$s neutrino beam spill, or an equivalent time window for external triggers.
Data is only saved if sufficient light is observed within this window.
This PMT trigger algorithm is estimated to be over 99.9\% efficient for the signal definition used in this analysis and reduces cosmic backgrounds by a factor of 10.
The PMT and TPC data are recorded for 1.6~ms prior to the trigger and 3.2~ms after the trigger, though the TPC data are later truncated to include 400~$\mu$s of time before the neutrino arrival and another 400~$\mu$s after the last possible time drifting electrons from beam interactions can arrive at the anode in order to reduce the data processing time.

The BNB operates at an average repetition rate of 5~Hz with approximately $4 \times 10^{12}$ protons in each 1.6~$\mu$s spill.
The protons exit the Fermilab Booster accelerator at an energy of 8~GeV where they impinge on an air-cooled beryllium target.
The resulting mesons are focused by a single magnetic horn and directed into a 50~m long decay pipe.
Decays of these focused mesons, as well as secondary decays of muons produced in meson decays, produce a neutrino beam with a broad energy spectrum with a mean of around 800~MeV and a long tail at higher energies.
The beam is over 99\% muon-flavor neutrinos, with around 10\% of the 99\% being muon antineutrinos.
The data used in this analysis were collected between February and July 2016, totalling 1.6$\times 10^{20}$ protons-on-target.
A total of 72 million external triggers are used for cosmic background estimation --- approximately double the number of beam triggers collected.

The MicroBooNE detector is 470~m from the beam target and slightly below the surface in the open-cut pit of the LAr Test Facility building.
There is no substantial shielding above the detector and, for the data used in this analysis, there was no external cosmic tagger though one has since been installed~\cite{Adams:2019bztCOSMICTAGGER}.

\section{Experiment Simulation} 
\label{sec:expt_sim}
Monte Carlo simulation is essential to provide accurate modeling of efficiency, resolution, backgrounds, and systematic uncertainties.  The \mub collaboration has developed a full suite of simulations for beam, detector, and material surrounding the detector.  The neutrino beam properties are simulated by a GEANT4-based algorithm~\cite{Allison:2016lfl} developed by MiniBooNE~\cite{AAAA_2009_MBflux}, which is valid for all Fermilab experiments on the BNB beam line.  
Neutrino interactions in the fiducial volume and all surrounding material are simulated by GENIE~\cite{Andreopoulos:2009rq} v2.12.4 (called ``GENIE v2'' in this article).  Propagation of particles through the detector volume are handled by GEANT4~\cite{Wright:2015xia}.
The software framework, LArSoft~\cite{Snider:2017wjd}, is used to simulate the detector response including light production and propagation, charge production and drift, and wire response, as well as the electronics response and digitization.
All these software packages are in common use for LArTPC neutrino experiments.

\subsection{Event Generation} 
\label{sec:evgen}
The GENIE generator simulates the interaction of all neutrino flavors at a wide variety of energies with all stable nuclei.  
GENIE is in common use among many neutrino experiments and has been well validated.
At the neutrino energies applicable to this experiment ($E_{\nu} <2\ \mathrm{GeV}$), the dominant interactions are through quasielastic (QE),  multi-nucleon processes (called by the general term $2p2h$, or two particle-two hole processes here), and pion production processes (with (RES) or without (NONRES) nucleon resonances).  Coherent pion production is included in the generator codes but doesn't contribute to the simulated samples for this measurement.
Interactions can occur via charged-current (CC) and neutral-current (NC) processes.

Quasielastic interactions are particularly sensitive to the nuclear model used.  The relativistic Fermi gas model with a high momentum nucleon-nucleon correlation tail~\cite{Bodek:1980ar} is used for all nuclei.
A fixed-value binding energy (29.5 MeV for argon) is applied.
Interaction with a single nucleon produces a lepton and a single nucleon which is then propagated through the residual nucleus.
The Llewellyn-Smith model~\cite{LlewellynSmith:1971zm} with \maqe=1.04 $\mathrm{GeV}$ is used for CC and NC interactions.
$2p2h$ processes involve interactions with two nucleons, producing two nucleons at the initial interaction vertex which are both then subject to final state interactions (FSI).
A component of these processes is also known as meson exchange currents (MEC).
The version of GENIE used in simulation uses an empirical $2p2h$ model~\cite{Katori:2013eoa} with parameters fit to \mb data~\cite{miniboone-ccqe}.  Nucleon resonance production is governed by the Rein-Sehgal model~\cite{Rein:1980wg} which includes a wide range of short-lived nucleon excited states.  The $\Delta(1232)$ state is the most important nucleon resonance for this experiment.
Small contributions from coherent pion production~\cite{Rein:2007}, non-resonant pion production, and deep inelastic scattering (DIS)~\cite{Bodek:2002ps} processes are also relevant.

All hadrons produced during event generation in GENIE are subject to FSI.  GENIE v2 uses a data-driven empirical model that is tuned to hadron-nucleus data~\cite{Dytman:2011zz}.  

Overlaid cosmic rays in the same detector readout as the neutrino interaction are simulated with CORSIKA~\cite{1998cmcc}, but cosmic backgrounds where there is no neutrino interaction in the detector are measured using data collected in between beam spills (``external'' triggers).

Particles produced by either GENIE or CORSIKA are transported through the detector by GEANT4.

\subsection{Detector Simulation}
\label{sec:detsim}

The simulation of the MicroBooNE detector uses the LArSoft framework.
This simulates the production of ionization charge and scintillation photons, followed by their transport through the detector.
Where possible, data-driven techniques are used to constrain the detector response including position-dependent wire responses~\cite{Adams:2018gbiPROCESS}, non-responsive channels\cite{Acciarri:2017sdeNOISE}, and the effect of a non-uniform electric field due to the build up of positive ions in the detector volume~\cite{uBUVLaser2019, uBcosmicSCE_2020}.
The production of signals on the wires uses a simple model whereby charge is assumed to either induce currents or collect on the nearest wire.  Known deficiencies in this model are considered as part of the uncertainty in detector modeling.

\section{Event Reconstruction} \label{sec:reconstruction}

\subsection{TPC Reconstruction}
The first step in event reconstruction is noise-filtering~\cite{Acciarri:2017sdeNOISE} and deconvolution~\cite{Adams:2018gbiPROCESS} of wire signals.
In this step the intrinsic response of the wires and electronics is deconvolved from the raw signals and wire waveforms to become a series of Gaussian peaks.
A threshold is applied to remove residual noise and the peaks are fit with Gaussian curves to form ``hits''.
These hits lie in a 2-dimensional plane corresponding to wire location and arrival time, with the position along the wire unknown.
Based on the known drift velocity and assuming hits originate from a neutrino interaction (which occurs in a known very narrow time window immediately after the event trigger), the position in the $x$-direction (corresponding to the drift direction) can be calculated.
At this point, hits are allowed to have unphysical x-positions (i.e. outside of the detector boundaries) which will be used later to tag cosmic activity.

The Pandora software package~\cite{Acciarri:2017hatPANDORA} is used for all further steps of the TPC reconstruction.
Hits are grouped into clusters, separately on each plane, based on proximity in time and space.
These clusters can be matched across planes using time information and knowledge of wire crossing points.

Once TPC clusters have been matched across planes, it is possible to fit particle trajectories through them, giving each 2D hit a 3D position in the detector.
Pandora contains distinct algorithms for fitting clusters that it deems track-like and shower-like.
For this analysis we do not rely on these algorithms, but consider all clusters to be track-like and rely on particle identification methods to remove showering particles such as electrons and photons.

Reconstructed tracks are not perfectly straight, but follow slightly curved paths due to multiple Coulomb scatterings (MCS).
The fact that particles undergo multiple scattering and that this effect is largest at low momentum is used in two ways in this analysis.
By comparing the average angular deflection observed as a function of position along the track to a prediction for a given particle momentum and direction, a likelihood can be assigned for that hypothesis.
By selecting the momentum value with the maximumum likelihood, we achieve a momentum resolution from MCS as low as a few percent for contained muons and up to 15 percent for exiting muons (dependent on the initial momentum and contained length)~\cite{Abratenko:2017nkiMCS}.
Additionally, by comparing the likelihood for both directions along the track, it is possible to differentiate between tracks that enter from outside and stop in the detector from those produced in the detector and then exit.

For contained tracks, particle identification is applied and the momentum is estimated from the total length of the track path (in general longer than the start-end distance) based on the continuous slowing down approximation (CSDA).
Using range, the proton momentum resolution is around 60~MeV/c across all measurement bins in this analysis.
For contained muons, the momentum resolution from range information alone is around 10\% below 0.3~GeV/c, dropping to below 5\% above 1~GeV/c.
For both protons and muons, the momentum reconstructed from range is more likely to be underestimated than overestimated, as tracking algorithms sometimes stop before the end of a track but rarely continue past the end.
For exiting muons where the momentum is reconstructed based on MCS, the resolution is more symmetric.

\subsection{Light Reconstruction}
The waveforms from each PMT are first formed into optical hits using a simple threshold algorithm.
Hits from many PMTs are combined to form ``flashes'' by integrating over a fixed time window of 8~$\mu$s.
The integration window is intended to be long enough to capture the both the fast and slow components of argon scintillation light.
Each flash is primarily characterised by the number of photoelectrons observed by each PMT and a start time.
As the PMTs form a 2D plane, PMTs each have a z-position and a y-position and flashes have a mean value and width in each of those dimensions calcualted from the PE-weighted mean and RMS position of the PMTs that register a response.

Because the flash time integral is longer than the beam spill, only one flash can be observed within the time window associated with the neutrino beam and this is referred to as the ``beam flash''.

\subsection{Charge-light Matching} \label{subsec:TPCPMTmatching}
Due to the long drift length, the low drift velocity in liquid argon, and the location of the detector on the Earth's surface, many cosmic-induced muons are observed within any readout of the TPC.
When a neutrino interaction occurs in the detector, the event activity is seen in conjunction with cosmic muons in the TPC data.
By using the fast timing information provided by the PMTs, the time of any activity can be reconstructed much more precisely to eliminate these cosmic backgrounds.
In order to do this, TPC activity must be matched to the PMT signals.
This matching is used in multiple places in this analysis.

Given a collection of TPC tracks which are believed to have originated from the same initial interaction (and thus were produced at the same time), a prediction can be made for how much light would be observed by each PMT.
This hypothesis for the number of photoelectrons seen by PMT $i$ is denoted $H_{i}$.
Although neutrinos all interact at approximately the trigger time, cosmic backgrounds can occur at any time.
It is therefore not safe to assume that the $x$-position of the tracks is correct. As the $x$-position impacts the predicted flash, such prediction is calculated for several $x$ positions: $H_i = H_i(x)$.

This prediction can be used to produce a ``matching likelihood'' between an observed and predicted flash given by:
\begin{equation}
-LL(x) = - \sum_{i=1}^{32} \ln \left (  P(O_i| H_i(x))  \right ),
\end{equation}
where $O_i$ is the PE measurement for PMT $i$, and $P$ is a Poisson pdf with parameter $H_i(x)$ evaluated at $O_i$.

After the minimization, a minimum point $x_\text{min}$ gives the position of the TPC object along $x$ and the collection with the smaller value of $-LL(x_\text{min})$ is identified as the neutrino candidate.

\begin{figure}
  \centering
  \includegraphics[width=0.45\textwidth]{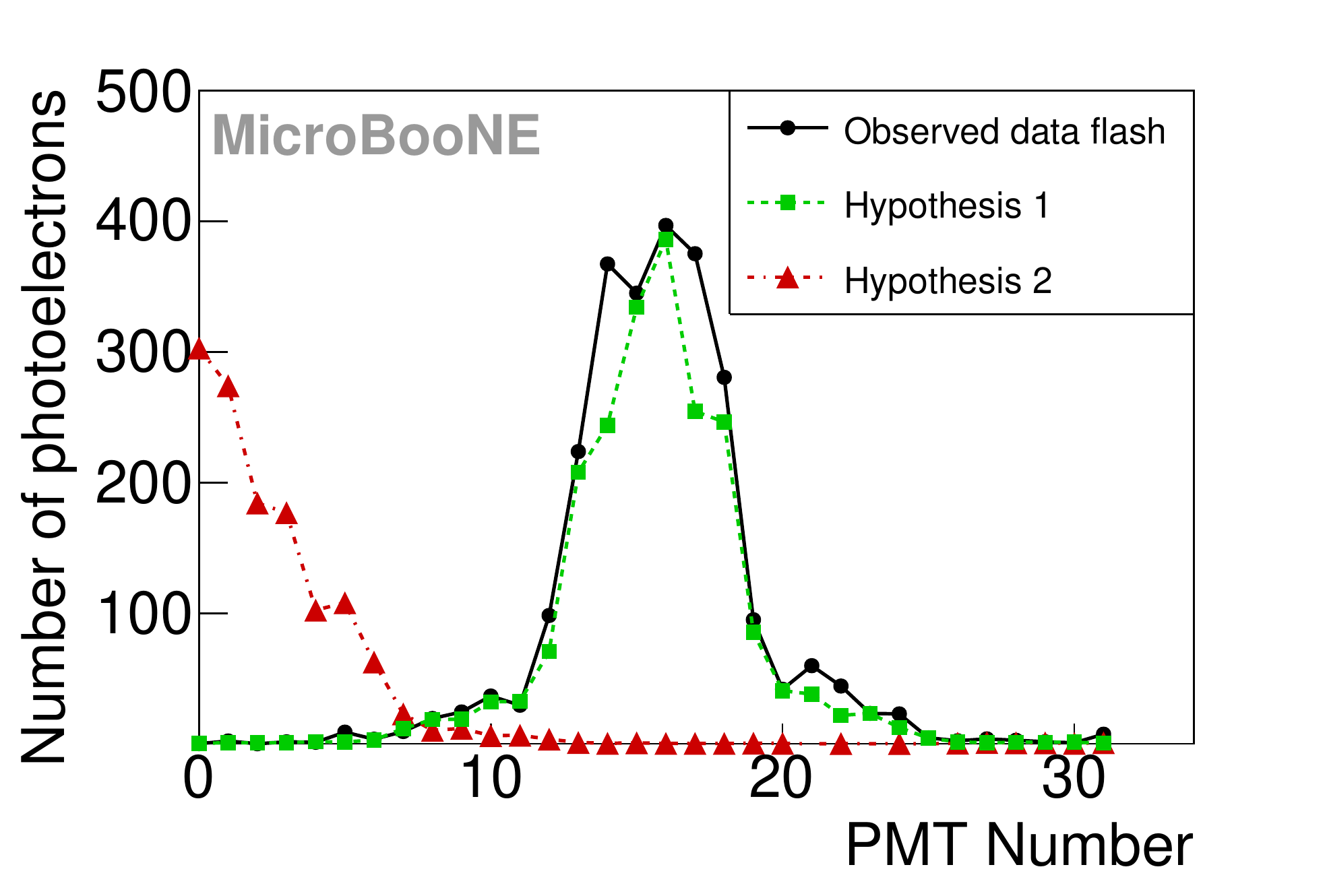}
  \caption{The observed flash photoelectrons (PE) distribution over the 32 PMTs in one data event and two hypothesis distributions from candidate TPC clusters.  The first hypothesis (shown in green) is selected as the best match to the observed flash. The PMT $z$-position generally increases with increasing PMT number.}
  \label{fig:flashmatch}
\end{figure}

Figure~\ref{fig:flashmatch} shows the observed flash and two hypotheses from one event.  In this case one hypothesis is clearly a much better match than the other.

\subsection{Cosmic Rejection}\label{subsec:cosmicrej}
Pandora operates in two ``passes''.
The first pass efficiently reconstructs cosmic-induced tracks, then tags the hits in those clusters that are identified as highly likely to be cosmic-induced, such that they can be ignored by the second pass which attempts complete reconstruction of a neutrino interaction.
A number of algorithms are used for this cosmic tagging.

Firstly, any cluster which contains hits reconstructed outside of the TPC boundaries in the drift direction is assumed to be cosmic-induced as it must have passed through the detector at a different time than the neutrinos.
Additionally, clusters are tagged as cosmic-induced if they come within 30~cm of both the top and bottom faces of the TPC.

The algorithms in section \ref{subsec:TPCPMTmatching} form the basis for optical-based cosmic rejection.
Clusters are removed if they are clearly inconsistent with the beam flash.  This means the distance from the hypothesis mean $z$-position to the measured $z$-position must be larger than the flash width, and at least one PMT must have a predicted number of PEs more than 3$\sigma$ above the measured PE response.

Then, a specific search is made for clusters which appear to pierce either the anode or cathode plane and either the top or bottom face of the TPC.
Clusters are moved in $x$ until one end touches either the anode or cathode, and based on this $x$-position and the known drift velocity we reconstruct the time that the particle must have traversed the TPC.
If there is a flash observed within 7~$\mu$s of this time that is not the beam flash, the identified flash $z$-position is consistent with the track's $z$-position, and the geometry is consistent with a downwards-going particle, the cluster is tagged as cosmic-induced.

Tracks that enter through the top and stop in the detector are tagged as cosmic-induced by tailored searches for a decay (Michel) electron or a Bragg peak at the contained end of the track.

Finally, for muon candidates that cross one detector boundary, and could therefore be assumed to be either entering or exiting, we perform a multiple-scattering fit for the muon momentum under both the assumption that it originates in the detector and exits, and the assumption that it originates outside the detector but stops in the detector.
If the likelihood for the incoming case is significantly better than the likelihood for the outgoing case, the track is tagged as cosmic-induced.

\subsection{Neutrino Reconstruction} 
The second reconstruction pass of Pandora is tailored to reconstruct neutrino interaction topologies as precisely as possible.
This second pass begins again from clustering hits but with a reduced set of hits that have not been tagged as belonging to cosmics in the first pass.
Clusters are then matched across planes and tracks formed from them.
Tracks that have a start or end point in close proximity to others are clustered together as neutrino interaction candidates with specific algorithms fitting the exact vertex position close to the point at which the tracks meet.

Shower-like clusters that point back to the initial neutrino vertex and could be from a neutral pion decay are grouped with the neutrino interaction, although then for this analysis they are fit as tracks.

\subsection{Particle Identification}

Particles traversing liquid argon lose energy in characteristic ways.
Electrons and photons form electromagnetic showers, while muons, charged pions, protons, and kaons all tend to move in approximately straight lines, losing energy primarily through ionization according to the Bethe-Bloch formula.
Particles that stop in the detector leave a characteristic Bragg peak that allows the identification of the particle species.
In MicroBooNE, neutrino interactions rarely produce kaons or particles heavier than protons that have ranges above the minimum length for tracking (approximately 6~mm).
Additionally, the resolution at which the energy loss can be measured is not sufficient to distinguish muons and pions with their very similar masses.
Because of this, particle identification in this analysis is reduced to determining whether a track is a proton or not.

\begin{figure}[ht!]
    \includegraphics[width=0.45\textwidth]{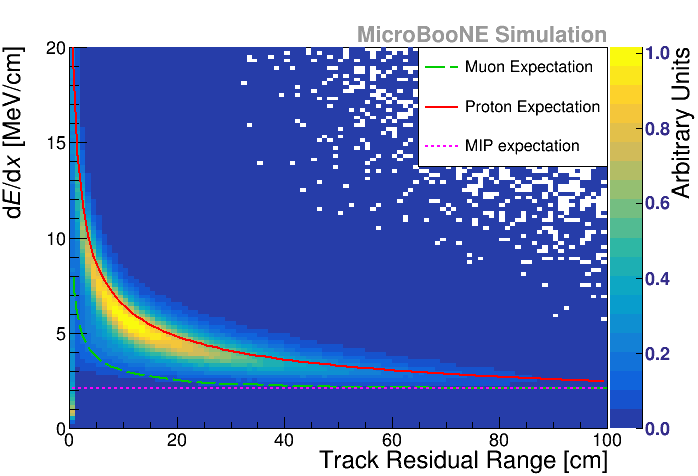}
    \caption{Scatter plot of $dE/dx$ vs. residual range in Monte Carlo events for non-muon tracks after proton identification cuts.  Curves for the mean expectation for protons, muons, and minimum ionizing particles are shown overlaid.}
   \label{fig:PID_2D}
\end{figure}

To calculate the energy loss as a function of track position, we take the hit charge and divide by the 3D distance between hits.
This charge is calibrated to account for varying wire response, electron attenuation from impurities in the drift region, and the impact of space charge~\cite{Adams:2019ssgCAL}.
Using the modified Box model - with parameters fit to MicroBooNE data~\cite{Adams:2019ssgCAL} - recovers the energy loss per unit length, $dE/dx$, at that position.
Using the Bethe-Bloch formula to predict the $dE/dx$ for a proton as a function of the residual range, or distance from the stopping point, we construct a discriminator, $\text{PID}_\text{prot}$, by comparing measurements over the last 30~cm of the track to this prediction (or over the full track, if less than 30~cm in length).
Low values of $\text{PID}_\text{prot}$ are proton-like.
In Fig.~\ref{fig:PID_2D}, a scatter plot of $dE/dx$ vs. residual range for selected proton tracks is shown for Monte Carlo events.
For most hits from proton candidates, good agreement with the expectation is seen for protons.
Some disagreement is visible at very low residual range.
This is because $dE/dx$ estimation is unreliable for the last hit on a track due to the uncertainty on the exact stopping point of the particle --- these hits are neglected when computing $\text{PID}_\text{prot}$.

\begin{figure}[ht!]
       \includegraphics[width=0.45\textwidth]{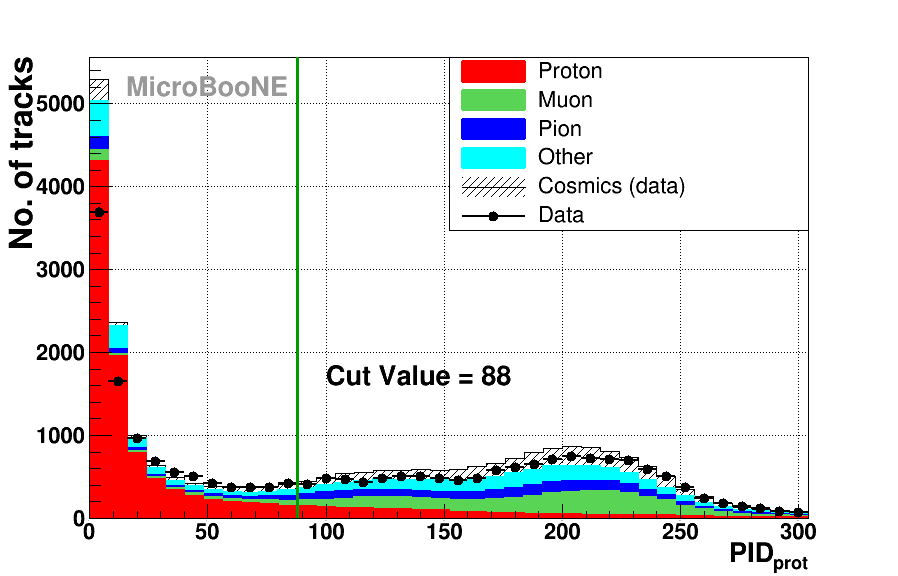}
    \caption{Values of $\text{PID}_\text{prot}$ for data and Monte Carlo events.  The Monte Carlo events are subdivided according to particle type and have been scaled to the data exposure of $1.6\times10^{20}$ P.O.T.} 
\label{fig:pid_prot}
\end{figure}

Values of $\text{PID}_\text{prot}$ for data and Monte Carlo are shown in Fig.~\ref{fig:pid_prot}.
Tracks with $\text{PID}_\text{prot}<88$ are accepted as protons.
The charge response has a wider distribution in data than simulation, due to the impact of induced charge on the collection wires which is not modeled (but is accounted for in detector uncertainties which is discussed in Sec.~\ref{sec:syst}).
This leads to a wider distribution of $\text{PID}_\text{prot}$ in data at very low values.
The impact of this effect is small near the cut value.

\FloatBarrier

\subsection{Particle Thresholds}
\label{sec:pid_thesholds}

The Pandora reconstruction package is able to form tracks from particles that produce hits on only a few wires, meaning they can have ranges of less than 1~cm.
In order to ensure accurate particle identification, we require at least 5 hits on the collection plane.
This leads to an absolute minimum track length of 1.2~cm.

For protons that do not travel parallel to the z-axis, the length required to produce 5 hits increases. Based on this, it is reasonable to expect that a 2~cm proton track can be reconstructed and identified. This leads to an expected momentum threshold of around 300~MeV/c. The efficiency for proton identification as a function of momentum
and the subsequent signal thresholds are discussed in Section~\ref{sec:event_sel}.

For muons and pions, detection thresholds are significantly lower --- around 30~MeV/c.  For this analysis, the ability to reconstruct and positively identify these particles is primarily driven by choices made in the event selection to reject backgrounds, rather than fundamental limitations of the detector, as discussed in Sec.~\ref{sec:event_sel}.

\section{Event Selection} 
\subsection{Summary of Event Selection}
\label{sec:event_sel}

There are three primary classes of background that this event selection strives to eliminate, while retaining efficiency for the  CC0$\pi$Np events of interest.
The largest two backgrounds are cosmic-induced --- a combination of events where there is a neutrino interaction in the cryostat but cosmic-induced tracks are selected instead (labeled as ``Cosmics (Overlay)'' in the figures) and events where there are only cosmic-induced tracks and no neutrino interaction in the cryostat (labeled as ``Cosmics (Data)'' in the figures).
The final background class is neutrino-induced backgrounds (labeled as ``$\nu$-induced Background'' in the figures), which consist of neutral current interactions, wrong-sign (antineutrino) interactions, interactions in the active detector but outside of the fiducial volume, and muon neutrino interactions that either do not produce a proton above threshold or that produce additional particles such as pions.

We begin with an inclusive selection of $\nu_{\mu}$ charged-current neutrino interactions~\cite{Adams:2019iqcINCL}.
This selection searches for interactions in a relatively restrictive fiducial volume (to mitigate the effects of unresponsive wires and space-charge distortions close to boundaries).
After the cosmic rejection and neutrino reconstruction algorithms described in Sec.~\ref{sec:reconstruction}, there are in general still several neutrino candidates in any event, most of which are of cosmic origin.
To select the best neutrino candidate from these, we again utilize the TPC-PMT matching.
The matching likelihood with the beam flash is calculated for all neutrino candidates and the candidate with the highest matching likelihood is selected.
The beam flash is required to have a magnitude of at least 50~PE, and a veto is placed on the presence of more than 20~PE of optical activity integrated over the 2.0~$\mu$s before the beam spill - a requirement that rejects michel electrons from cosmics immediately before the beam.

The selected neutrino candidate is then verified for additional consistency with the beam flash by requiring that the flash-matching best likelihood doesn't occur far away from the x-position assumed for a neutrino interaction.
Additionally, the hypothesis flash position in z must be within 75~cm of the measured flash position.

This selection then classifies the muon candidate as the longest track in the set of tracks associated with the neutrino interaction candidate and applies additional quality checks on this track candidate, such as requiring that the charge deposition along the track is consistent with a mimimally ionising particle and that the spatial distribution of hits is consistent with a track-like, rather than shower-like, particle.
Finally, for contained muon candidates, the momentum reconstructed from range and from a multiple scattering fit are required to be within 0.2~GeV/c of one another.

For the further selection of CC$0\pi Np$ events from this inclusive pre-selection, we retain the same muon candidate (the longest track) and then consider all other tracks to be proton candidates.
All proton candidates must be at least 10~cm from of the edge of the TPC in order to identify them as protons and measure their momenta.

\begin{figure}[h]
  \centering
  \includegraphics[width=0.45\textwidth]{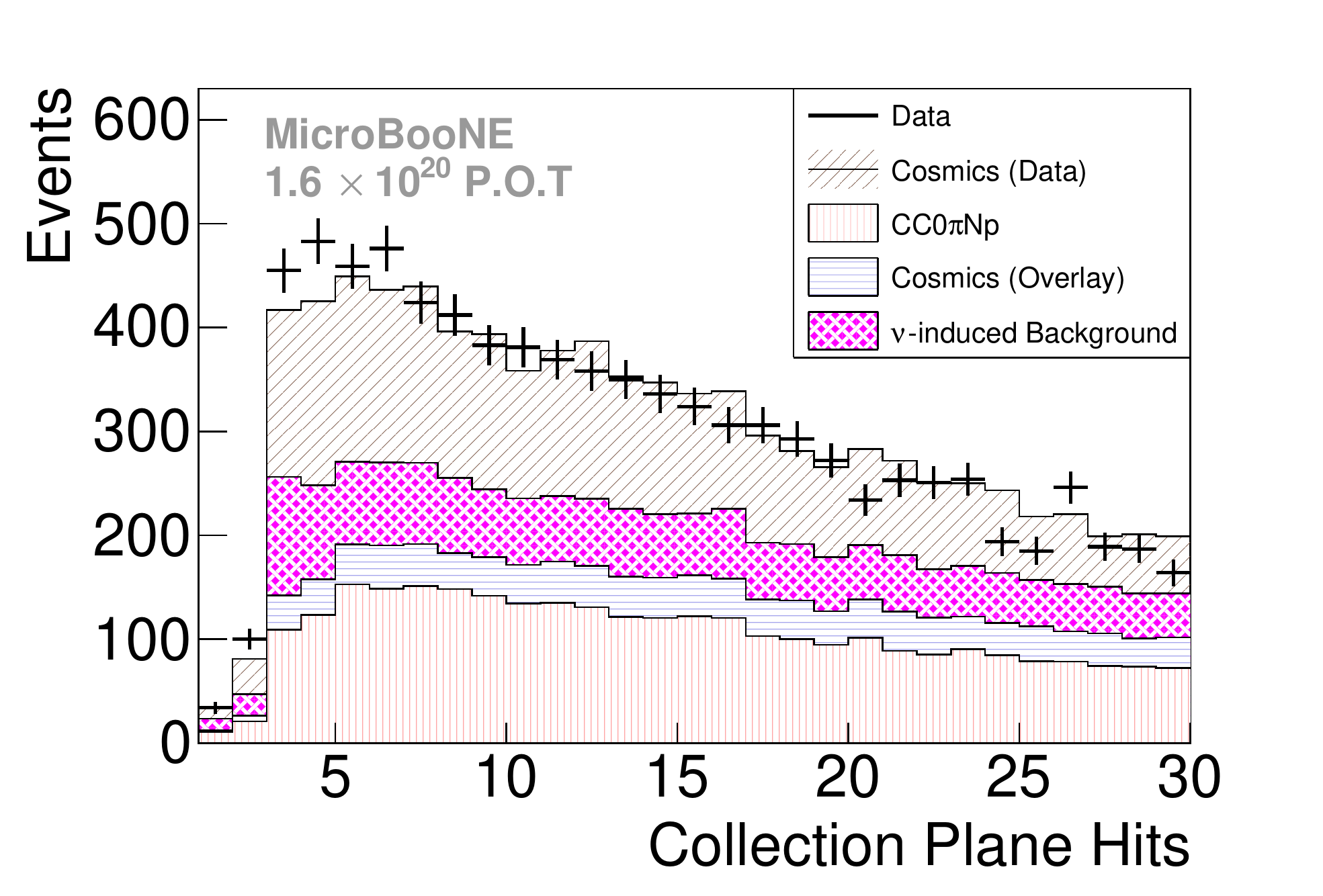}
  \caption{The number of collection-plane hits on the leading proton candidate, for contained proton candidates after the pre-selection with no PID requirement.  Tracks with no collection plane information are not included. The distribution is truncated at 30 for clarity and the Monte Carlo events have been scaled to the data exposure of $1.6\times10^{20}$ P.O.T.}
  \label{fig:NhitsBeforeThresh}
\end{figure}

The longest proton-candidate track is required to have at least 5 hits on the collection plane and a proton-like value of $\text{PID}_\text{prot} < 88$. Figure \ref{fig:NhitsBeforeThresh} shows the distribution of the number of collection plane hits for proton candidates before these two requirements. It is then required to be reconstructed with a range-based momentum above the threshold of 300~MeV/c.

Secondary proton candidates are then considered.
The PID method employed becomes less reliable with fewer than 5 hits, so any additional proton candidate with fewer than 5 hits on the collection plane cannot be positively identified as a proton. 
Due to the tiny number of particles that make such short tracks at the vertex, these tracks are overwhelmingly produced by protons.
For this reason, secondary proton candidates with fewer than 5 collection plane hits are assumed to be protons while those with 5 or more collection plane hits are required to have a proton-like value of $\text{PID}_\text{prot}$.

\begin{figure}
  \centering
   \includegraphics[width=0.45\textwidth]{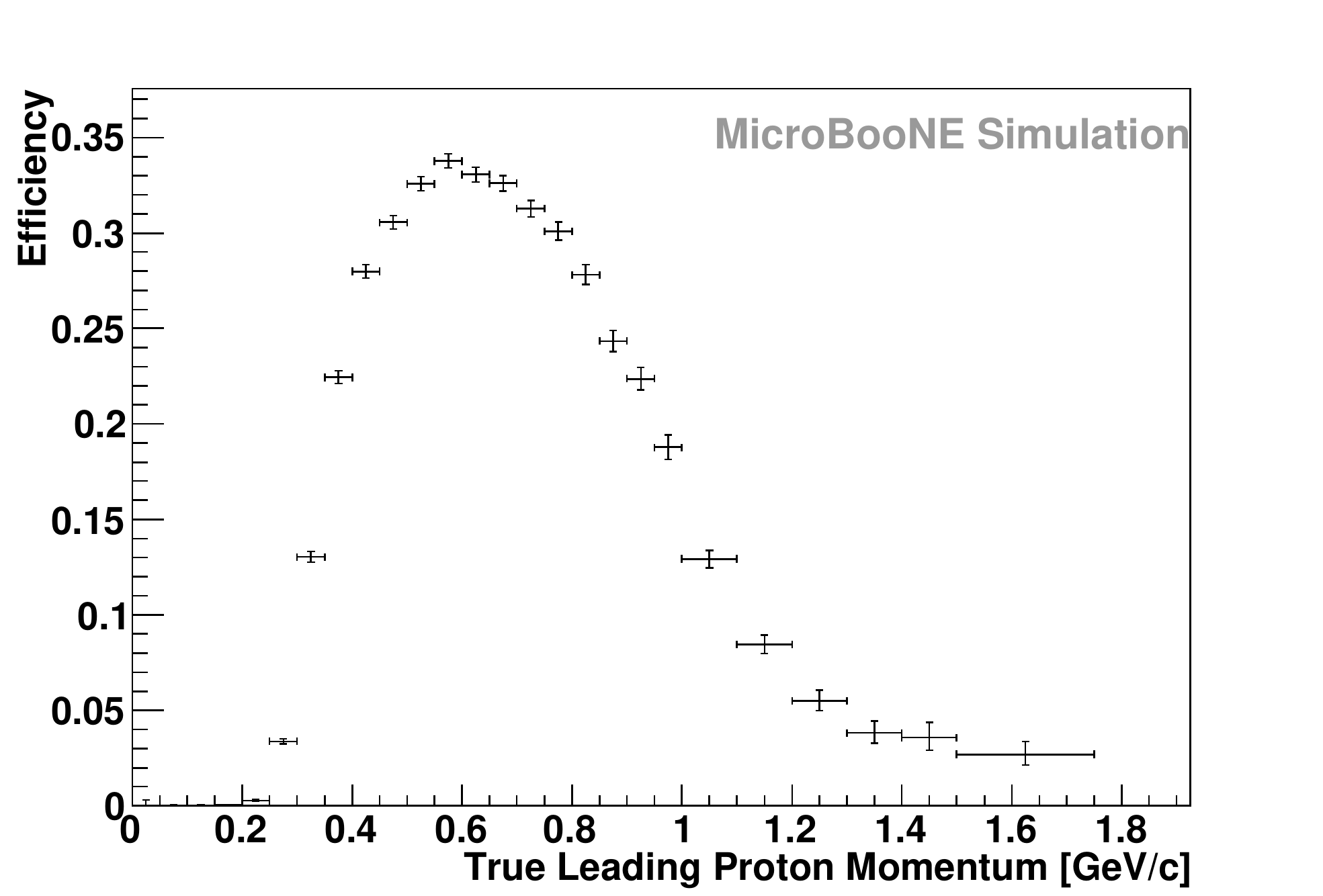}
  \caption{The efficiency of this event selection for simulated CC$0\pi Np$ events with no threshold requirement, as a function of proton momentum.  The efficiency drops significantly for protons below 300~MeV/c or above 1200~MeV/c, so we exclude them from our signal definition.}
  \label{fig:protonMomEff}
\end{figure}

The efficiency of this event selection as a function of leading proton momentum is shown in Fig.~\ref{fig:protonMomEff}, for a signal definition that has no explicit particle thresholds.
As expected from particle ID studies discussed in Sec.~\ref{sec:pid_thesholds}, the efficiency drops rapidly below 300~MeV/c. Based on where the efficiency at low proton momentum remains above 5\%, we include a threshold of 300~MeV/c on the leading proton in the signal definition and require the proton candidate to have a reconstructed momentum above 300~MeV/c.
A complete discussion of the efficiency of this selection is included in Sec.~\ref{sec:efficiency}.

Using the same 5\% efficiency requirement, the muon candidate is required to have a reconstructed momentum greater than 100~MeV/c, and the leading proton candidate is required to have a reconstructed momentum less than 1.2~GeV/c.
These last requirements, along with the leading proton momentum threshold, form part of the definition of the signal for this analysis.
They are required primarily to avoid a region of phase space where the proton forms a longer track than the muon, leading to the proton being misidentified as the muon candidate, and then failing the muon PID requirements for the longest track.
Additionally, at high momentum protons are contained less often and have a high probability of re-interacting and not forming a Bragg peak. 

\subsection{Event Selection Performance}
\label{sec:event_sel_perform}

\begin{figure}
    \centering
    \includegraphics[width=0.45\textwidth]{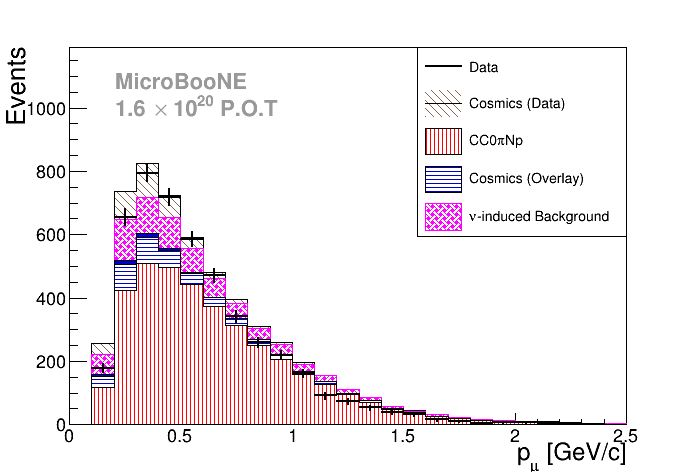}
    \caption{Momentum distribution of the muon candidate. The Monte Carlo events have been scaled to the data exposure of $1.6\times10^{20}$ P.O.T.}
    \label{fig:mumom}
\end{figure}

\begin{figure}
    \centering
    \includegraphics[width=0.45\textwidth]{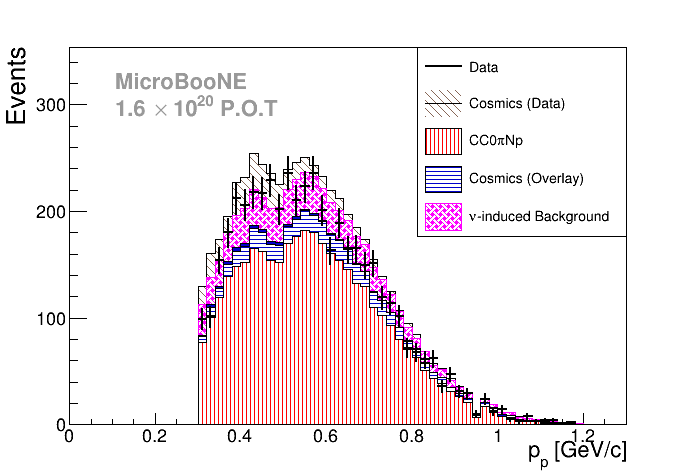}
    \caption{Momentum distribution of the leading proton candidate. The Monte Carlo events have been scaled to the data exposure of $1.6\times10^{20}$ P.O.T.}
    \label{fig:pmom}
\end{figure}

\begin{figure}
    \centering
    \includegraphics[width=0.45\textwidth]{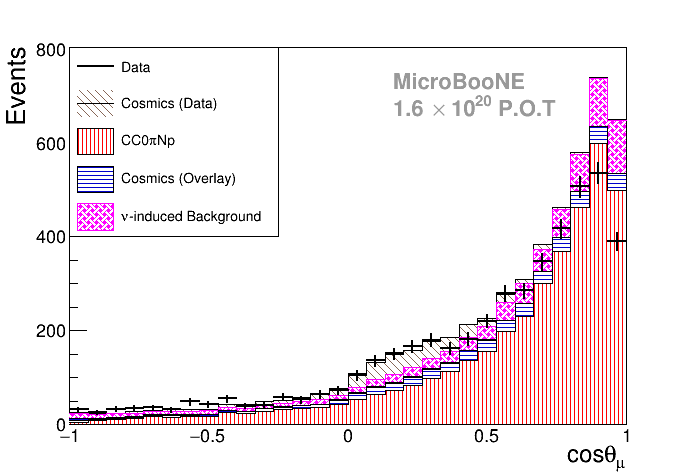}
    \caption{Cosine of the track polar angle ($\theta$) with respect to the beam for the muon candidate. The Monte Carlo events have been scaled to the data exposure of $1.6\times10^{20}$ P.O.T.}
    \label{fig:muangle}
\end{figure}

\begin{figure}
    \centering
    \includegraphics[width=0.45\textwidth]{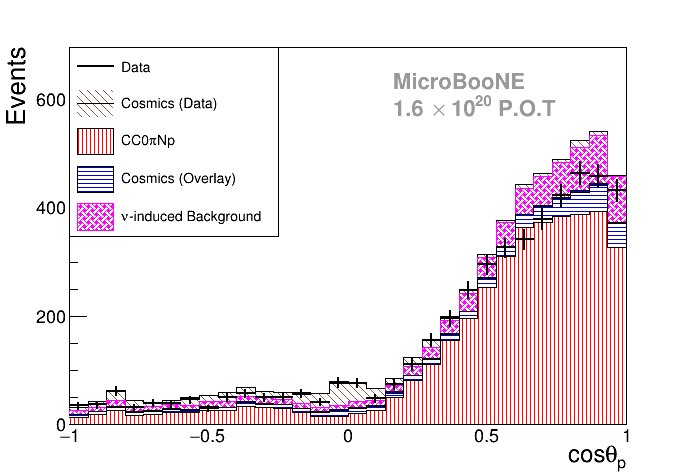}
    \caption{Cosine of the track polar angle ($\theta$) with respect to the beam for the leading proton candidate. The Monte Carlo events have been scaled to the data exposure of $1.6\times10^{20}$ P.O.T.}
    \label{fig:pangle}
\end{figure}

\begin{figure}
    \centering
    \includegraphics[width=0.45\textwidth]{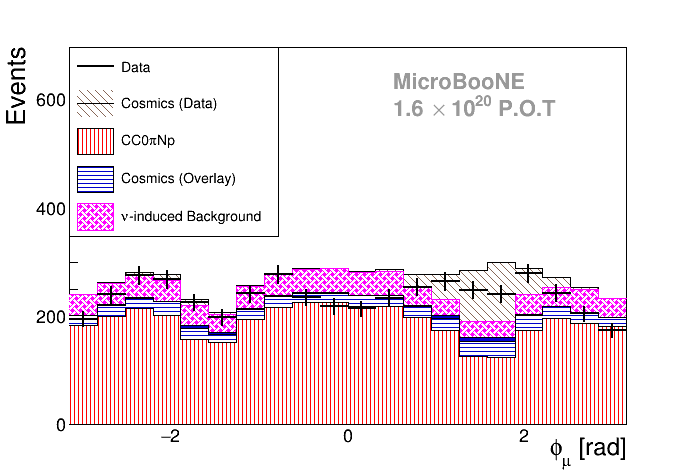}
    \caption{Track azimuthal angle $\phi$ with respect to the beam ($\pi/2$ is upward going) for the muon candidate.  The Monte Carlo events have been scaled to the data exposure of $1.6\times10^{20}$ P.O.T.}
    \label{fig:muanglephi}
\end{figure}

\begin{figure}
    \centering
    \includegraphics[width=0.45\textwidth]{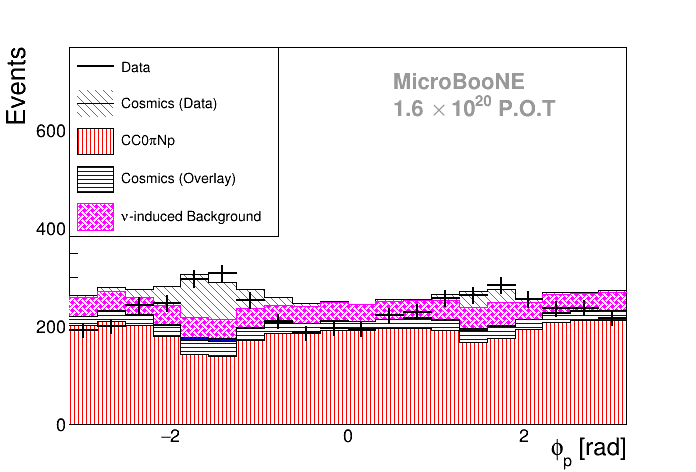}
    \caption{Track azimuthal angle $\phi$ with respect to the beam ($\pi/2$ is upward going) for the leading proton candidate. The Monte Carlo events have been scaled to the data exposure of $1.6\times10^{20}$ P.O.T.}
    \label{fig:panglephi}
\end{figure}

Within the phase space limits of this analysis, the event selection has an average efficiency of 29\% and a purity of 71\% (76\% excluding cosmics measured in data).
Table \ref{tab:eventRates} shows the number of final selected data events and the predicted backgrounds.
The neutrino-induced backgrounds make up 57\% of background events and are approximately evenly split between events where the interaction occurred outside of the fiducial volume boundaries and those where a pion was produced but either not reconstructed or misidentified.
A small number of the selected events are neutral current or wrong-sign (antineutrino) backgrounds.
The dominant cause of out-of-fiducial neutrino backgrounds is the space-charge effect that distorts reconstructed positions close to detector boundaries.
All neutrino-induced backgrounds are estimated from simulation, as is the uncertainty associated with subtracting them.

\begin{table}[ht]
    \centering
    \begin{tabular}{|c|c|} \hline
         sample & number of events \\ \hline
         Data & 4736 \\
         Cosmics (Data) & 393.1 \\
         Signal & 3773 \\
         Cosmics (Overlay)  & 416.0  \\
         $\nu$-induced Background & 750.8 \\ \hline
    \end{tabular}
    \caption{The number of selected data events, scaled off-beam data events, and scaled MC signal and background events.}
    \label{tab:eventRates}
\end{table}

The muon momentum, $p_\mu$, is calculated from range when the track is fully contained and by MCS fitting when it escapes.  
The proton momentum, $p_p$, is always calculated from range as the proton is always required to be contained in the \mub  detector.
The cosine of the muon (proton) polar angle $\theta_\mu$ ($\theta_p$) is determined with respect to the neutrino beam direction, and the angle $\phi$ is computed as the angle around the beam direction with $\phi=0$ aligned with the positive $x$-axis and $\phi=\pi/2$ aligned with the positive $y$-axis.
Finally, we determine the $\mu$-$p$ opening angle, $\theta_{\mu p}$.

\begin{figure}
    \centering
    \includegraphics[width=0.45\textwidth]{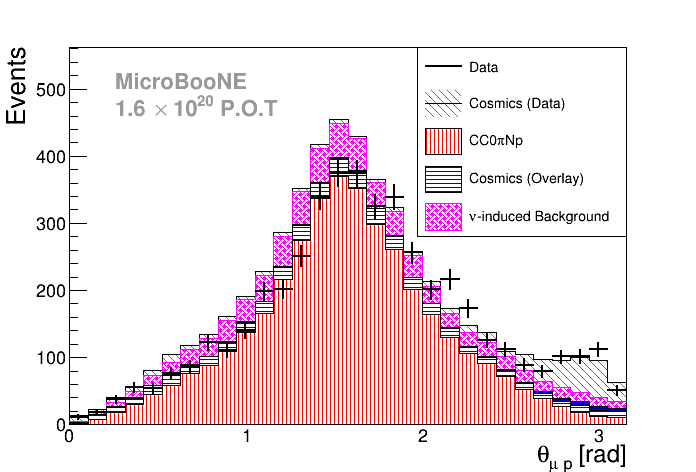}
    \caption{Distribution of angle between the muon candidate and the leading proton candidate. The Monte Carlo events have been scaled to the data exposure of $1.6\times10^{20}$ P.O.T.}
    \label{fig:thetamup}
\end{figure}

Figures~\ref{fig:mumom}-\ref{fig:panglephi} show the event distributions as a function of the reconstructed momentum, polar angle, and azimuthal angle for the muon and leading proton candidates. 
Figure~\ref{fig:thetamup} shows the reconstructed 3D opening angle between the two tracks.
There are a few discrepancies between the data and the Monte Carlo simulation prediction (GENIE v2). 
The most striking of these discrepancies shows up at the most forward muon angles in Fig.~\ref{fig:muangle}.
This region is dominated by low $Q^2$ events which is an area where older models are known to be deficient.
A similar discrepancy at low $Q^2$ was also observed in the MicroBooNE CC inclusive sample.~\cite{Adams:2019iqcINCL}
Additionally, at $\phi$ of 0,$\pi$, and $-\pi$ for both the muon and proton candidates a small data deficit is seen - this is due to the effects of induced charge on neighboring wires, which is not simulated but is accounted for in our detector uncertainties.
In general, the MC here does a reasonable job at describing the data and as such is adequate for estimating the efficiencies and background rates for this analysis.
Differences between the data and a variety of predictions are discussed in detail in Sec.~\ref{sec:results}.

Residual cosmic backgrounds appear in a small region of angular phase space with the muon and proton close to back-to-back that corresponds to an approximately upward-going muon and an approximately downward-going proton.
In these events, a cosmic muon entering the detector from the top is reconstructed as two tracks.  This topology is rare, but the exposure to cosmic muons is large, meaning that a number of these instances remain.
The same regions show a drop in signal efficiency because tracks at those angles do not cross many collection plane wires making PID challenging.

In the final data sample, 72\% of events contain only one reconstructed proton, with 22\% containing two reconstructed protons, 5\% containing three reconstructed protons, and 0.1\% being observed to produce four visible protons.

\section{Cross-Section Extraction}
\label{sec:xsect_def}

These cross-section measurements are presented in ``reconstructed'' quantities.  This avoids the ill-posed problem of trying to correct for detector smearing to present a result in true kinematics (known as unfolding~\cite{DAgostini:1994fjx}).
Instead, theoretical predictions must be smeared to compare to the data; this process called forward-folding and has been used by  T2K~\cite{Koch2019}.

To simplify the procedure for others wishing to compare predictions with our data, a single smearing matrix is provided for each measurement.  Resulting systematic uncertainties in the resolution and efficiency are estimated and included with the final data. This means that the data are corrected for an effective efficiency, as a function of a reconstructed variable.

The smearing matrix is calculated as

\begin{equation}
S_{ij} = N_{ij}^{\text{sel}} / N_{j}^{\text{sel}},
\end{equation}

\noindent
where $N_{ij}^{\text{sel}}$ is the number of selected events in reconstructed bin $i$, which come from true bin $j$, and $ N_{j}^{\text{sel}}$ is the total number of selected events from true bin $j$.
Defined in this way, a binned histogram multiplied by this smearing matrix leads to a smeared distribution with the same normalization.

In order for the data to be compared to such a prediction, they must be corrected for inefficiency.  We define an effective efficiency, $\tilde{\epsilon}_i$ which can be applied as a function of reconstructed variables:

\begin{equation}
\label{eq:eff_smear}
\tilde{\epsilon}_i = \frac{ \sum_{j=1}^{M} S_{ij}N^\text{sel}_j}{ \sum_{j=1}^{M} S_{ij}N^\text{gen}_j}, \quad 
\end{equation}

\noindent
where $N^\text{sel}_j$ is defined above, $N^\text{gen}_j$ is the number of generated signal events in true bin $j$, and $M$ is the total number of true bins used.

It is known that there are limitations to this method.  The primary limitation here is that uncertainties on the resolution are not fully encapsulated, but the uncertainty on the resolution in this particular measurement is negligible.

To determine the final differential cross sections, both signal and background event rates, as well as efficiencies, are binned as a function of each variable; the single differential cross section in bin $i$ is then calculated as (using $p_\mu$ as an example):

\begin{equation}
\label{eq:xsec_differential}
\begin{split}
\left ( \frac{d\sigma}{dp_\mu} \right )_i &= \frac{N_i - B_i}{\tilde{\epsilon}_i \cdot N_\text{target} \cdot \Phi_{\nu_\mu} \cdot (\Delta p_\mu)_i}, \\
\end{split}
\end{equation}

\noindent
where $N_i$, $B_i$, and $\tilde{\epsilon_i}$ are the the number of candidate events (Sec.~\ref{sec:event_sel}), the expected number of background events (Sec.~\ref{sec:event_sel}), and the efficiency smeared by resolution in reconstructed bin $i$, respectively. 
$N_{target}$ is the number of argon atoms in the fiducial volume ($2.61 \times 10^{31}$~\cite{Adams:2019iqcINCL}), and $\Phi_{\nu_\mu}$ is the integrated flux of muon neutrinos passing through the detector fiducial volume ($1.17 \times 10^{11}  \text{cm}^{-2}$ for this dataset~\cite{Adams:2019iqcINCL}). 
$(\Delta p_\mu)_i$ is the bin width for bin $i$ in the $p_\mu$ distribution, which was optimised based on a consideration of available statistics and detector resolution.
The other variables ($\cos(\theta_\mu)$, $p_p$, $\cos(\theta_p)$, and $\cos(\theta_{\mu p})$) are handled in a similar way.

To determine if the cross section extraction technique was dependent upon the use of GENIE v2, a sample of simulated events was generated with an alternate GENIE model. The alternate GENIE sample was then treated as observed data, and cross sections extracted using the described technique in the previous paragraphs. The study showed that the cross section extraction is not dependent upon the model used for making the measurement.

Systematic uncertainties (discussed in Sec.~\ref{sec:syst}) on the efficiency and resolution propagate to uncertainties on the final measurement through this procedure, allowing a simple comparison with a smeared theory prediction.

\label{sec:efficiency}

The effective efficiencies and smearing matrices are crucial inputs to the cross section, and are produced from Monte Carlo simulation.
The efficiencies and smearing matrices showed minimal model dependence in fake data tests and in our systematic uncertainty evaluation.

\begin{figure}
    \centering
    \includegraphics[width=0.45\textwidth]{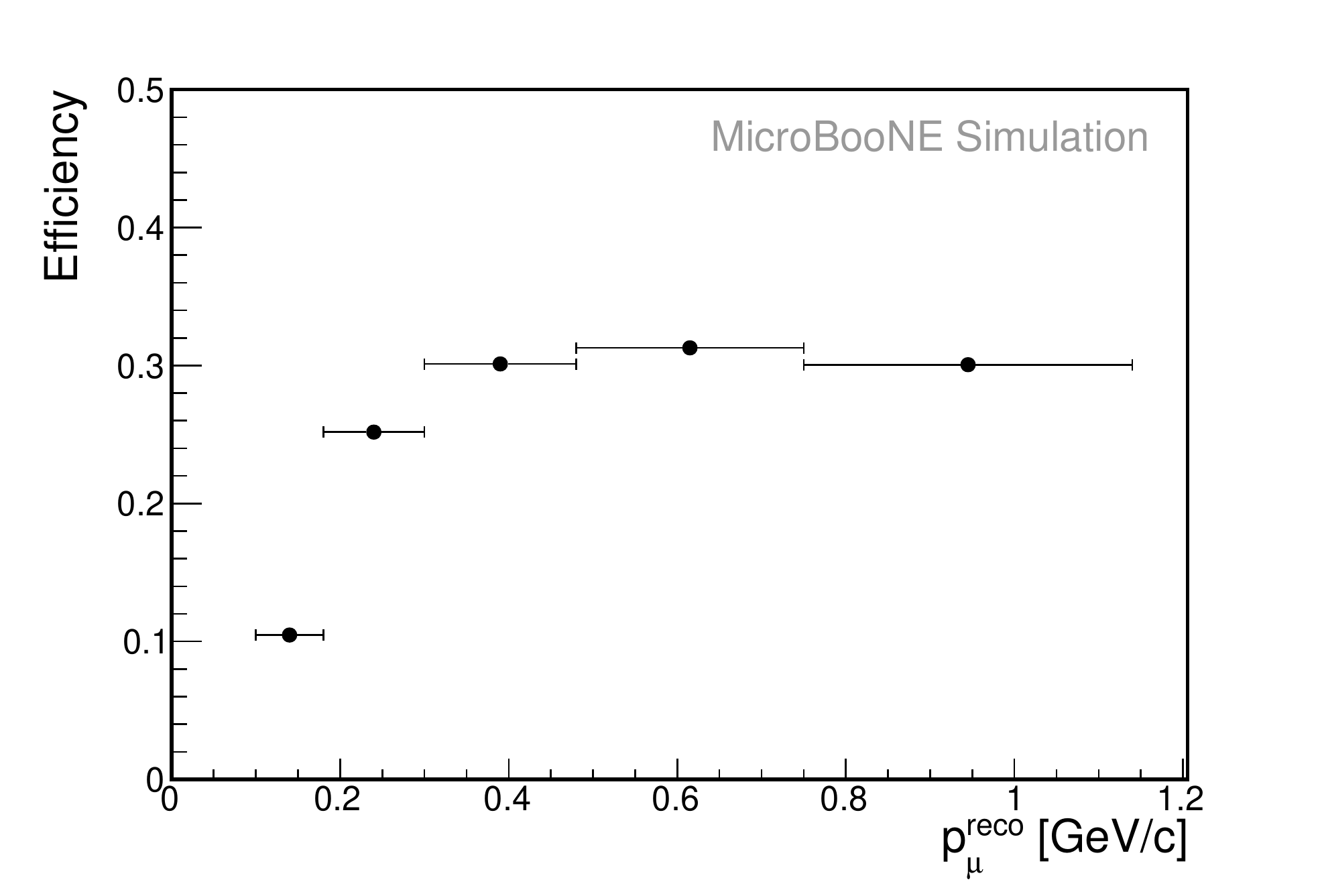}
    \caption{Efficiency as a function of reconstructed $p_\mu$ in simulated CC$0\pi Np$ events.  Statistical error bars are too small to be seen.}
    \label{fig:eff_mumom}
\end{figure}

As an example, the efficiency as a function of muon momentum is shown in Fig.~\ref{fig:eff_mumom} and the equivalent smearing matrix is shown in Fig.~\ref{fig:migmat_mumom}.  Plots for the other primary variables and the smearing matrices are found in the Appendix.  
Efficiency values peak at roughly 35\% for all variables; this limit is primarily caused by the need to suppress cosmic-ray background and identify stopping protons.
Good efficiency is seen for proton with a momentum above 300 MeV/c (see Fig.~\ref{fig:eff_pmom}). The appendix contains figures that demonstrate the relatively high efficiency of wide angle tracks, giving the analysis $4\pi$ coverage of outgoing particles from interactions.

\begin{figure}[h!]
    \centering
    \includegraphics[width=0.45\textwidth]{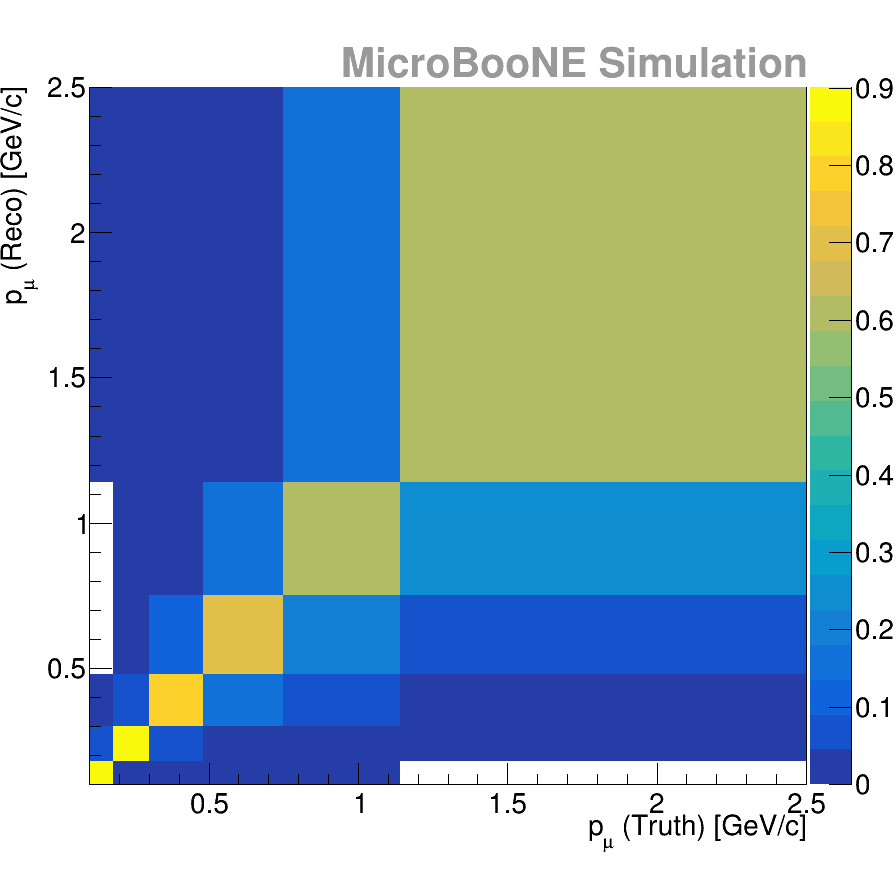}
    \caption{Migration matrix between true and reconstructed bins for reconstructed muon momentum ($p_\mu$) in simulated CC$0\pi Np$ events.}
    \label{fig:migmat_mumom}
\end{figure}

\section{Systematic Uncertainties} 
\label{sec:syst}
Four categories of systematic uncertainties were considered in the analysis of the data: detector modeling, neutrino interaction modeling, neutrino beam flux predictions, and modeling of secondary hadronic interaction following the primary neutrino interaction.

\subsection{Methodology} 
To determine the impact of an uncertainty, two different formalisms were used. Most uncertainties were evaluated by applying an event-based reweighting factor to the primary simulated neutrino interaction events, and determining the impact on the measured result. For other uncertainties, the generation of re-simulated samples was necessary. 
For the neutrino interaction modeling, beam flux predictions, and secondary hadronic interaction modeling, the systematic uncertainty was determined by applying reweighting factors to simulated events. The use of these weights allows for the complete sampling of the systematic phase space and construction of correlated uncertainties between differential cross-section bins. In the case of the GENIE uncertainty parameters, some parameters have nonlinear dependencies and therefore must be varied simultaneously. Detector modeling uncertainties were estimated by a series of re-simulated samples based on parameters discussed later.
In all cases, covariance matrices are produced capturing the bin-to-bin correlations between uncertainties (a non-negligible effect in these data), which must be used for detailed comparisons with the data.

Uncertainties determined with reweighting factors were evaluated in either a ``multisim" or a ``unisim" approach. The multisim approach varies all parameters from an uncertainty category together and applies the product of all reweighting factors to simulated events as a single weight. On the other hand, the unisim or “1$\sigma$” approach varies each parameter by a shift up or down of one standard deviation individually. The final analysis uses the multisim approach for neutrino interactions, beam flux uncertainties, and secondary hadronic interactions using 1000 sampled universes for each set of uncertainties. 

After determining the systematic uncertainty from each of the four categories, the four uncertainties are then added in quadrature as they are assumed to be fully uncorrelated to each other. The covariance matrices are available in the supplemental material and data release.
The uncertainties as a function of our measured variables are shown in Figs.~\ref{fig:mu_mom_syst} ($p_\mu$), \ref{fig:mu_angle_syst} ($\cos(\theta_\mu$)), \ref{fig:p_mom_syst} ($p_p$), \ref{fig:pangle_syst} ($\cos(\theta_p$)), and \ref{fig:mup_angle_syst} ($\theta_{\mu p}$).
Each figure shows the four main sources of systematic uncertainty, as well as the statistical uncertainty and total uncertainty for each bin.

\subsection{Detector Modeling Uncertainties}
\label{sect:syst_detector}
A variety of effects related to the ability to analyze events in LAr were examined, using actual data whenever possible.
To account for uncertainties in the modeling of ionization yield and drifting of electrons, uncertainties were determined for the following parameters: space charge effect, drift-electron diffusion both transverse and longitudinal, drift-electron lifetime, and drift-electron recombination. 
We vary the magnitude of the space charge effect based upon early measurements of the spatial distortions at the edge of the TPC, extrapolated to the bulk field.
The nominal drift-electron transverse and longitudinal diffusion parameters are set to be the central value of measurements in liquid argon and the variations are based upon the spread of those measurements~\cite{LI2016160}.
The drift-electron lifetime was changed from the nominal value (100~ms) to the lowest measured lifetime of any run period of the detector operations (10~ms) and scaled to the 10\% of the data represented by that low lifetime.  Drift-electron recombination uncertainty was estimated by comparing the nominal simulation which models electron-ion recombination using the modified box model with parameters fit to ArgoNeuT data~\cite{Acciarri_2013} to the variation which substitutes the Birks model~\cite{birks_model} with parameters tuned to ICARUS data~\cite{ICARUS_recomb}.

The shapes of signals on the detector anode wires has uncertainties from dynamically induced charge, intrinsic anode wire and electronics response, and the existence of unresponsive channels. Dynamically induced charge is the creation of electronic signals on anode wires other than the wire closest to the path of drifting electrons, including collection plane wires. In the default simulation this effect is not included, and the variation adds simulation of induced signals on the 10 closest wires. Anode wire response was measured in data~\cite{Adams:2018gbiPROCESS}, and the nominal width is increased based upon the 1$\sigma$ uncertainty of that measurement.  The map of unresponsive channels was used to completely remove any response from those channels which don't respond in data. 
  
To account for uncertainties in the response of the detector to light, uncertainties were determined for light yield, photo-electron noise in PMTs, and light production outside of the TPC. The light yield was varied to cover the differences in scintillation light production in liquid argon based upon modeling different particles types instead of exclusively electrons. The photo-electron noise in the PMTs was varied by $\pm$1$\sigma$ of the measured yield from data. The production of light outside of the TPC may be incorrectly modeled so a sample with light yield outside the TPC increased by 50\% was generated.

The uncertainties from detector modeling are estimated using the unisim method.  For each of the listed detector modeling parameters, the same simulated GENIE v2 neutrino sample of events is reprocessed through GEANT4 with the modified detector parameter. The uncertainty for that parameter is determined by taking the difference in the measured cross section using the nominal measurement or the modified sample and treating this difference as a 1$\sigma$ Gaussian uncertainty for the cross section. For parameters that have both an upward and a downward variation,
we conservatively take the variation that leads to the largest uncertainty and treat this as a 1$\sigma$ uncertainty on the cross-section measurement. The various detector modeling uncertainties are then added in quadrature to determine the total detector uncertainty. 
The largest single uncertainty - leading to an uncertainty of 18\% on the integrated cross section and up to 40\% at $\cos \theta_{\mu}^{\text{reco}} = 0$ - comes from the modeling of dynamically induced charge on the induction wires of the anode. 
These induced signals can interfere with the ``primary'' signal, leading to cancellation of induction plane signals and smearing of collection plane signals, reducing the tracking efficiency in some regions of angular phase space and degrading the charge resolution.
This effect is more successfully modeled in later detector response simulations and will be less significant in future measurements. The detector uncertainty is on average approximately 30\% but can be as high as 50-60\% in some specific regions of phase space.

\subsection{Neutrino Interaction Uncertainties}
\label{sect:syst_genie}

\begin{figure}
    \centering
    \includegraphics[width=0.45\textwidth]{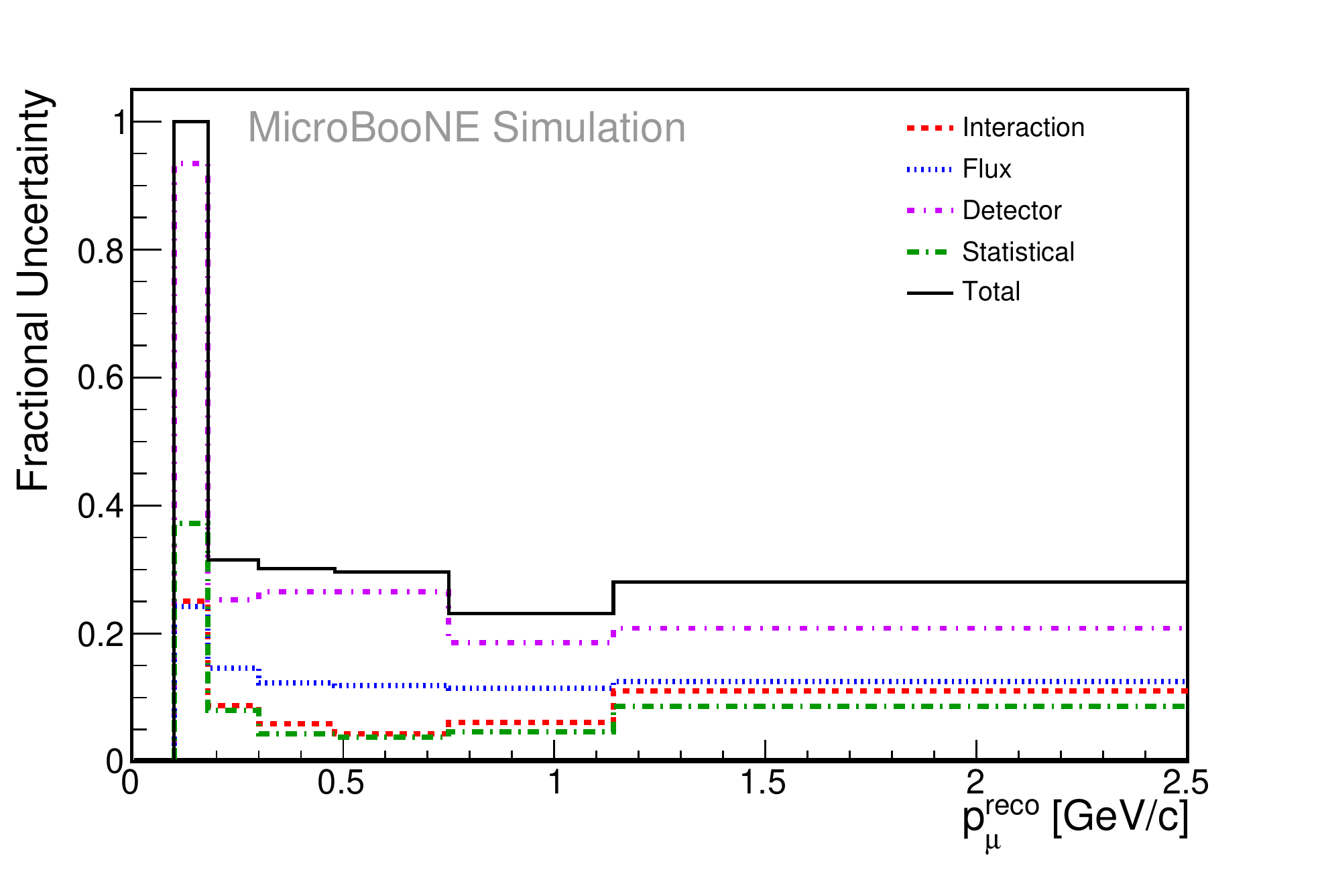}
    \caption{Uncertainties for muon candidate momentum ($p_{\mu}$).}
    \label{fig:mu_mom_syst}
\end{figure}

\begin{figure}
    \centering
    \includegraphics[width=0.45\textwidth]{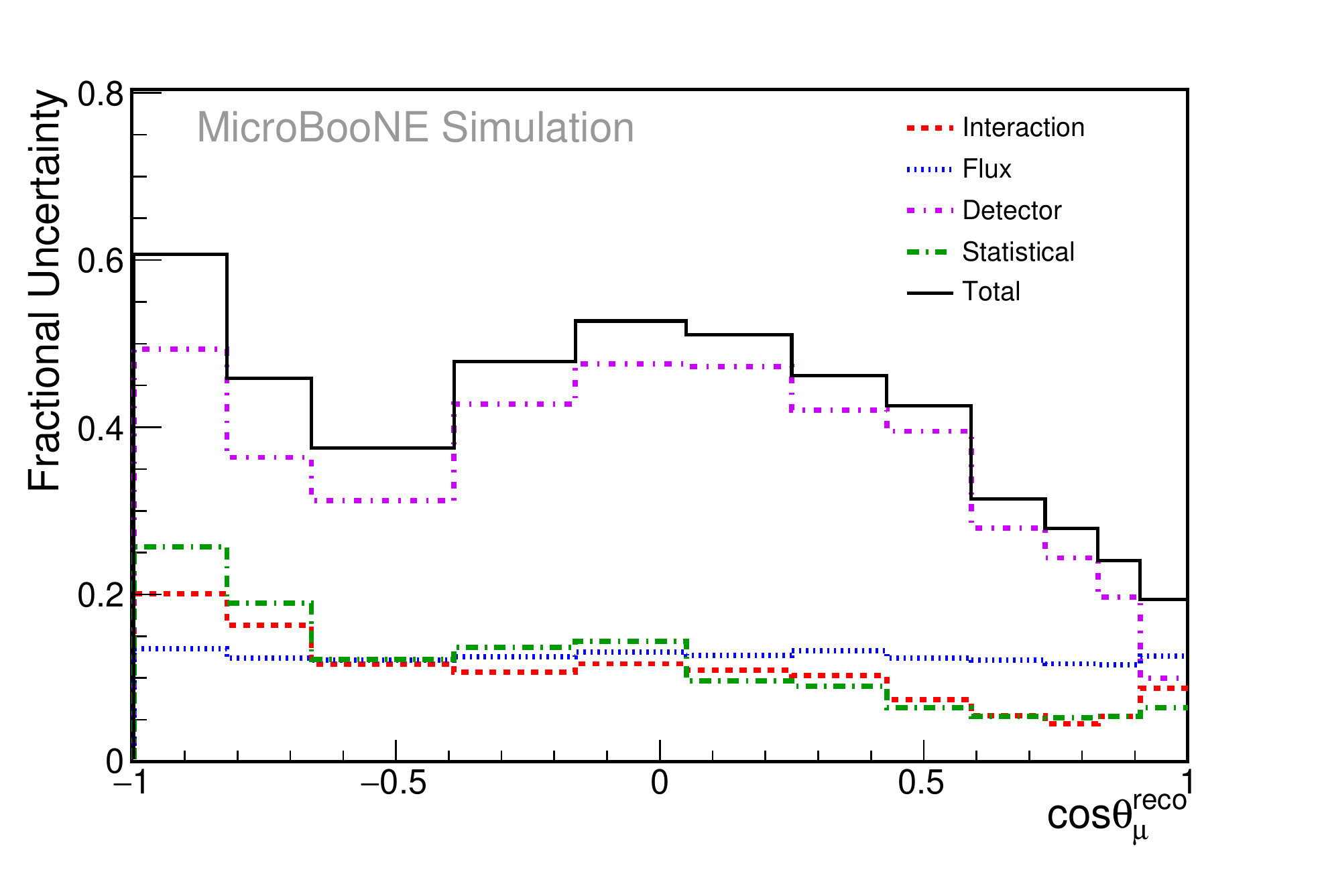}
    \caption{Uncertainties for cosine of the muon polar angle ($\cos \theta_{\mu}$).}
    \label{fig:mu_angle_syst}
\end{figure}

\begin{figure}
    \centering
    \includegraphics[width=0.45\textwidth]{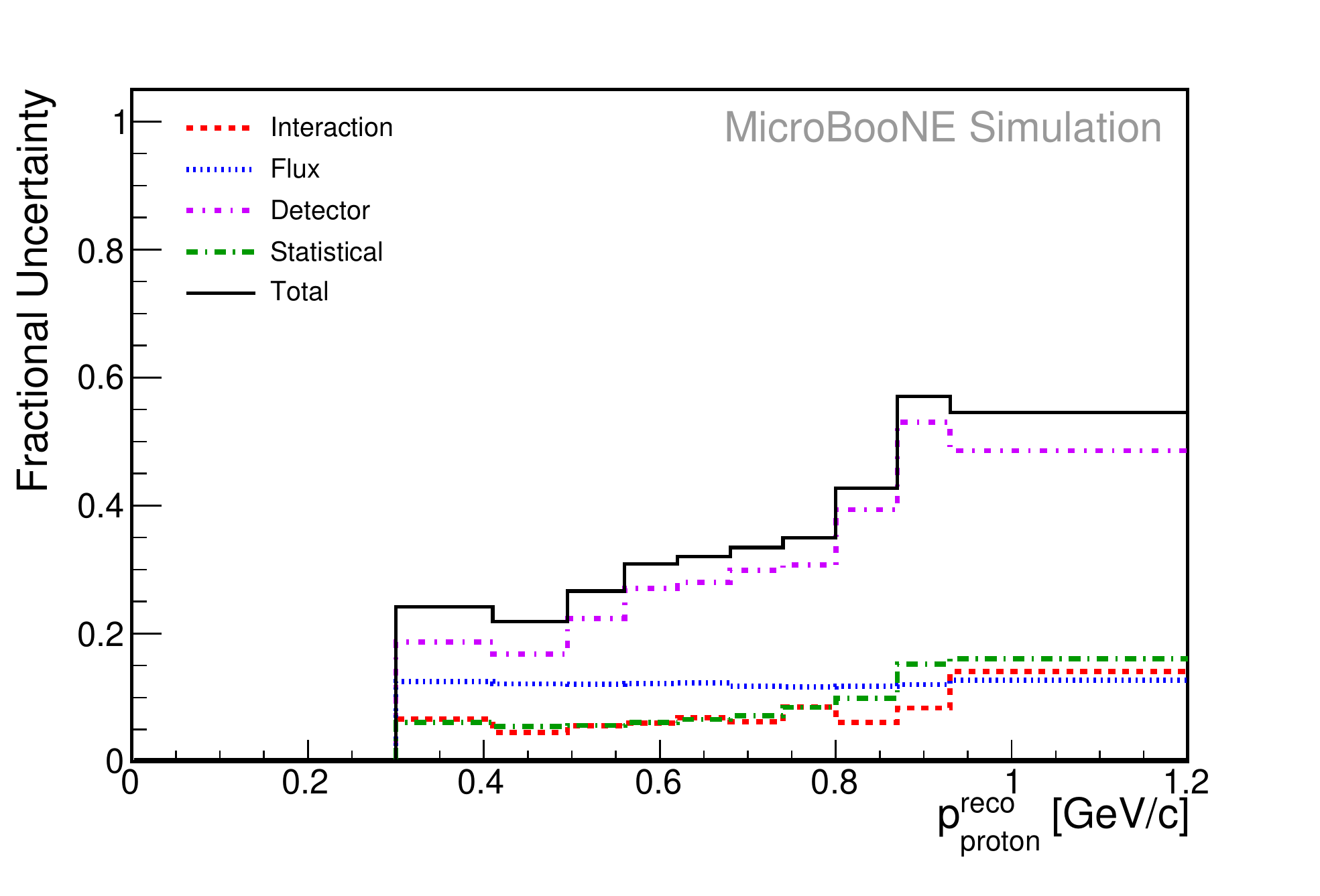}
    \caption{Uncertainties for leading proton candidate momentum ($p_{p}$).}
    \label{fig:p_mom_syst}
\end{figure}

\begin{figure}
    \centering
    \includegraphics[width=0.45\textwidth]{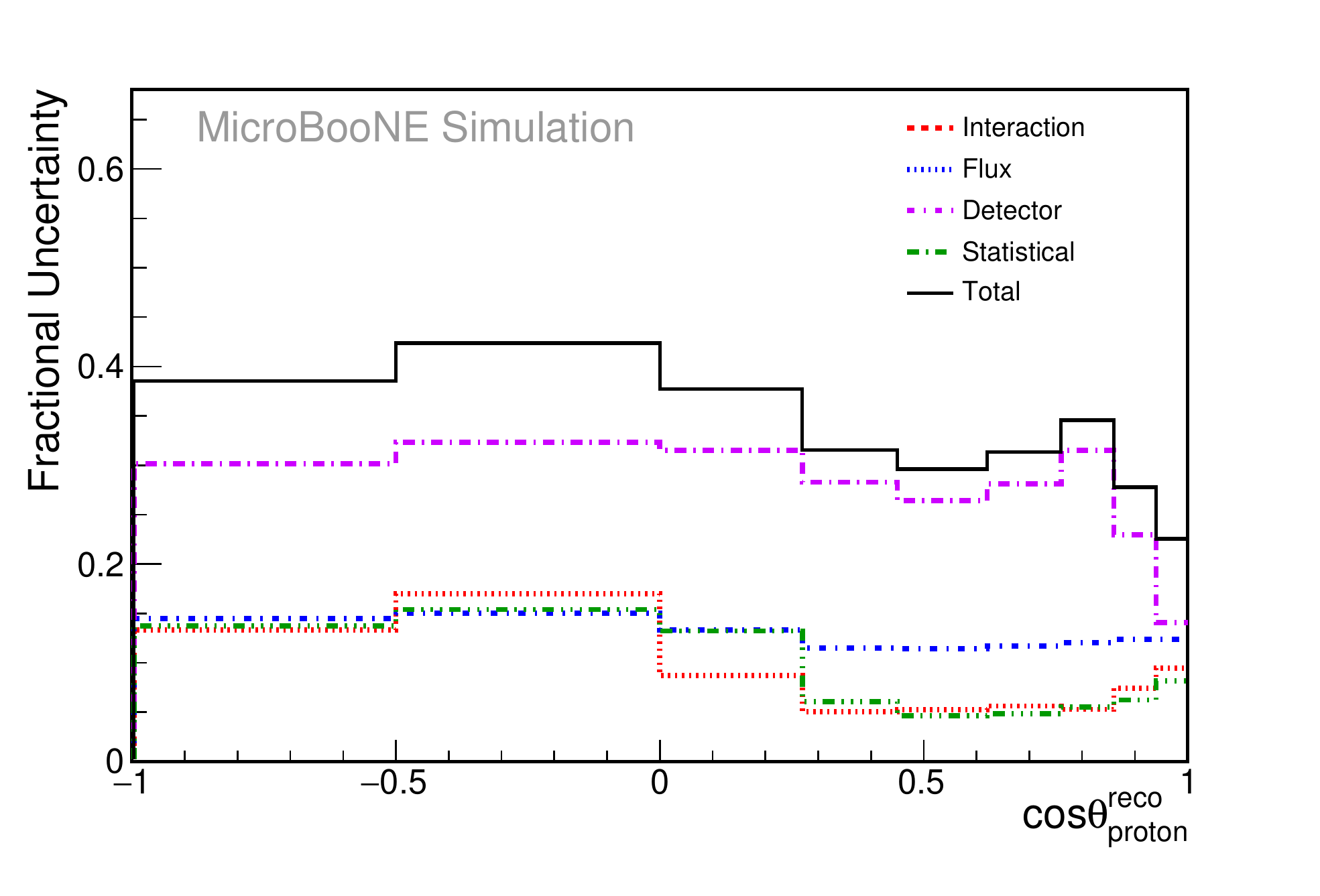}
    \caption{Uncertainties for cosine of the polar angle of the leading proton candidate ($\cos \theta_{p}$).}
    \label{fig:pangle_syst}
\end{figure}

\begin{figure}
    \centering
    \includegraphics[width=0.45\textwidth]{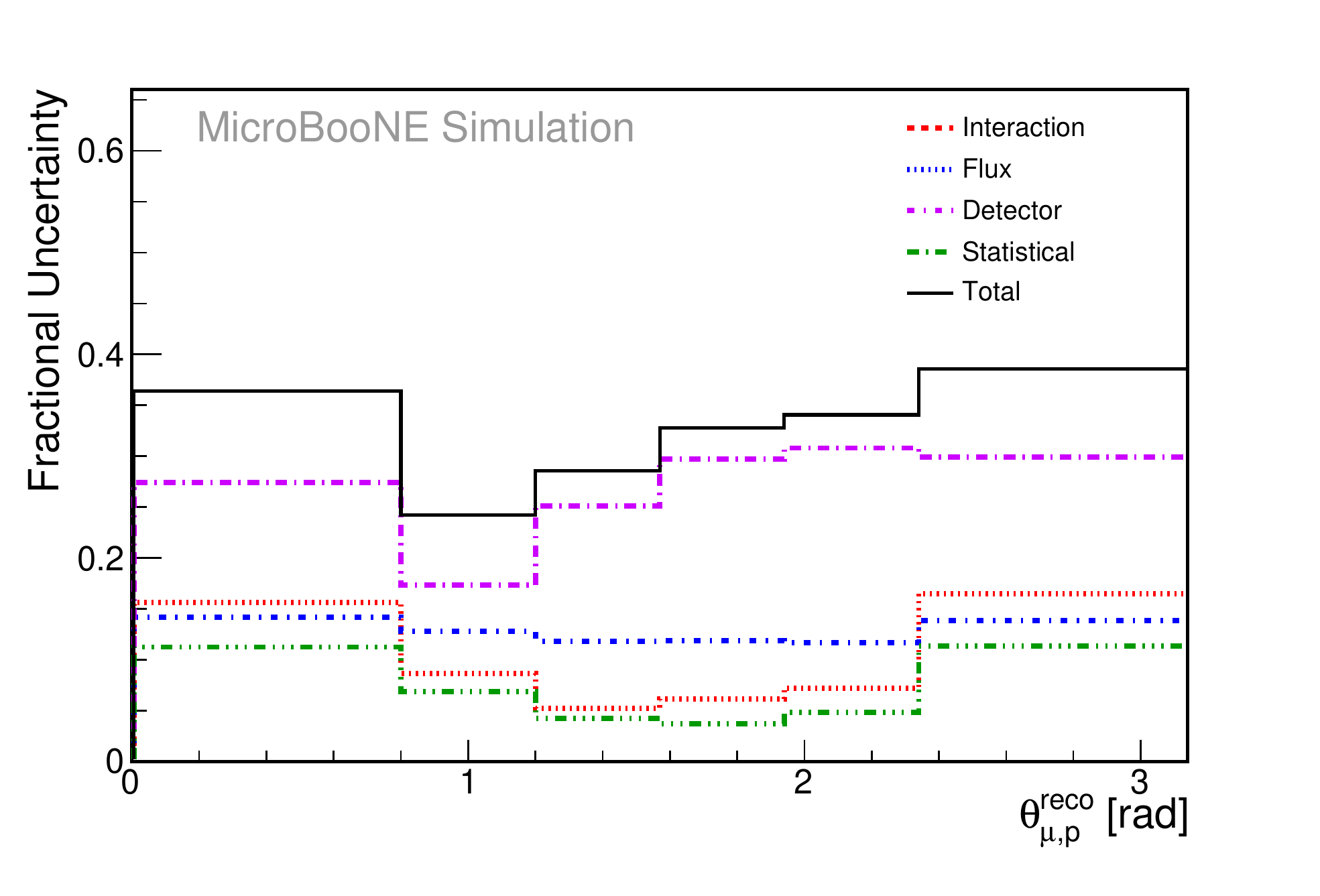}
    \caption{Uncertainties for the opening angle between muon candidate and leading proton candidate ($\cos \theta_{\mu p}$).}

    \label{fig:mup_angle_syst}
\end{figure}

GENIE provides a reweighting framework for the evaluation of the uncertainties from parameters contained in the GENIE models.  The GENIE authors provide estimated uncertainties for various parameters which are mostly determined from fits to external data. For a given parameter $P$, GENIE provides a modified event weight based upon uncertainty $\sigma_P$ that is applied to the sample events and creates a systematically varied distribution. The full list of parameters varied within the GENIE framework can be found in \cite{Andreopoulos:2015wxa}. Beyond the uncertainties on parameters provided in the GENIE model, two additional uncertainties in the neutrino interaction modeling were considered for this measurement. 
We reweight events as a function of true energy transfer and true momentum transfer, according to the ratio of our nominal model (GENIE v2, see Sec.~\ref{sec:evgen}) and the Valencia  model~\cite{Nieves:2004wx,Gran:2013kda}.
This is done separately for CCQE and $2p2h$ 
events, and each uncertainty is added in quadrature to the multisim GENIE uncertainties to obtain the total uncertainty from neutrino interaction modeling.
This sum is labeled as ``Interaction" in Fig.~\ref{fig:mu_mom_syst}-\ref{fig:mup_angle_syst}.
The impact of these uncertainties on the detection efficiency or resolution is very small; the largest impact from these uncertainties is due to changes in background components which we do not constrain.
As the dominant backgrounds are out-of-fiducial-volume and overlaid cosmics, which both scale linearly with the total event rate, the dominant uncertainties are those that change the total rate of predicted neutrino interactions in the MicroBooNE detector.
Parameters that change the shapes of signal or background kinematics, including final state interactions, do not lead to substantial uncertainty on the final measurement.
The neutrino interaction uncertainty is approximately 6\% for most bins of differential cross section with the largest contribution to uncertainty in the lowest bin of muon momentum at 25\%.

\subsection{Beam Flux Uncertainties}
\label{sect:syst_beam_flux}
There are two major sources of systematic uncertainty in predicting the neutrino beam flux: hadron production rates and horn focusing effects.
The hadron production rate uncertainties cover the rate of particles produced from protons striking the horn target with variations in $\pi^+$, $\pi^-$, $K^+$, $K^-$, and $K^0$.
These variations are addressed using the multisim approach.
The uncertainty from focusing horn current measurements is estimated using a unisim approach where the beamline is re-simulated with the horn current set at a value 1$\sigma$ away from the nominal.
The multisim hadron production rate uncertainty and the unisim horn current uncertainty are then added in quadrature to give a total beam flux uncertainty. 
The overall uncertainty on the neutrino flux is dominated by an approximately 11\% uncertainty on the absolute normalization, with small uncertainties on the flux shape~\cite{AAAA_2009_MBflux}.
For this reason, the largest impact these uncertainties have on the final cross section measurement is a fully correlated normalization uncertainty of 11\%.
Second order effects from the background subtraction increase this to 15\% in low-purity bins and one bin as high as 25\%.

\subsection{Secondary Hadronic Interaction Modeling}
\label{sect:syst_had_int}
Protons and charged pions can scatter, both elastically and inelastically, in the detector through hadronic interactions with nuclei. These interactions can lead to the production of additional particles or large angle changes in particle trajectories that may lead to reconstruction algorithms failing to form a single, well-reconstructed track.
Interactions such as these can impact the signal efficiency or the background rejection rate, and uncertainties on the rates of these hadronic interactions are propagated to uncertainties on the cross-section measurement. Studies were performed separately for elastic and inelastic interactions, and elastic-scatter uncertainties were found to have negligible impact.

GEANT4 is used to propagate all hadrons through the detector medium based on a semi-classical cascade model~\cite{Wright:2015xia}.  Events with inelastic hadronic interactions are reweighted independently for interactions containing protons, positive pions, and negative pions. For each particle type, the total inelastic cross section is varied around its mean by 30\%, independent of the particle's energy. The pion interaction rate has a negligible impact on the analysis - less than 1\%, while the proton interaction probability does have an impact on the event selection efficiency.
The probability that a proton undergoes an inelastic interaction somewhere along its length increases with trajectory length, so it follows that the impact of proton reinteraction uncertainties is largest at high proton momenta. The uncertainty from proton reinteractions is 8\% at the highest proton energies. 
Events with high proton momemtum tend to have high energy transfer from the lepton, meaning this uncertainty at high proton momentum impacts the measurement of low muon momenta and backwards muon angles.
When this effect is averaged over other variables, the estimated uncertainty from secondary hadronic interaction modeling is 2\%.

In Figs. \ref{fig:mu_mom_syst} - \ref{fig:mup_angle_syst} these uncertainties are added in quadrature to the detector uncertainty.

\section{Results}
\label{sec:results}
\begin{figure}[h!]
    \centering
    \includegraphics[width=0.45\textwidth]{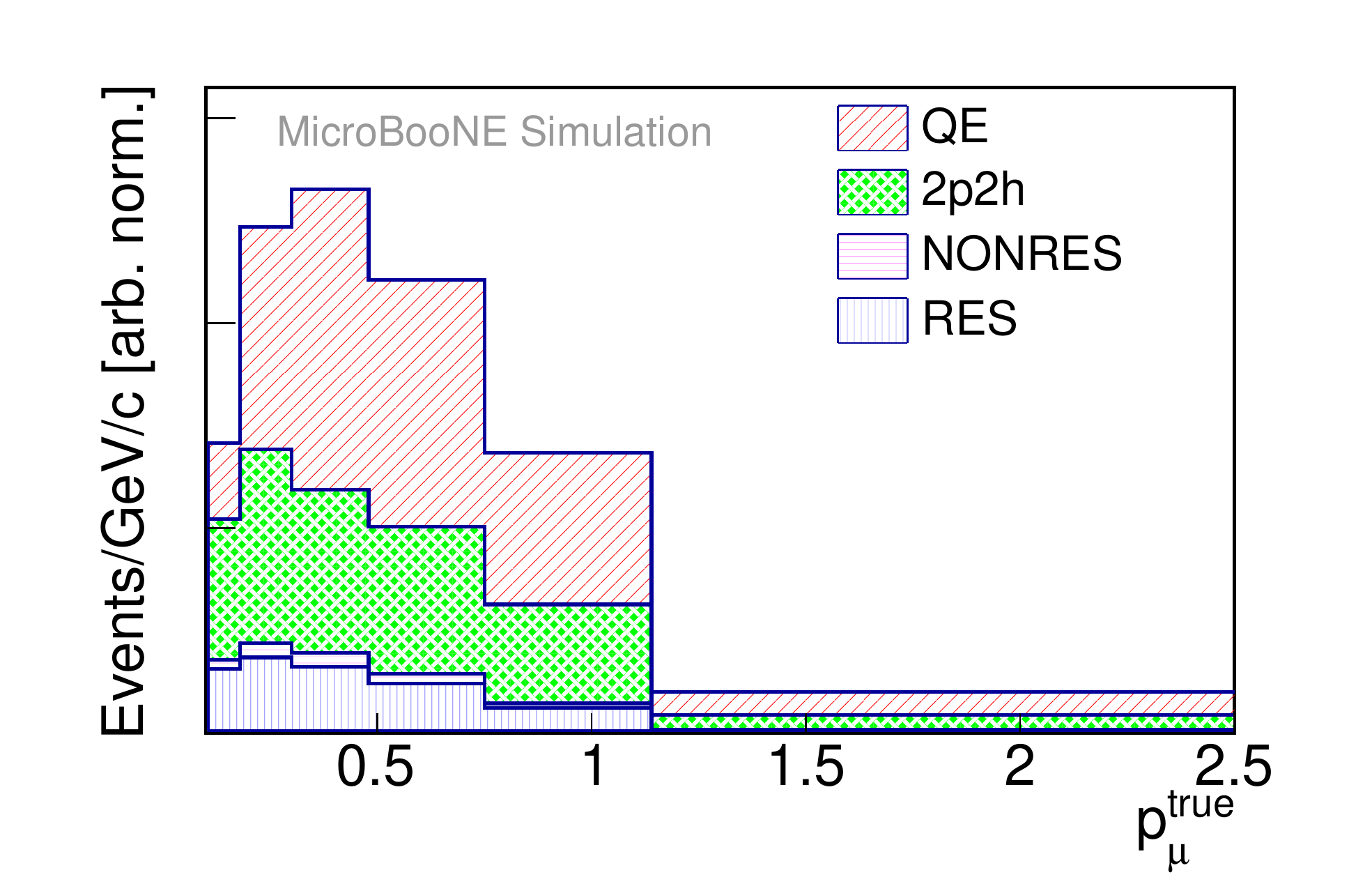}
    \caption{Breakdown of interaction components of cross section as a function of $p_{\mu}$ according to GENIE v2. Interaction types of CCQE, $2p2h$, and pion production are shown; the pion production events are further divided into resonant (RES) and nonresonant (NONRES) channels.  As a result of the signal definition, there are no coherent pion production events in the final simulation sample.  The number of events per bin is shown with arbitrary normalization.}
    \label{fig:xsec_breakdown_mumom}
\end{figure}

\begin{figure}[h!]
    \centering
    \includegraphics[width=0.45\textwidth]{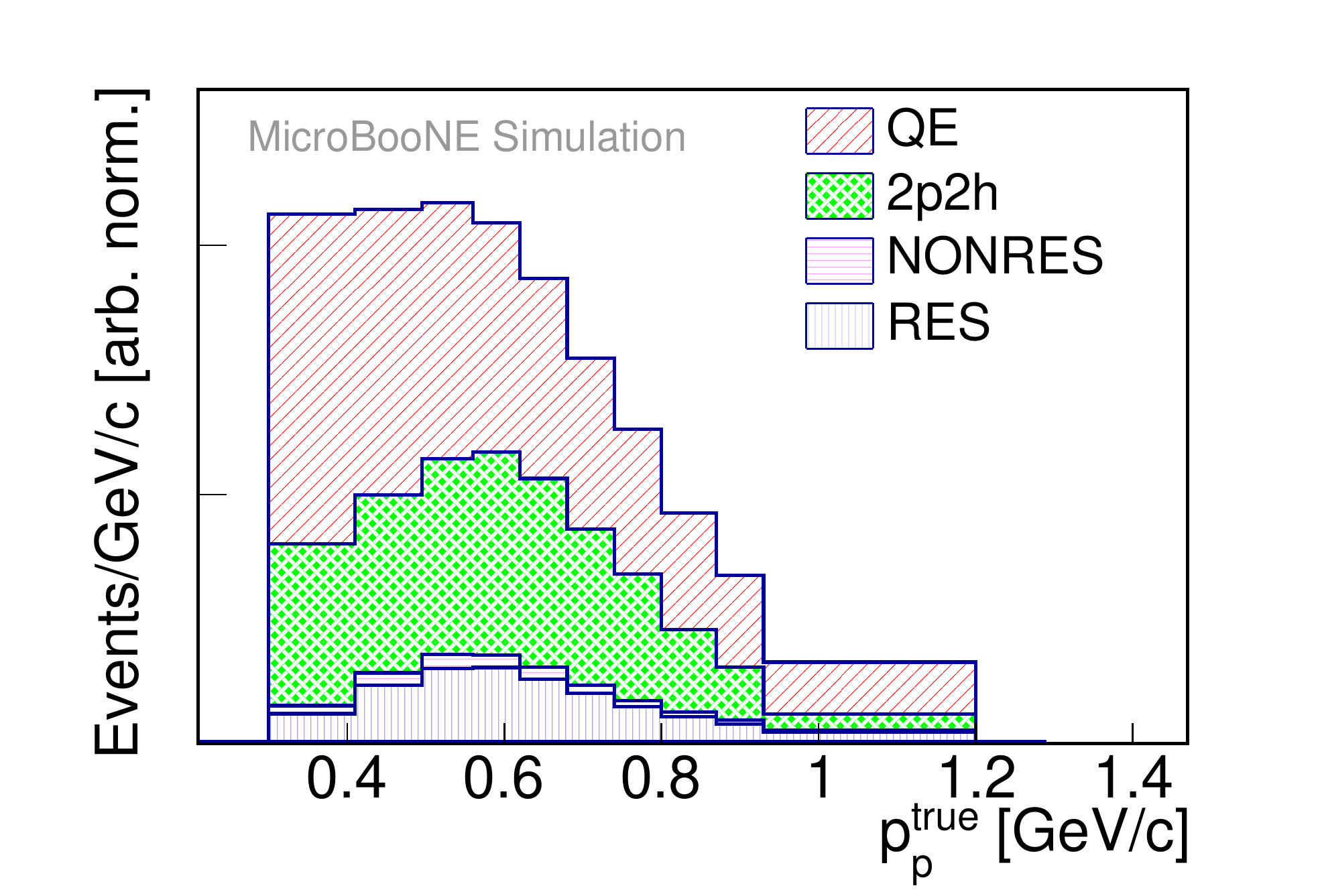}
    \caption{Breakdown of interaction components of cross section as a function of $p_{p}$ according to GENIE v2.   The number of events per bin is shown with arbitrary normalization.}
    \label{fig:xsec_breakdown_pmom}
\end{figure}

\begin{figure}[th!]
    \centering
    \includegraphics[width=0.45\textwidth]{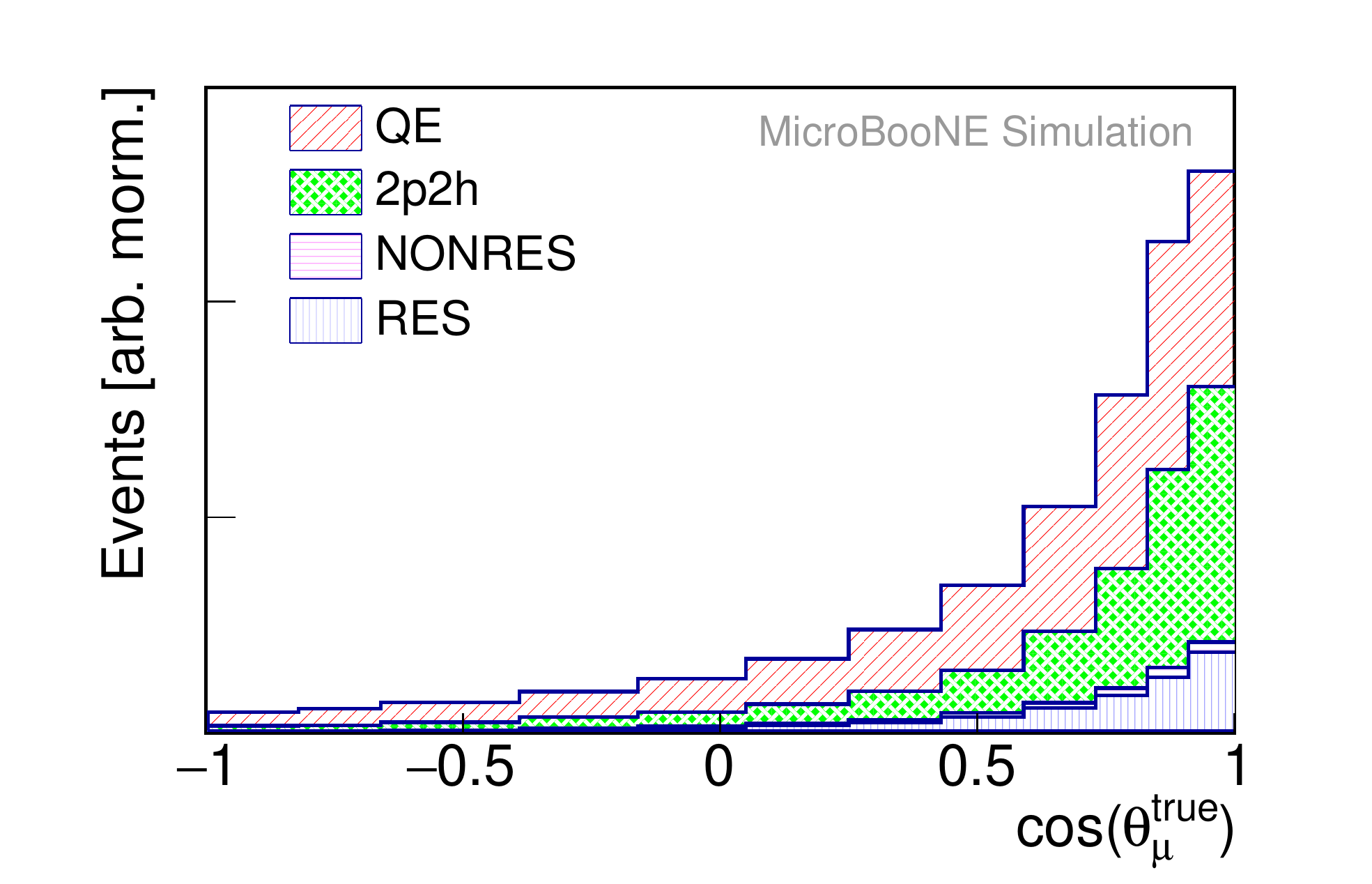}
    \caption{Breakdown of interaction components of cross section as a function of $\cos\,\theta_{\mu}$ according to GENIE v2. The number of events per bin is shown with arbitrary normalization.}
    \label{fig:xsec_breakdown_muangle}
\end{figure}

\begin{figure}[th!]
    \centering
    \includegraphics[width=0.45\textwidth]{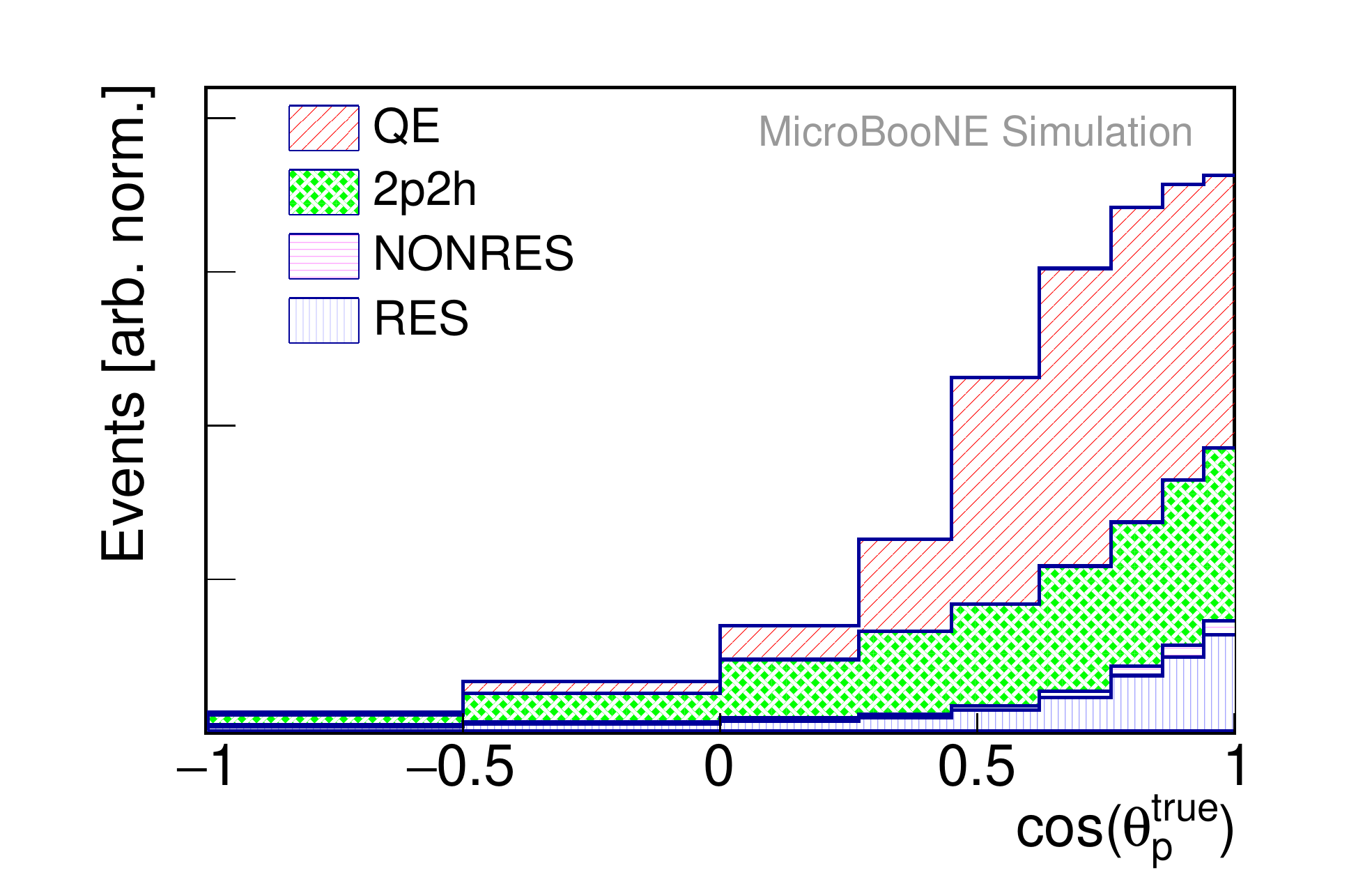}
    \caption{Breakdown of interaction components of cross section as a function of $\cos\,\theta_{p}$ according to GENIE v2. The number of events per bin is shown with arbitrary normalization.}
    \label{fig:xsec_breakdown_pangle}
\end{figure}

\begin{figure}[h!]
    \centering
    \includegraphics[width=0.45\textwidth]{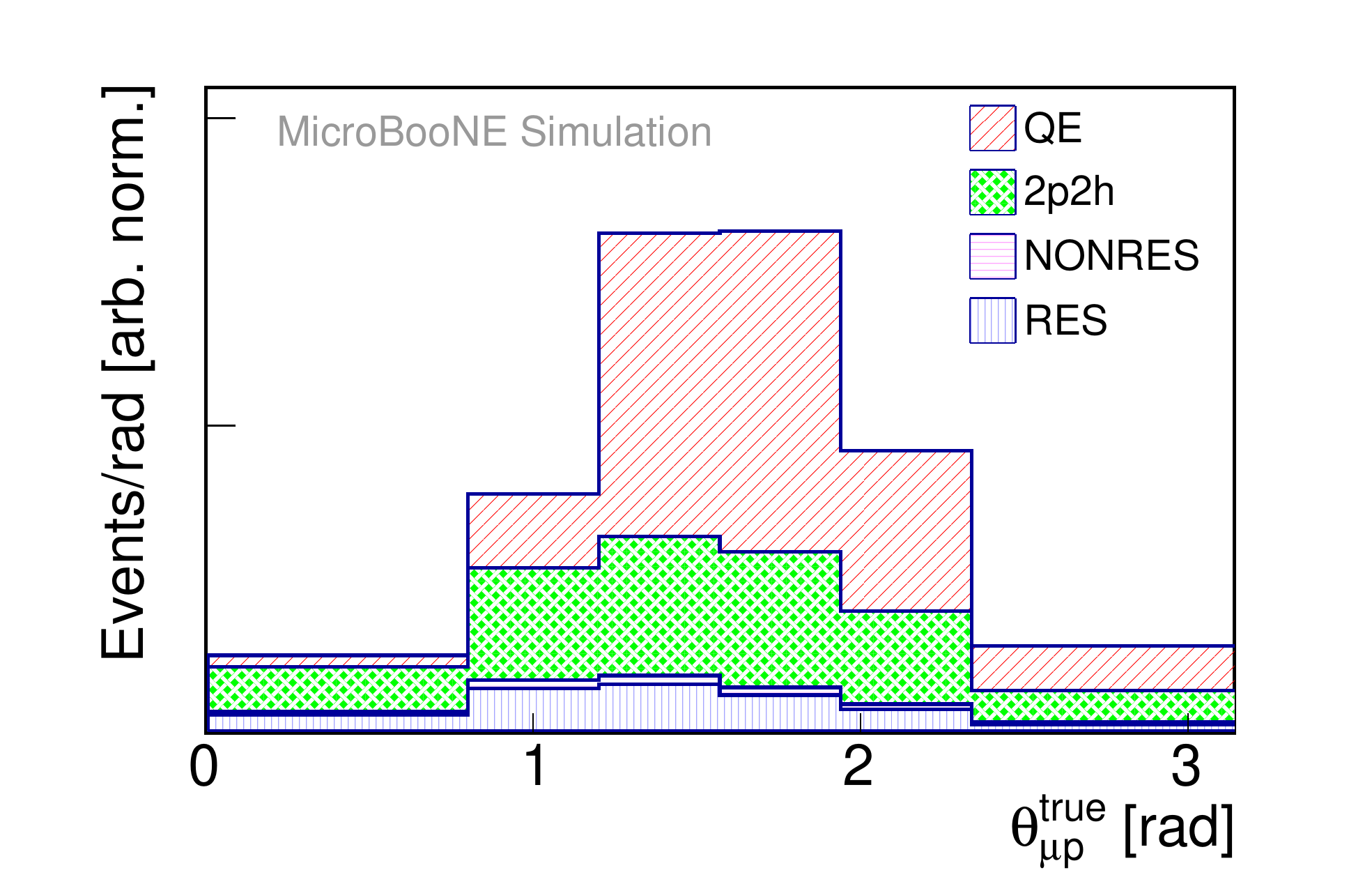}
    \caption{Breakdown of interaction components of cross section as a function of $\theta_{\mu, p}$ according to GENIE v2. The number of events per bin is shown with arbitrary normalization.}
    \label{fig:xsec_breakdown_thetamup}
\end{figure}

The signal definition for this measurement includes contributions from three processes: CCQE, $2p2h$, and pion production followed by pion absorption in the same nucleus.
All of these processes can be further modified by Fermi motion and intranuclear rescattering.
Although there is no attempt to isolate any of them,
the Monte Carlo events that survive the analysis criteria can provide insight into how these processes are distributed in the interaction variables provided.
The relative size and shape of these processes is shown in Figs.~\ref{fig:xsec_breakdown_mumom}-\ref{fig:xsec_breakdown_thetamup} for GENIE v2, see Sec.~\ref{sec:evgen} for details), the model used for backgrounds and systematic uncertainty determination.
True CCQE is predicted to be the largest component, contributing 50\% of the total and it is the largest fraction in most bins.
There are no contributions from other sources such as coherent pion production or production of other mesons.

The subdivision of simulated events as a function of muon momentum is shown in Fig.~\ref{fig:xsec_breakdown_mumom}.
GENIE predicts all underlying processes to have a similar shape in this variable.
At low proton momentum in Fig.~\ref{fig:xsec_breakdown_pmom}, the CCQE contribution is large and growing as $p_p$ decreases.
Simulations show that events at both low $p_p$ and $\theta_{\mu p}$ are predominantly due to FSI.
For both muon and proton cos($\theta$) spectra, very forward angle bins have a large component of $2p2h$ events.
Although CCQE scattering is the dominant reaction mechanism at opening angles of about 90$\degree$ as expected from the 2-body kinematics, the small- and large-angle regions are an interesting combination of interactions which will have to be understood through further theoretical and experimental studies.

\subsection{Model Comparisons} 
\label{sec:models}

Models of neutrino interactions are improving rapidly and experiments must make a choice of models to use for efficiency and background estimation at the time of an analysis.  
For this MicroBooNE measurement, GENIE v2 is used.  
However, data are compared with GENIE v3.0.6 with tune G18\_10a\_02\_11a (labelled as ``GENIE v3''), NuWro 19.02.1~\cite{Golan:2012wx}, NEUT 5.4.0.1~\cite{Hayato:2009zz}, and GiBUU 2019~\cite{Buss:2011mx}.  
As described in Sec.~\ref{sec:xsect_def}, each calculation was folded with our smearing matrix before comparing with data.

GENIE, NuWro, and NEUT are developed by separate groups who often implement the same models in different ways.
In this case, each code uses the local Fermi gas (LFG) momentum distribution and a binding energy correction which can be constant (GENIE and NEUT) or derived from a potential (NuWro) for sampling the struck nucleon properties.  
Each code uses a similar CCQE model, but apply different empirical RPA corrections.  
GENIE also applies a Coulomb correction~\cite{Nieves:2004wx} on the outgoing muon.
All modern codes contain a multinucleon mechanism.  
GENIE v3, NuWro, and NEUT all use the Valencia $2p2h$ model~\cite{Gran:2013kda}.
The new models are built to agree with \mb data~\cite{miniboone-ccqe}, which uses the same beamline as \mub but a hydrocarbon target.

Both NEUT and GENIE use the Kuzmin-Lubushkin-Naumov~\cite{Kuzmin:2003ji} and Berger-Sehgal\cite{Berger:2007rqRES} model for  resonance and the Berger-Sehgal model for coherent~\cite{Berger:2008xsCOH} scattering.
Implementation differs in the cutoff in total hadronic energy in the nucleon rest frame $W$ between resonance and DIS models, e.g. fixed cutoff value of 1.9 GeV (GENIE) and 2.0 GeV (NEUT).  NuWro is a little different, using the Adler-Rarita-Schwinger model~\cite{Graczyk:2007xk} for the $\Delta$(1232) resonance and a smooth transition to DIS at 1.6 GeV.
All use form factors fit to data.  
In addition, each code uses a similar extrapolation of DIS processes to $\pi N$ threshold~\cite{Yang:1999xg,Bodek:2002ps} to model non-resonant processes.
Hadrons produced at the neutrino interaction vertex undergo FSI which are governed by models based on the impulse approximation with nuclear medium corrections. Here, FSI includes both pion and nucleon interactions with most generators using separate models to simulate the two types of interaction.
NEUT and NuWro use the Salcedo-Oset model~\cite{Salcedo:1987md} for pions which includes nuclear medium effects.  
GENIE uses a more empirical model~\cite{Katori:2013eoa} for pions which has no nuclear medium corrections.  
NuWro and GENIE use nuclear medium corrections for nucleon FSI~\cite{Pandharipande:1992zz} while the NEUT code is strictly impulse approximation for nucleon FSI.

GiBUU~\cite{Buss:2011mx} was developed from a first principles approach as opposed to the greater emphasis on empirical models with other generators.
Although the interactions covered are very similar to the generators described above, the details are different as all of the components were developed as a coherent package.
Instead of a semi-classical approximation for propagating hadrons as is used in the models described above, a transport equation is solved~\cite{Buss:2011mx}.
Nucleon momenta come from an LFG distribution, then hadrons propagate through the residual nucleus in a nuclear potential that is consistent between initial and final state.
The CCQE interaction is handled with a spectral function that has separate momentum and energy dependence.
All resonances propagate as particles and interact according to best available models.   
Version 2019 is used here; a recent publication~\cite{Mosel:2017ssx} shows comparisons of this version with the T2K $CC0\pi$ data~\cite{Abe:2018pwo}.

\subsection{Cross-Section Results} 
\label{sec:xsect}

\begin{figure}[th!]
    \centering
    \includegraphics[width=0.45\textwidth]{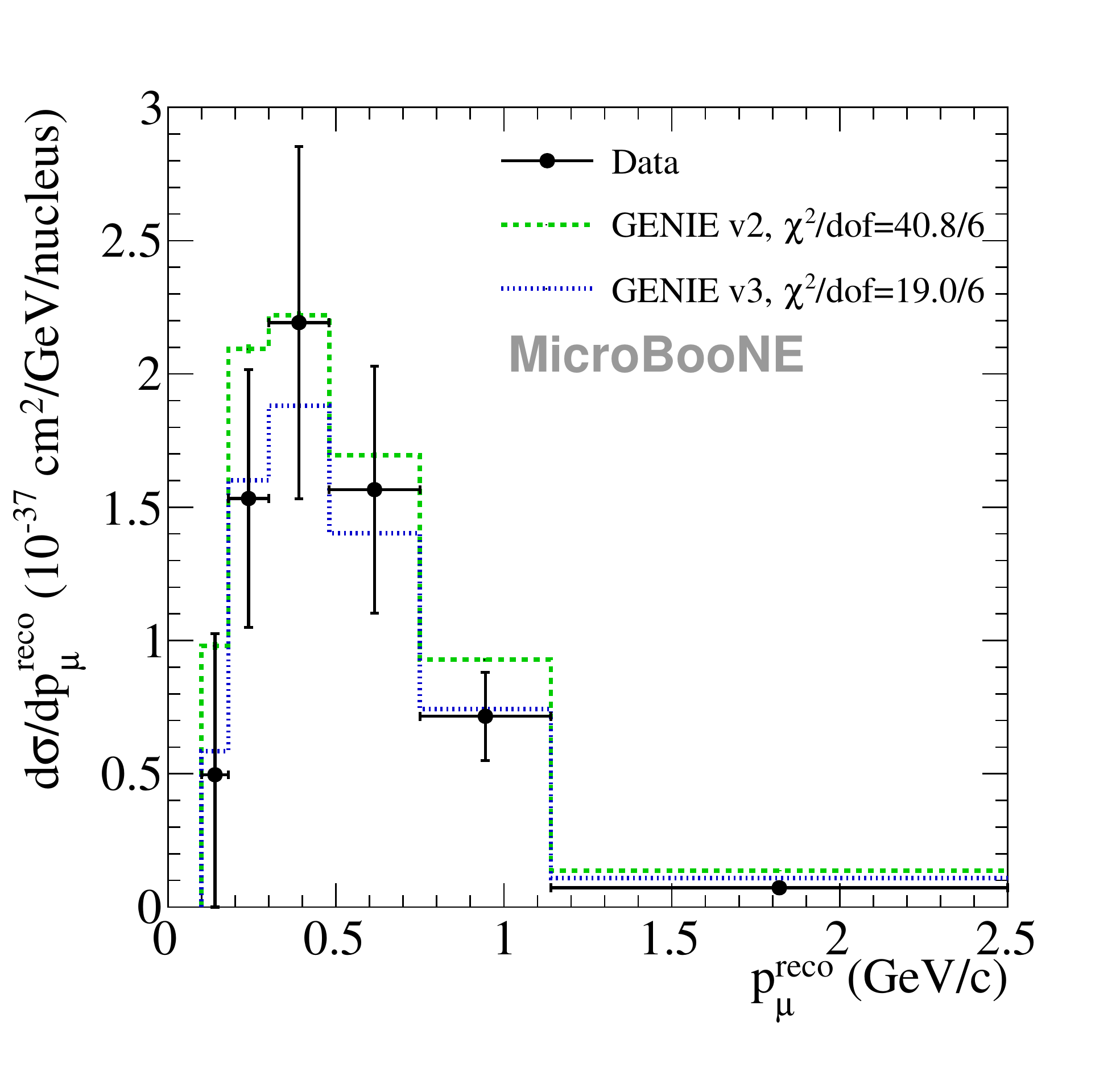}
    \caption{Measured cross section as a function of $p_{\mu}$(GeV) compared with GENIE v2 and v3 (see Sec.~\ref{sec:models} for details) from a data exposure of $1.6\times10^{20}$ P.O.T.  Error bars include all contributions from  statistical and systematic sources in the final analysis.}
    \label{fig:xsec_mumom_genie}
\end{figure}

\begin{figure}[h!]
    \centering
    \includegraphics[width=0.45\textwidth]{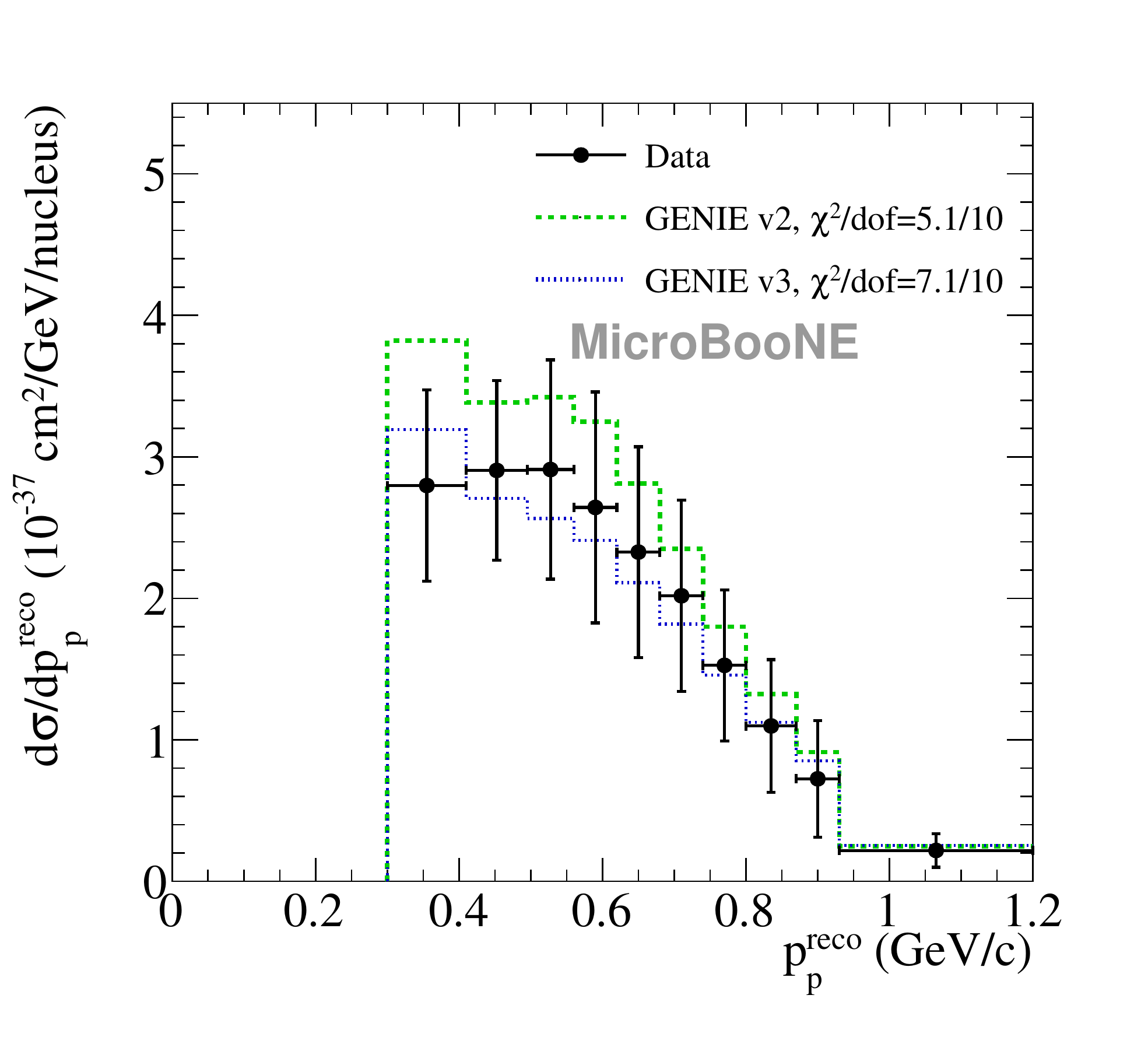}
    \caption{Measured cross section as a function of $p_{p}$ compared with GENIE v2 and v3 (see Sec.~\ref{sec:models} for details).}
    \label{fig:xsec_pmom_genie}
\end{figure}

\begin{figure}[h!]
    \centering
    \includegraphics[width=0.45\textwidth]{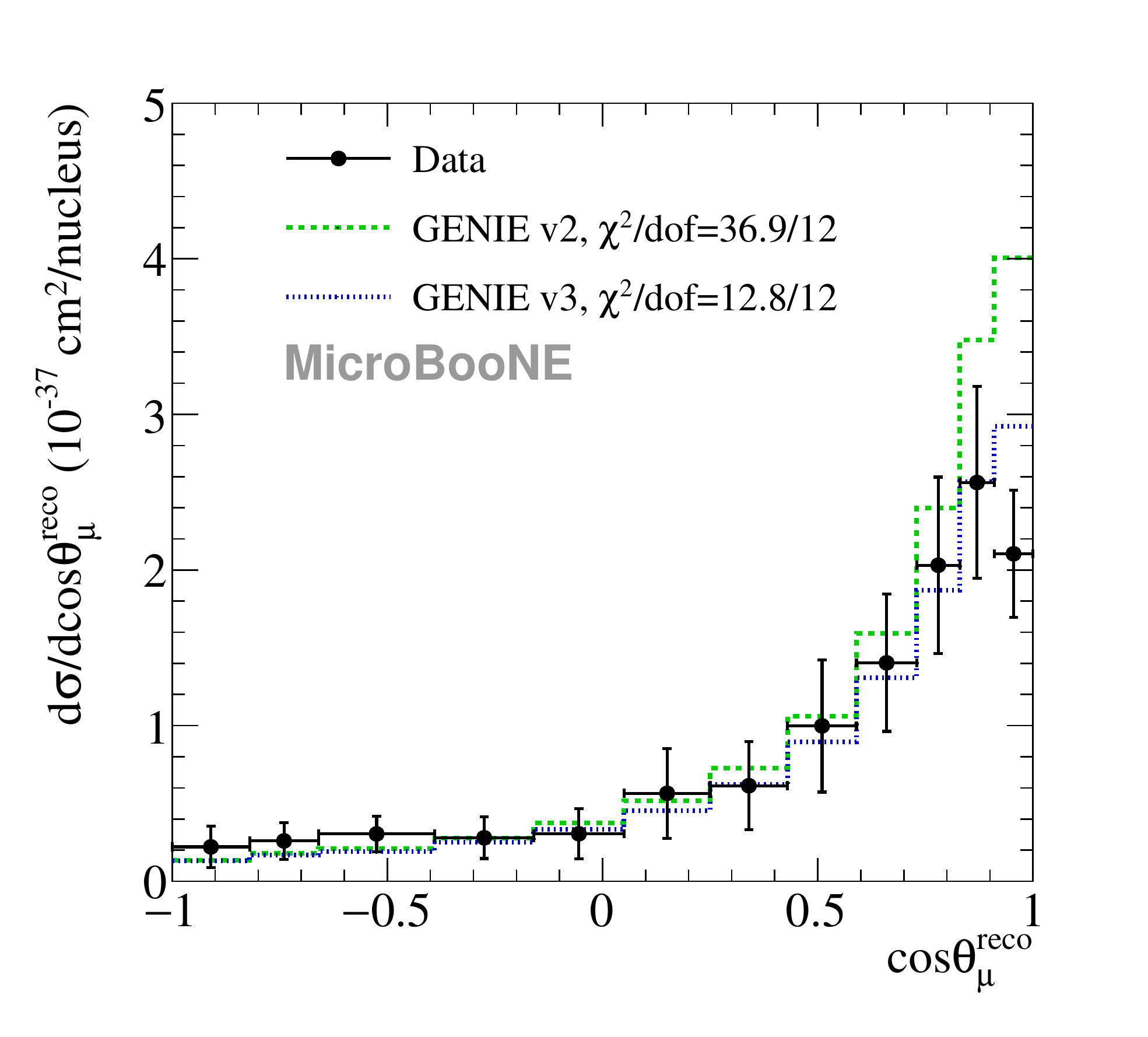}
    \caption{Measured cross section as a function of $\cos\,\theta_{\mu}$ compared with GENIE v2 and v3 (see Sec.~\ref{sec:models} for details).}
    \label{fig:xsec_muangle_genie}
\end{figure}

\begin{figure}[h!]
    \centering
    \includegraphics[width=0.45\textwidth]{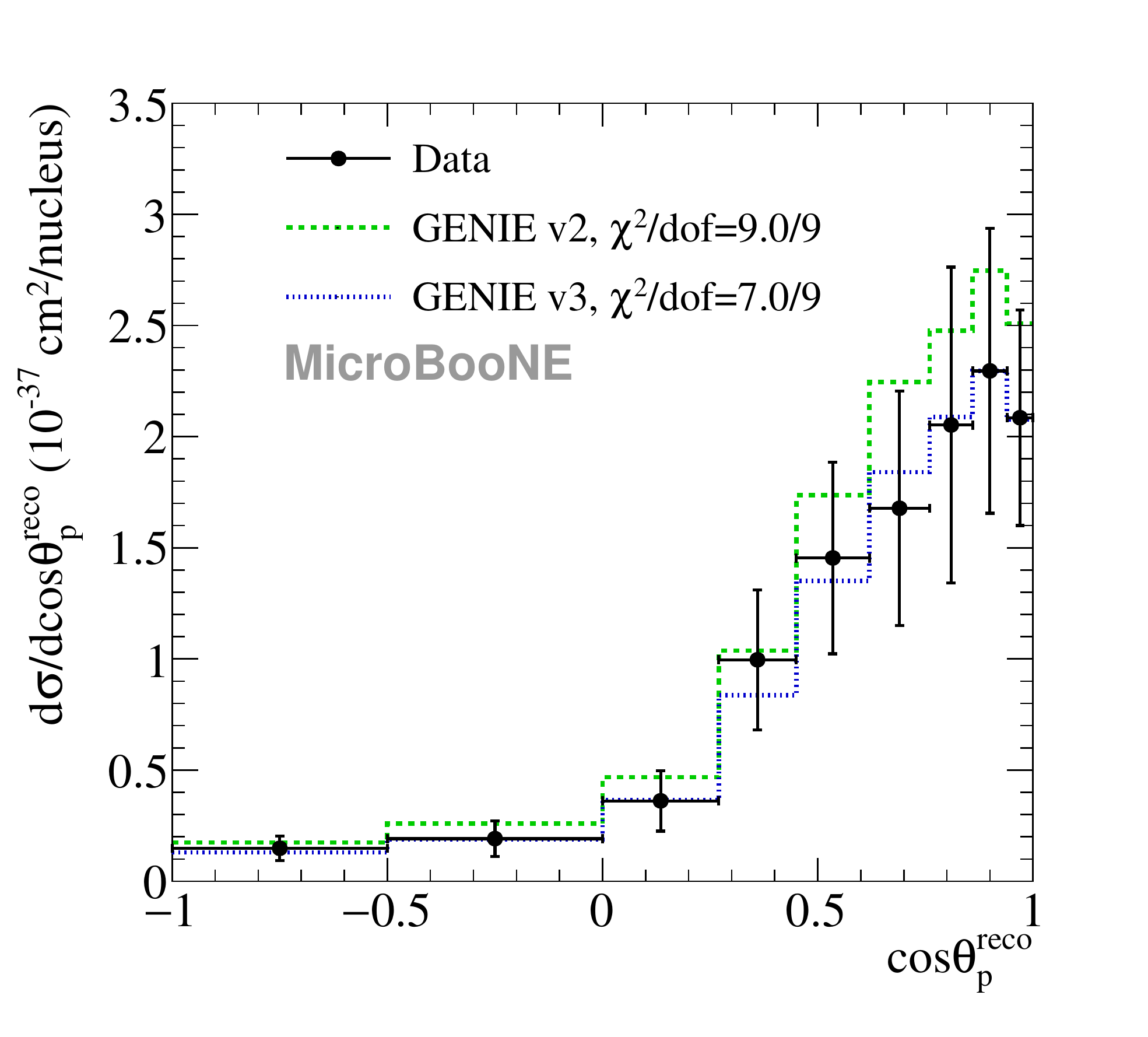}
    \caption{Measured cross section as a function of $\cos\,\theta_{p}$ compared with GENIE v2 and v3 (see Sec.~\ref{sec:models} for details).}
    \label{fig:xsec_pangle_genie}
\end{figure}

\begin{figure}[h!]
    \centering
    \includegraphics[width=0.45\textwidth]{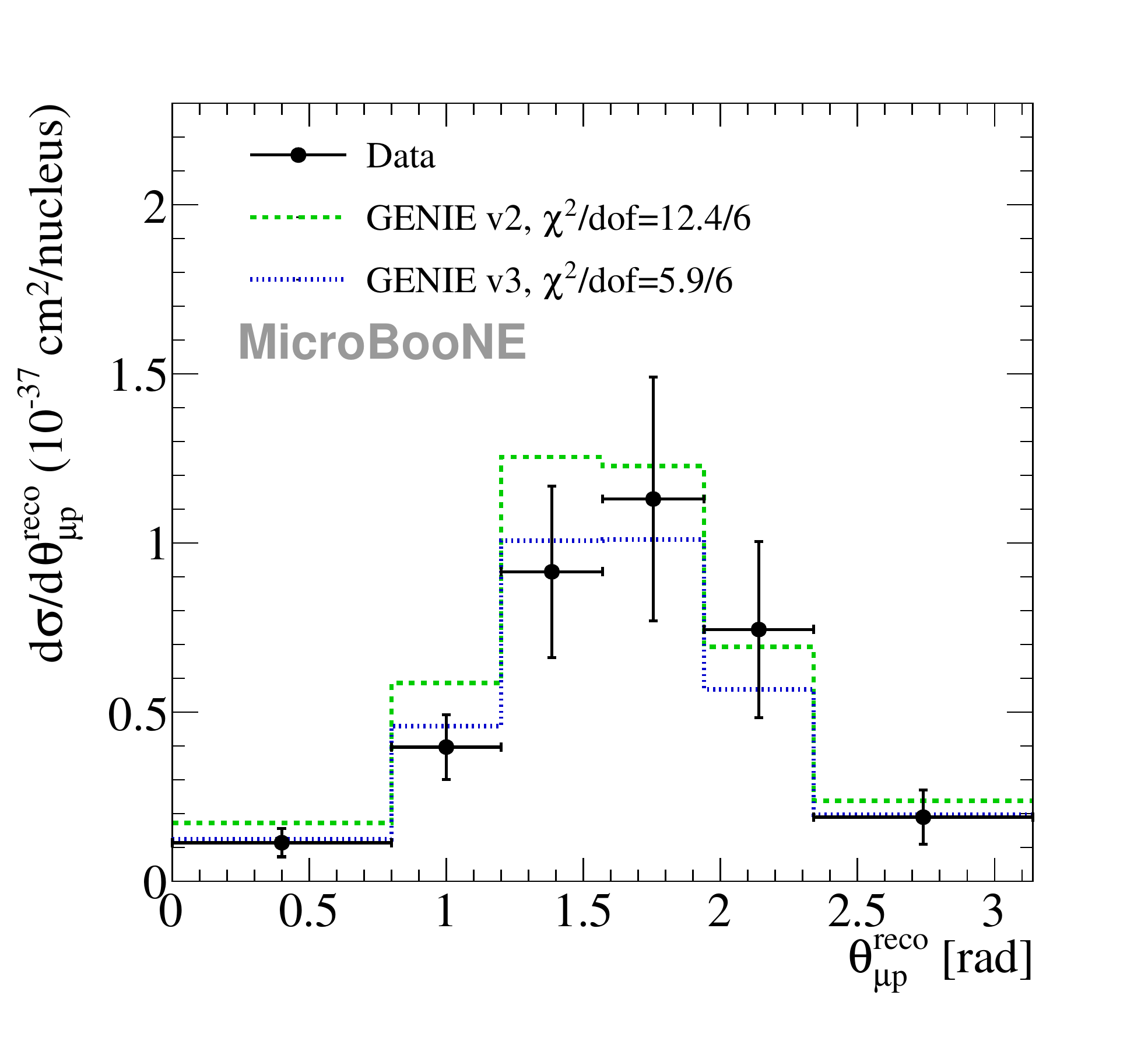}
    \caption{Measured cross section as a function of $\theta_{\mu, p}$ compared with GENIE v2 and v3 (see Sec.~\ref{sec:models} for details).}
    \label{fig:xsec_thetamup_genie}
\end{figure}

The first set of comparisons between the extracted CC0$\pi$Np cross section to both GENIE v2 and GENIE v3
is shown in Figs.~\ref{fig:xsec_mumom_genie}-\ref{fig:xsec_thetamup_genie}.  The purpose is to show the evolution of generator models from the time the analysis was started to the time it was finished.  For the muon momentum and polar angle, 
the overall magnitude in GENIE v3 is about 20\% smaller than v2 and in better agreement with these data even though no tuning to these data was performed.  A significant change is also seen in the shape of these inclusive distributions, particularly for forward muon angles where effects such as $2p2h$ and nucleon-nucleon correlations (e.g. the RPA correction) are important, and at low proton momentum where both FSI effects and the contribution of $2p2h$ are dominant.  These are effects that are widely considered to be at the forefront of interest for modelers and it is clear that the updated models in GENIE provide a better description of these data.

Calculations of $\chi^2$ using the covariance matrices derived for this measurement allow a more quantitative  assessment.  
There is an improvement of at least 25\% and over 50\% in some cases for the calculated $\chi^2$ when changing from GENIE v2 to GENIE v3.
The exception is $p_p$ where GENIE v3 is better at low momentum, but the overall $\chi^2$ is lower for GENIE v2.


\begin{figure}[h!]
    \centering
    \includegraphics[width=0.45\textwidth]{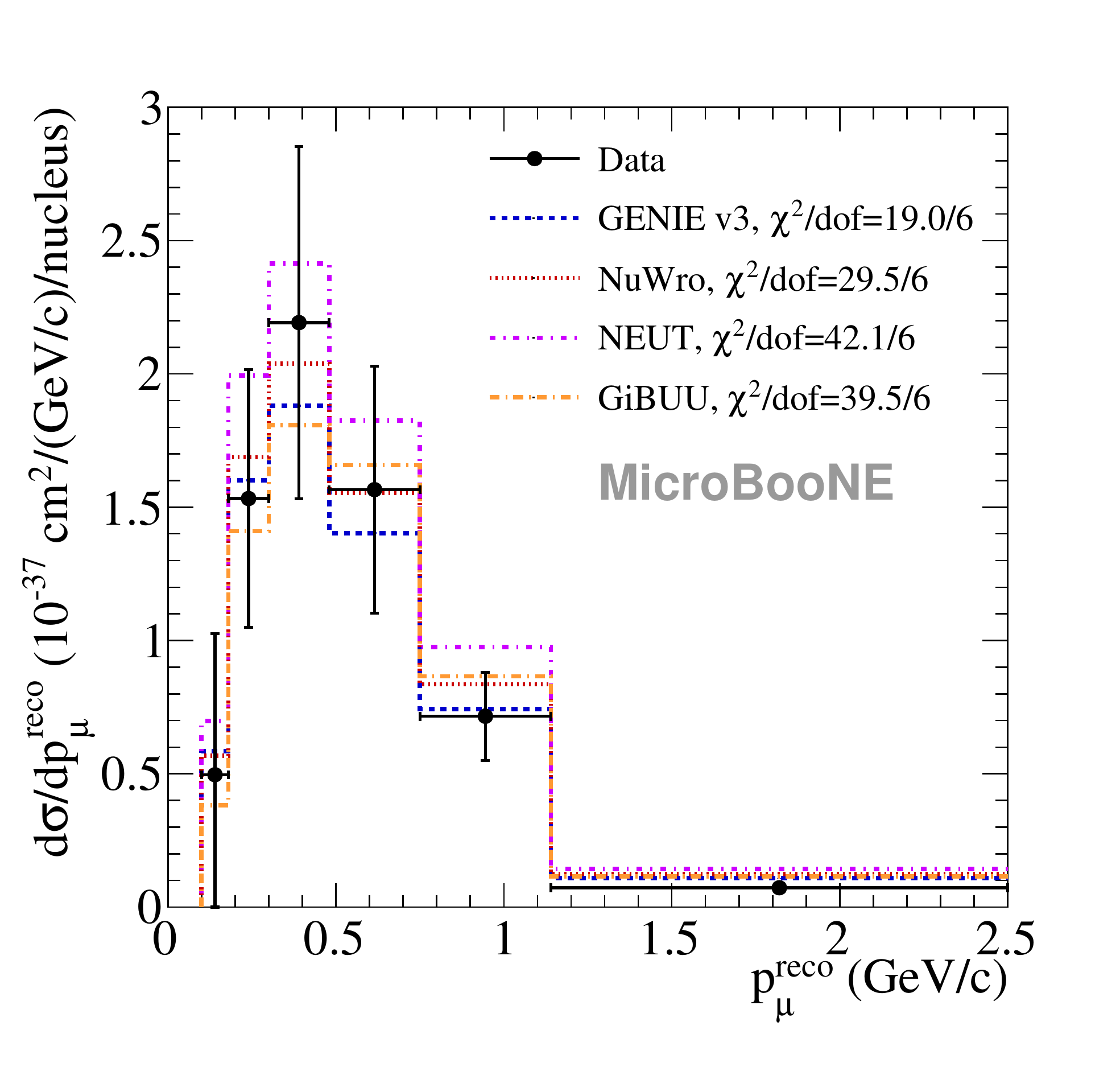}
    \caption{Measured cross section as a function of $p_{\mu}$ compared with GENIE v3, NuWro, NEUT, and GiBUU (see Sec.~\ref{sec:models} for details) from a data exposure of $1.6\times10^{20}$ P.O.T.  Error bars include all contributions from statistical and systematic sources in the final analysis.}
    \label{fig:xsec_mumom_modern}
\end{figure}
\begin{figure}[h!]
    \centering
    \includegraphics[width=0.45\textwidth]{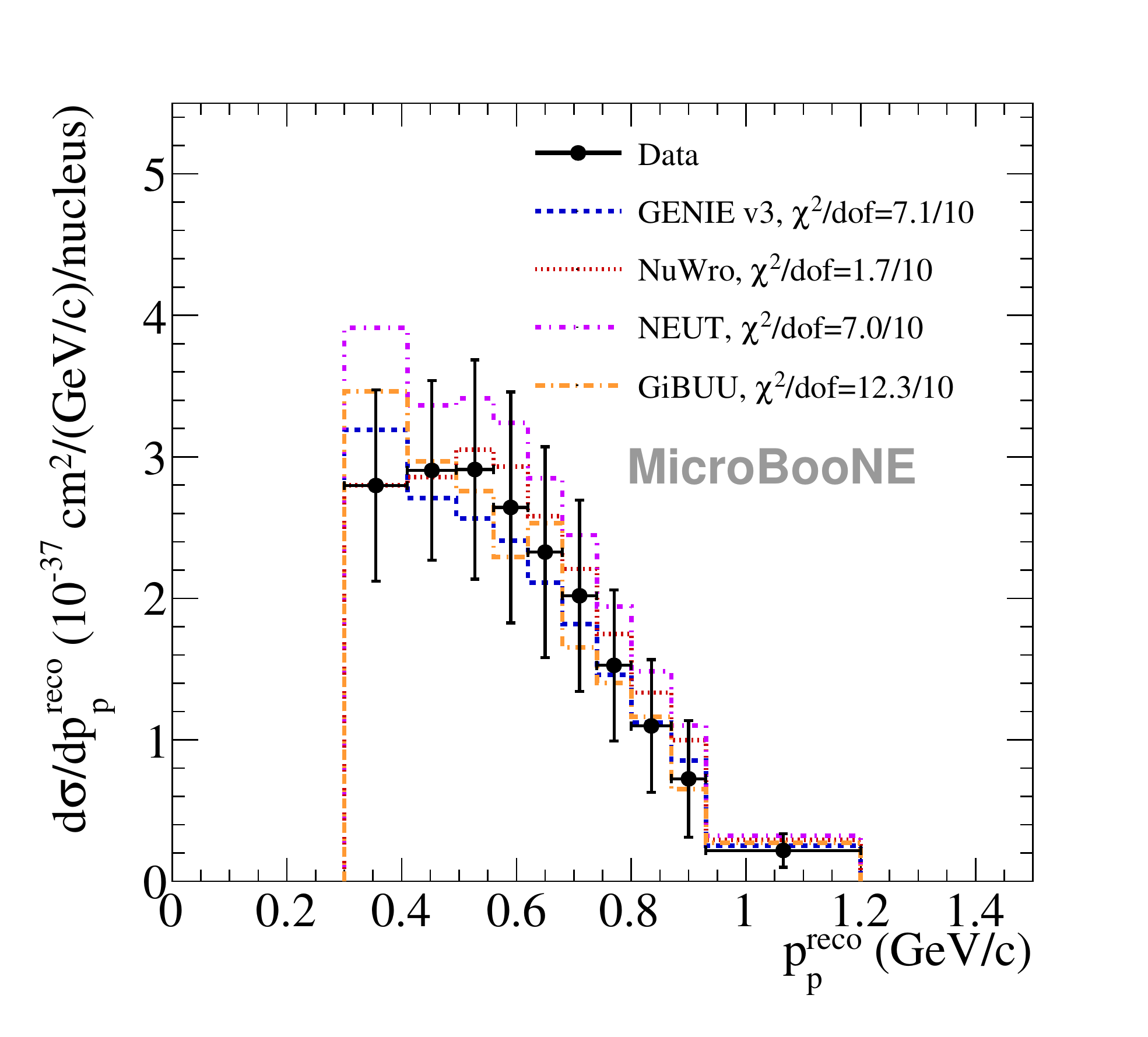}
    \caption{Measured cross section as a function of $p_{p}$ of the leading proton candidate compared with various models.  See Fig.~\ref{fig:xsec_mumom_modern} and Sec.~\ref{sec:models} for more details.}
    \label{fig:xsec_pmom_modern}
\end{figure}

\begin{figure}[ht]
    \centering
    \includegraphics[width=0.45\textwidth]{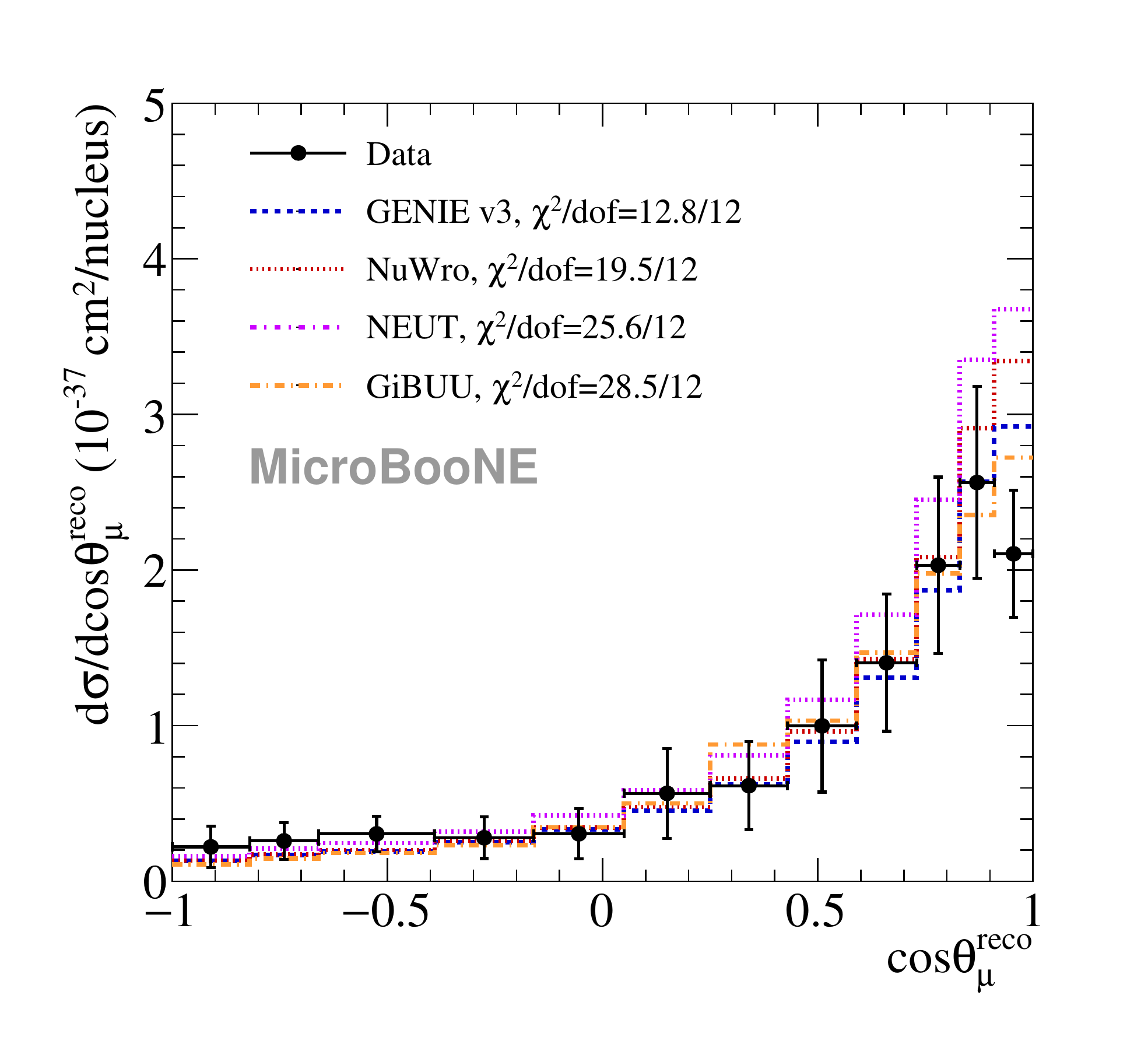}
    \caption{Measured cross section as a function of $\cos\theta_{\mu}$ compared with various models.  See Fig.~\ref{fig:xsec_mumom_modern} and Sec.~\ref{sec:models} for more details.}
    \label{fig:xsec_muangle_modern}
\end{figure}

\begin{figure}[ht!]
    \centering
    \includegraphics[width=0.45\textwidth]{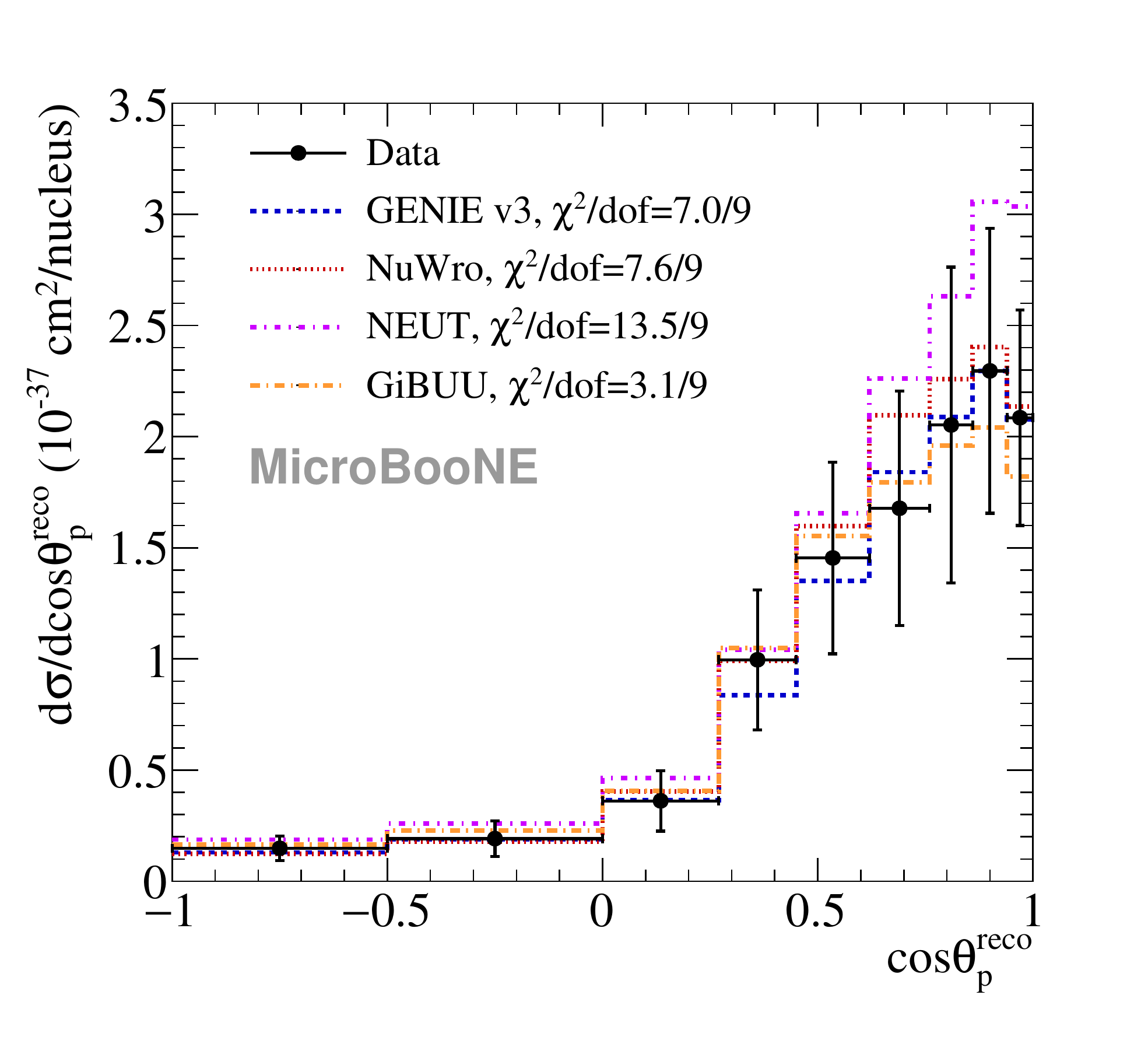}
    \caption{Measured cross section as a function of $\cos\theta_{p}$ of the leading proton candidate compared with various models.  See Fig.~\ref{fig:xsec_mumom_modern} and Sec.~\ref{sec:models} for more details.}
    \label{fig:xsec_pangle_modern}
\end{figure}

\begin{figure}[ht]
    \centering
    \includegraphics[width=0.45\textwidth]{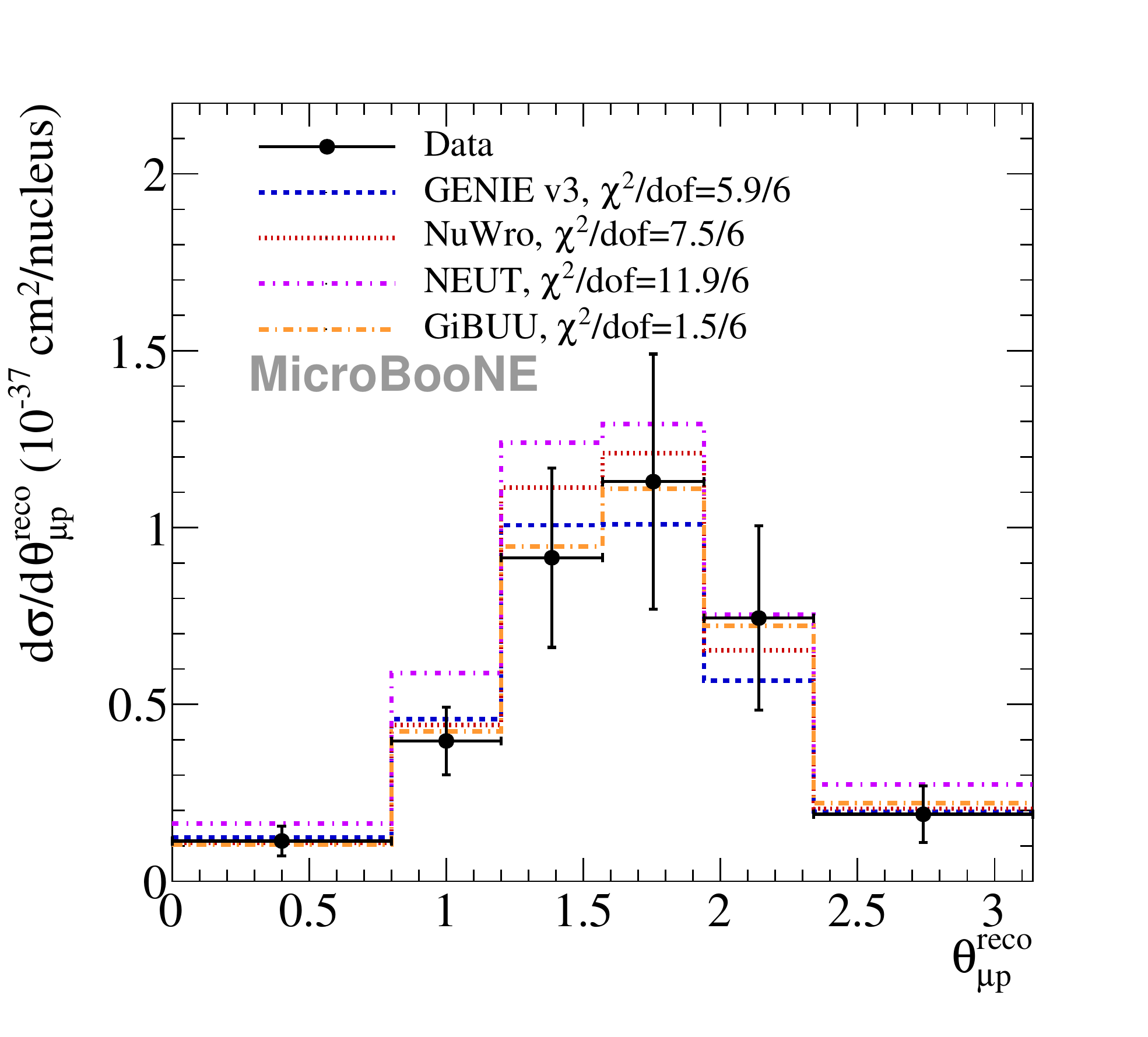}
    \caption{Measured cross section as a function of $\theta_{\mu, p}$ compared with various models.  See Fig.~\ref{fig:xsec_mumom_modern} and Sec.~\ref{sec:models} for more details.}
    \label{fig:xsec_thetamup_modern}
\end{figure}

Figures~\ref{fig:xsec_mumom_modern}-\ref{fig:xsec_thetamup_modern} contain comparisons of the data with GENIE v3, NuWro, NEUT, and GiBUU.  
These codes had similar evolution in models as was seen in GENIE for Figs.~\ref{fig:xsec_mumom_genie}-\ref{fig:xsec_thetamup_genie}, but here only the most recent results are shown.
Although agreement between calculations varies widely, the strongest sensitivity appears to be at the most forward muon and proton angles and lowest proton momenta.  

The muon momentum distribution (Fig.~\ref{fig:xsec_mumom_modern}) has the largest $\chi^2$ values in comparison with the calculations, ranging from about 3-7 units of $\chi^2$ per degree of freedom (dof).
These large $\chi^2$ values are driven by the highest momentum bin, which has a relatively small uncertainty and is in tension with all predictions.

The leading proton momentum distribution (Fig.~\ref{fig:xsec_pmom_modern}) in an inclusive spectrum is seldom seen in the literature for neutrino experiments.
Data at low leading proton momentum is most sensitive to FSI and nuclear effects and where new sensitivity is shown in this work.
At the lowest momenta in this measurement ($p_p<500$ MeV/c), the models show considerable variation, with NEUT furthest from the data, followed by GiBUU, GENIE v3, and finally NuWro predicting the lowest bin almost perfectly. 
For $\chi^2/dof$, all calculations have values of around 1 or lower.

At forward angles, the muon polar angle cross sections (Fig.~\ref{fig:xsec_muangle_modern}) have shown sensitivity in the past~\cite{Abe:2018pwo,Ruterbories:2018gub} because this is where nuclear effects such as nucleon-nucleon correlations are strongest.  
Model results vary by about 30\% at forward angles.
It is interesting that none of the calculations have the turnover at the most forward muon angle bin that is seen in the data.
At the beam energies of this measurement, both muons and protons are dominantly produced at forward angles due to the Lorentz boost.  The data at negative values of $\cos(\theta_\text{proton}^\text{reco})$ are particularly interesting as Monte Carlo simulations show that the protons at backward angles are almost totally due to FSI.  The proton polar angle cross section and comparison with model calculations is shown in Fig.~\ref{fig:xsec_pangle_modern}.  
According to Monte Carlo simulation, both muons and protons at forward angles are dominated by the CCQE interaction channel.
GiBUU has the highest $\chi^2$ value for the muon angle and the lowest $\chi^2$ for proton angle.  

The opening angle between the muon and the leading proton (Fig.~\ref{fig:xsec_thetamup_modern}) can show different features for CCQE and other mechanisms because it is more strongly peaked at about 90$\degree$ for CCQE and flatter for the other mechanisms that are expected to contribute to these data (see Fig.~\ref{fig:xsec_breakdown_thetamup}).  
The measured cross section in the lowest and highest angular bins is approximately 10\% of the cross section at the peak.
Simulations indicate the small opening angle data is populated entirely by events with FSI.
All the calculations have the two alternate mechanisms, $2p2h$ and RES, and tend to follow the data.  
The peak position in the calculations shows large variation.  Simulations with GENIE v3 show the peak position of the data is sensitive to FSI and binding energy effects.
The largest $\chi^2/dof$ value is for NEUT (2.0) and the smallest value is for GiBUU (0.25).

In general NEUT and GENIE v2 both overpredict the cross section, with the other predictions closer to the measured data.
However the $\chi^2/dof$ values are sensitive to shape as well as normalization, showing that in some variables NEUT predicts the shape much better than some other generators.

\section{Conclusions}

New muon neutrino cross section data for the CC$0\pi Np$ 
interactions on argon from the \mub  experiment are presented.
The simultaneous presentation of these distributions (covering a wide phase space) in muon and leading proton momentum and angle is a first for neutrino interactions in liquid argon.
The signal definition was chosen to minimize model dependence and allow straightforward theoretical comparisons.
We specify that at least one proton above 300 MeV/c momentum must be detected, there are no protons with momentum greater than 1,200 MeV/c, and the muon momentum must be greater than 100 MeV/c.
No containment requirement is demanded of the muon track; if the muon is not contained, its momentum is determined by MCS~\cite{Abratenko:2017nkiMCS}.
The signal definition has no limits on proton or muon angle.
Particle identification cuts ensure that protons in the final selections stop inside the detector volume and very few of them interact in the detector volume.
We present distributions for the muon momentum and polar angle, leading proton momentum and polar angle, and the angle between the muon and leading proton.
These data have a low proton threshold compared other neutrino experiments, and the high statistics measurement down to 300~MeV/c allows improved model testing.
The loose signal requirement increases the statistical precision over a large phase space enabling a more precise measurement of kinematic shapes.

Comparisons between GENIE v2 (used for background determination, efficiencies and systematic uncertainties) and GENIE v3 (Sec.\ref{sec:models}) in Figs.~\ref{fig:xsec_mumom_genie}-\ref{fig:xsec_thetamup_genie} show significant improvement for GENIE v3 in the ability to describe these data.

Further comparisons are made to modern versions of neutrino-interaction models often used in predicting and simulating neutrino events for neutrino experiments.
While GiBUU has paid more attention to theoretical details of a calculation in a nuclear environment,
the other event generators have adopted models that are often similar to each other.
As a result, the differences in their ability to match the data stem primarily from subtle implementation and tuning differences.
Difficulties for models to describe the muon forward angle spectrum in the MINERvA~\cite{Ruterbories:2018gub} CC0$\pi$ data, the MicroBooNE CCinclusive measurement~\cite{Adams:2019iqcINCL}, and the MicroBooNE CCQE measurement~\cite{uB_CCQE_2020} are now also seen in this measurement (see Fig.~\ref{fig:xsec_muangle_modern}).
Most of the calculations predict a larger cross section in the two most forward bins by between 1$\sigma$ and 2$\sigma$.
Although the prediction from GiBUU is closest to the data in these bins than the other predictions, it has the largest overall $\chi^2$ for this distribution when considering the full shape.
In general the shape is more constrained than the normalization in the covariance matrices for these data.

In addition to difficulties in describing the muon angular spectrum, interaction models are challenged to predict the rate of production of low energy protons from a heavy target.
A large variation in the calculations of proton momentum below roughly 600~MeV/c (175.4 MeV kinetic energy) is seen with most of the results overpredicting the cross section.
This is a stringent test of various components of the model, particularly proton FSI.
While NEUT has the worst agreement below 600 MeV/c, GiBUU has the largest $\chi^2$ for this distribution (1.2 per dof).
NuWro has the best agreement in the low momentum bins and the lowest $\chi^2$ value (0.7 per dof).

The opening angle ($\theta_{\mu p}$) distribution (Fig.~\ref{fig:xsec_thetamup_modern}) tests the underlying reaction mechanism in the most detailed way.  
Since the spectrum peaks at values of about $\pi/2$ radians, this is a strong indication that CCQE interactions dominate this data set, as expected.
There is a large variation in $\chi^2$ among the calculations with GiBUU having the best (0.25 per dof) and NEUT the worst (2.0 per dof) values.

The largest $\chi^2$ values are seen for $p_\mu$.  
The largest contribution comes from the highest momentum bin and agreement with data in the peak region is acceptable.

These data significantly enhance the ability to assess understanding of the neutrino interaction in argon at energies of roughly 1 GeV.
Overall, there is no dramatic disagreement of  calculations with these data.
However, problems previously seen with calculations for carbon targets that predict larger cross sections than observed in the forward muon angular spectrum are confirmed here for argon.
This work provides new information about the difficulties in describing the forward scattering region and low energy proton production in argon.
All calculations describe the overall trend of the distributions with widely varying ability to describe the details accurately. 
An overarching conclusion is that newer calculations which have shown improved agreement with light target data (CH or CH$_2$) also describe data from a heavier target (Ar) with similar accuracy.
\FloatBarrier

\begin{acknowledgments}
This document was prepared by the MicroBooNE collaboration using the resources of the Fermi National Accelerator Laboratory (Fermilab), a U.S. Department of Energy, Office of Science, HEP User Facility. Fermilab is managed by Fermi Research Alliance, LLC (FRA), acting under Contract No. DE-AC02- 07CH11359. MicroBooNE is supported by the following: the U.S. Department of Energy, Office of Science, Offices of High Energy Physics and Nuclear Physics; the U.S. National Science Foundation; the Swiss National Science Foundation; the Science and Technology Facilities Council of the United Kingdom; and The Royal Society (United Kingdom).

\end{acknowledgments}

\appendix

\section{Supplementary Information - Efficiencies, Migration Matrices}
\begin{figure}[h!]
    \centering
    \includegraphics[width=0.45\textwidth]{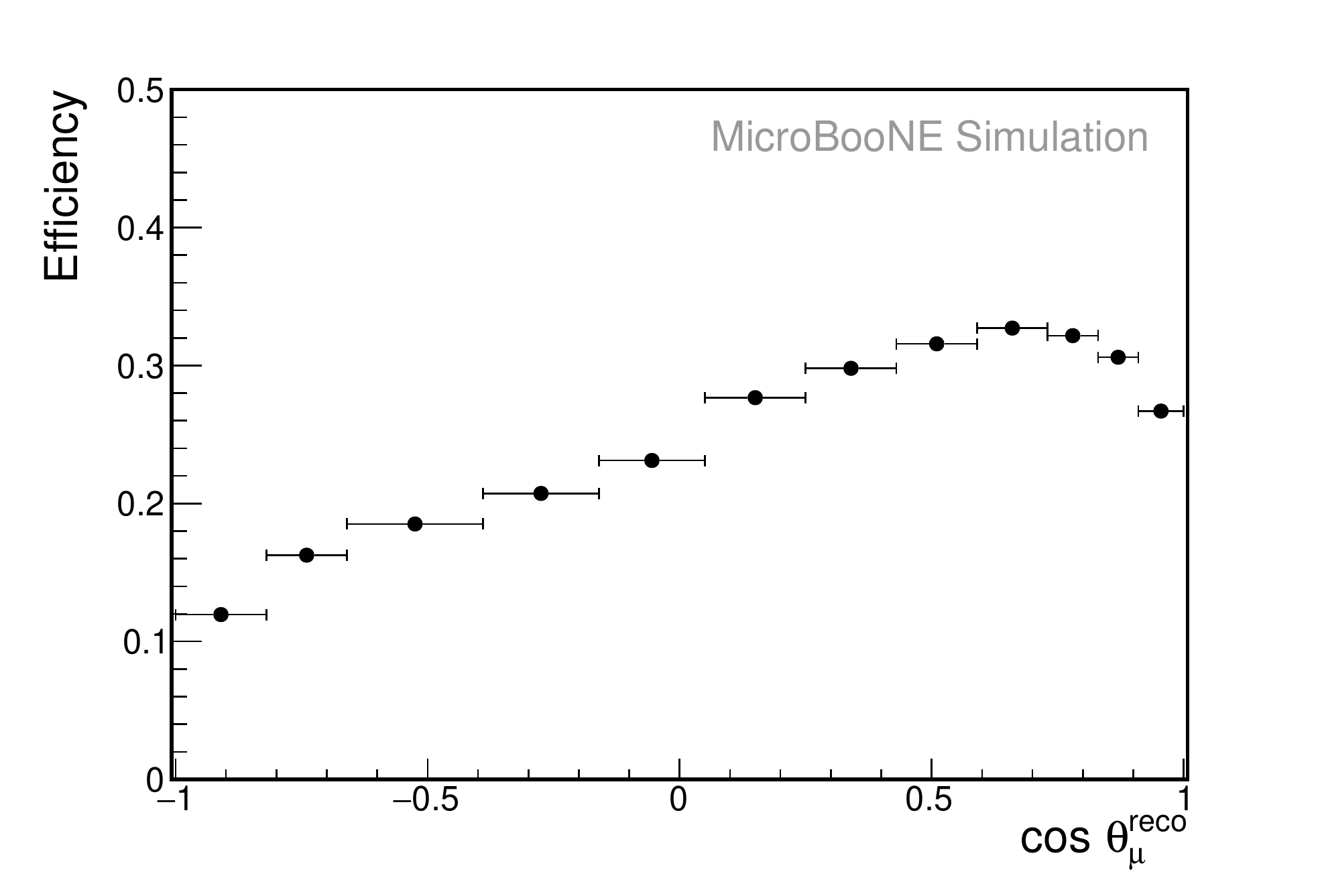}
    \caption{Efficiency as a function of reconstructed muon polar angle ($\cos(\theta_\mu)$).  Statistical error bars are too small to be seen.}
    \label{fig:eff_muangle}
\end{figure}

\begin{figure}[h!]
    \centering
    \includegraphics[width=0.45\textwidth]{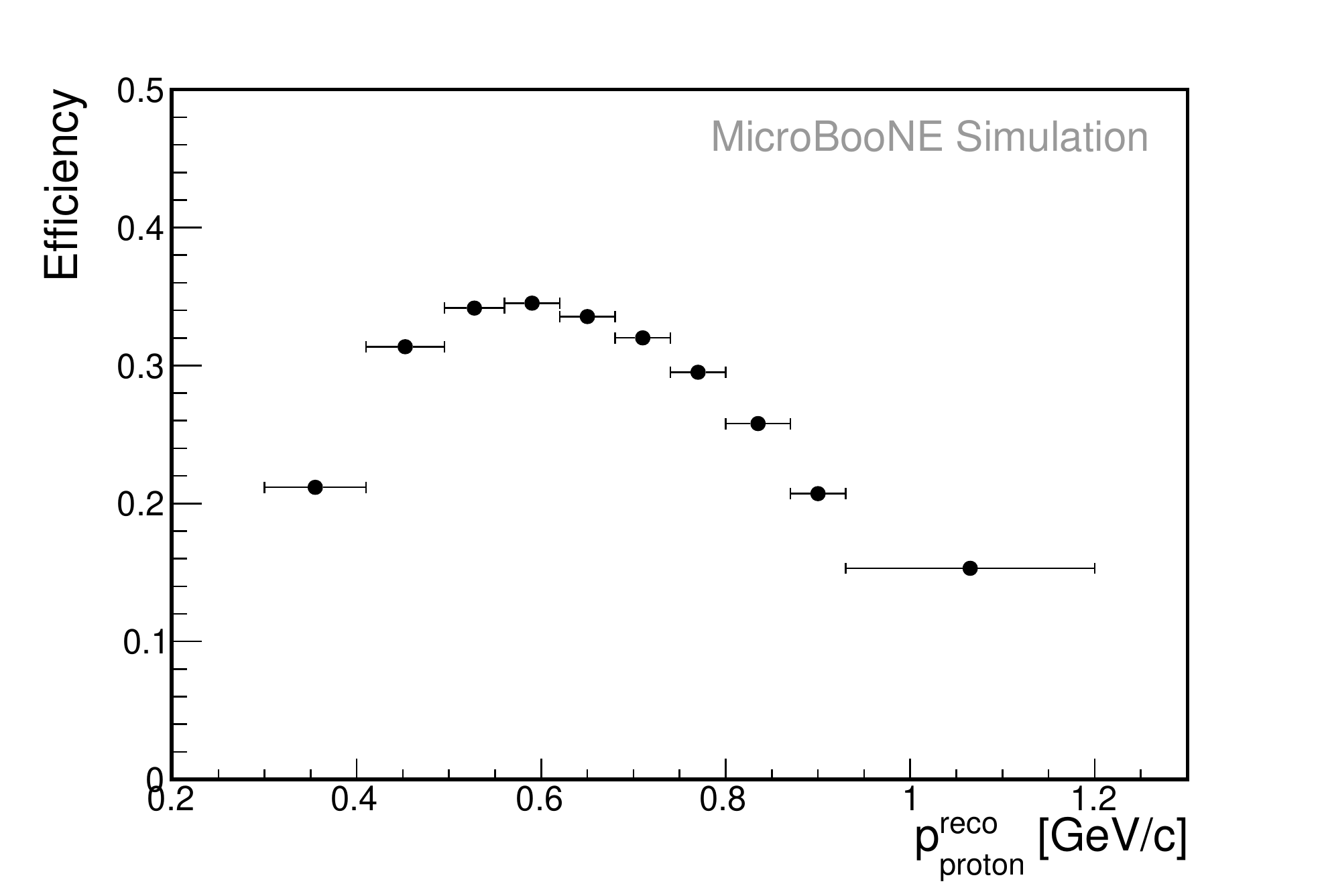}
    \caption{Efficiency as a function of reconstructed proton momentum ($p_p$) in simulated CC$0\pi Np$ events.}
    \label{fig:eff_pmom}
\end{figure}

Sec.~\ref{sec:efficiency} contained the efficiency plot for muon momentum.  
Efficiency plots for muon polar angle (Fig.~\ref{fig:eff_muangle}), leading proton momentum (Fig.~\ref{fig:eff_pmom}),
leading proton polar angle (Fig.~\ref{fig:eff_pangle}), and muon-leading proton opening angle (Fig.~\ref{fig:eff_thetamup}) are shown below.
The Monte Carlo predictions have been smeared with the migration matrices described in Sec.~\ref{sec:xsect_def}.
Each distribution peaks at roughly 35\% efficiency with no bin having an efficiency less than 10\%.

\begin{figure}
    \centering
    \includegraphics[width=0.45\textwidth]{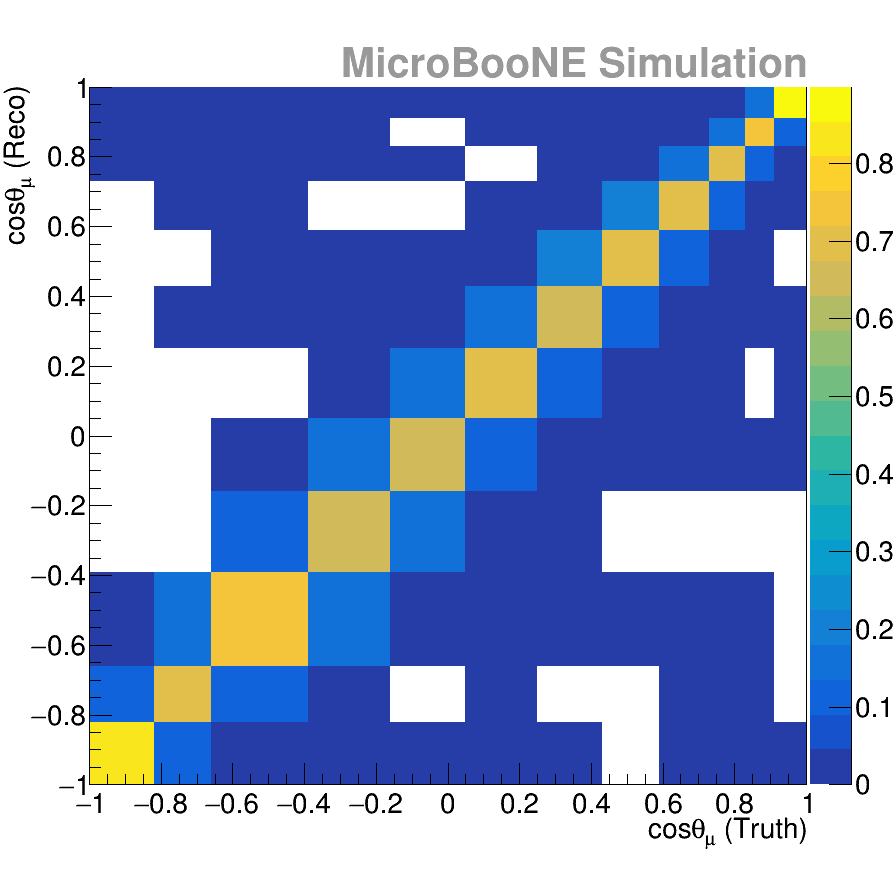}
    \caption{Migration matrix between true and reconstructed bins for $\cos(\theta_\mu)$ in simulated CC$0\pi Np$ events.}
    \label{fig:migmat_muangle}
\end{figure}

\begin{figure}[h!]
    \centering
     \includegraphics[width=0.45\textwidth]{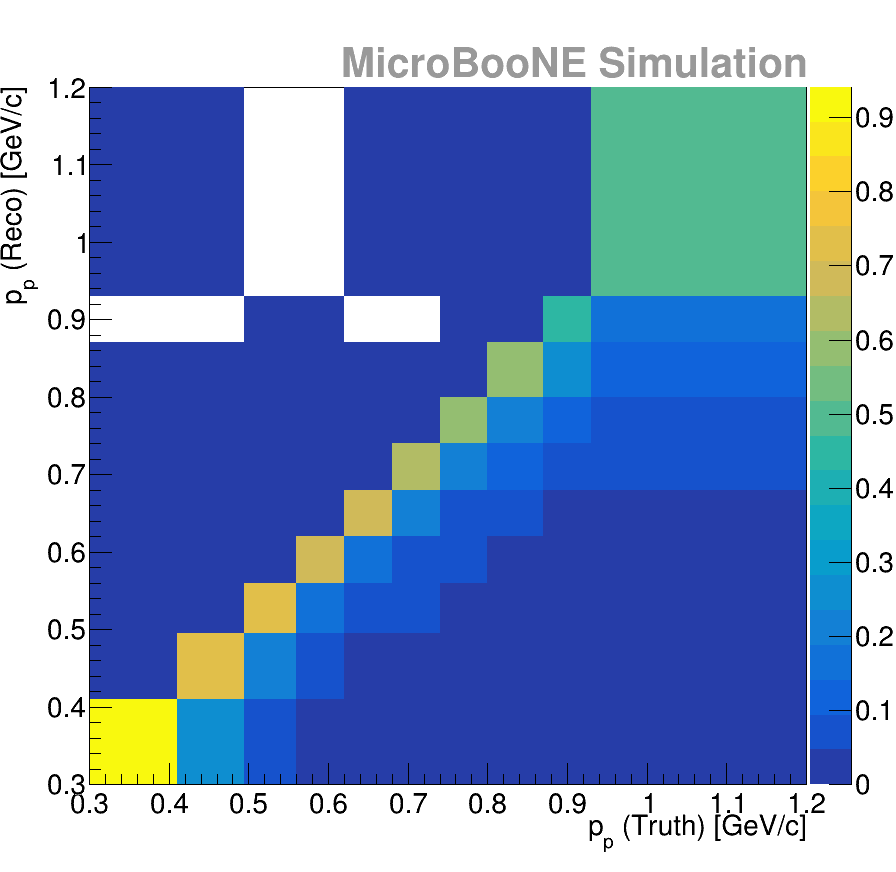}
    \caption{Migration matrix between true and reconstructed bins as a function of reconstructed proton momentum in simulated CC$0\pi Np$ events.}
    \label{fig:migmat_pmom}
\end{figure}

Plots of migration matrices for muon polar angle (Fig.~\ref{fig:migmat_muangle}), leading proton momentum (Fig.~\ref{fig:migmat_pmom}), leading proton polar angle (Fig.~\ref{fig:migmat_pangle}), and muon-proton opening angle (Fig.~\ref{fig:migmat_thetamup}) are also shown below.
These matrices are all close to diagonal, showing that the bin choices correspond to the resolutions of the observed quantities.


\begin{figure}[h!]
    \centering
    \includegraphics[width=0.45\textwidth]{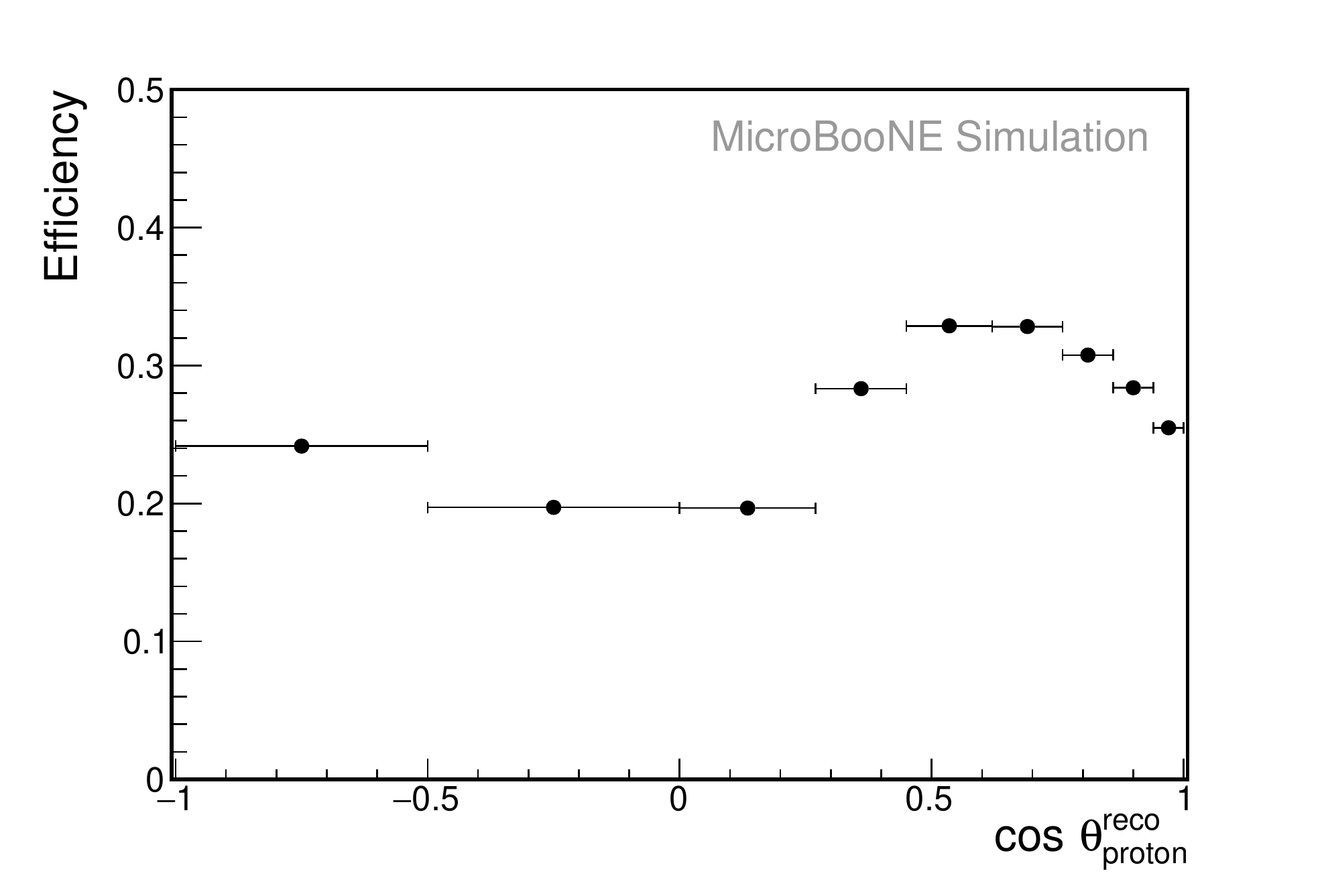} 
    \caption{Efficiency as a function of reconstructed proton polar angle $\cos(\theta_p)$ in simulated CC$0\pi Np$ events.}
    \label{fig:eff_pangle}
\end{figure}

\begin{figure} [h!]
    \centering
    \includegraphics[width=0.45\textwidth]{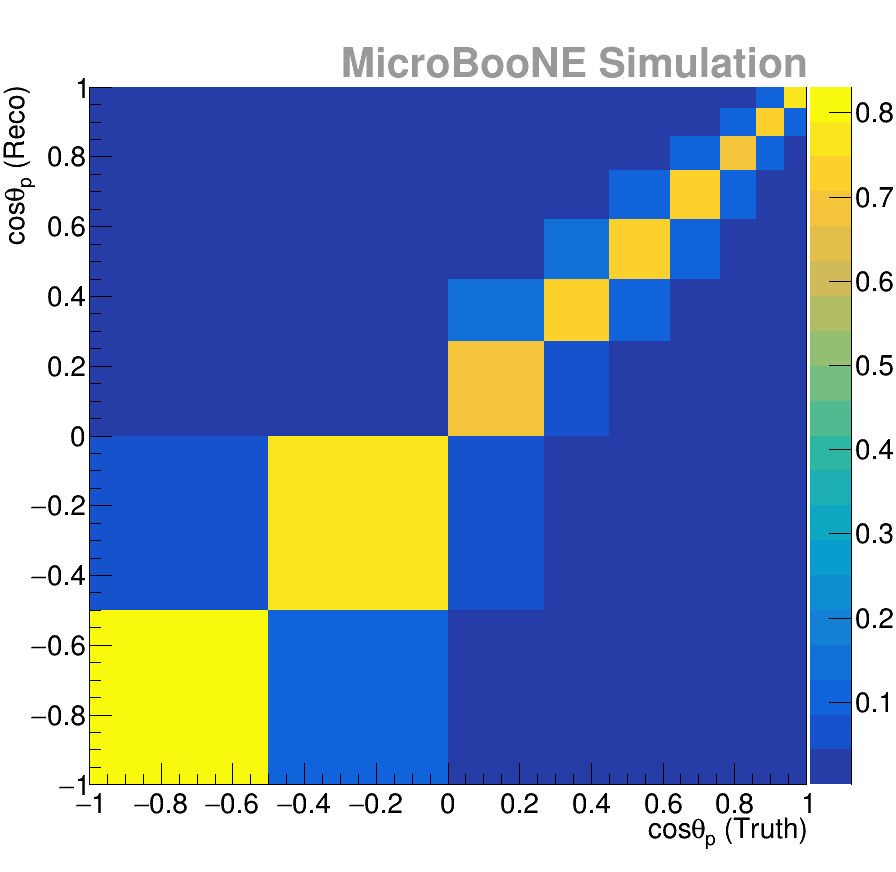}
    \caption{Migration matrix between true and reconstructed bins for reconstructed proton polar angle ($\cos(\theta_p)$) in simulated CC$0\pi Np$ events.  }
    \label{fig:migmat_pangle}
\end{figure}

\begin{figure}[h!]
    \centering
    \includegraphics[width=0.45\textwidth]{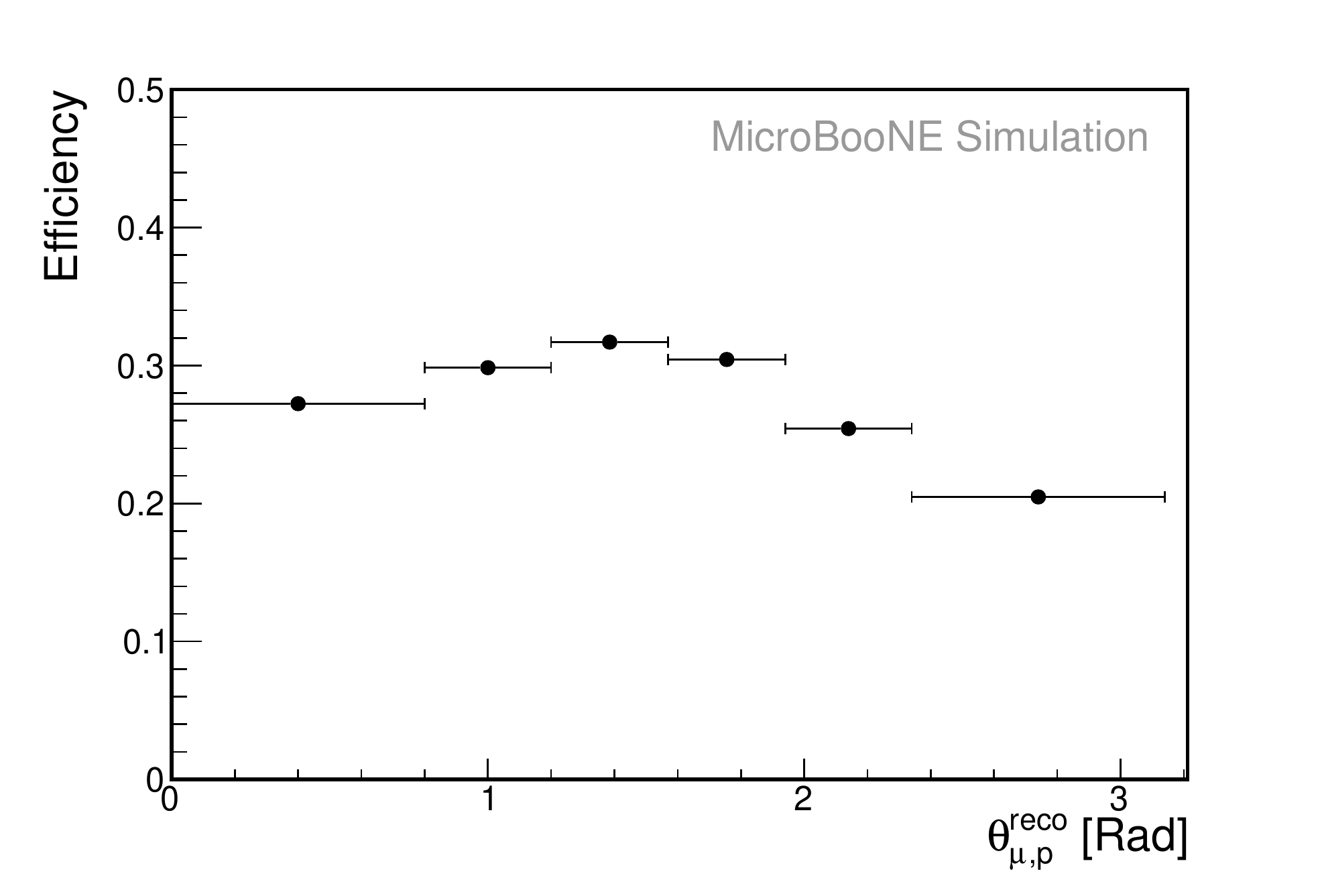}
    \caption{Efficiency as a function of reconstructed muon-proton opening angle ($\theta_{\mu p}$) in simulated CC$0\pi Np$ events.}
    \label{fig:eff_thetamup}
\end{figure}
\begin{figure}
    \centering
    \includegraphics[width=0.45\textwidth]{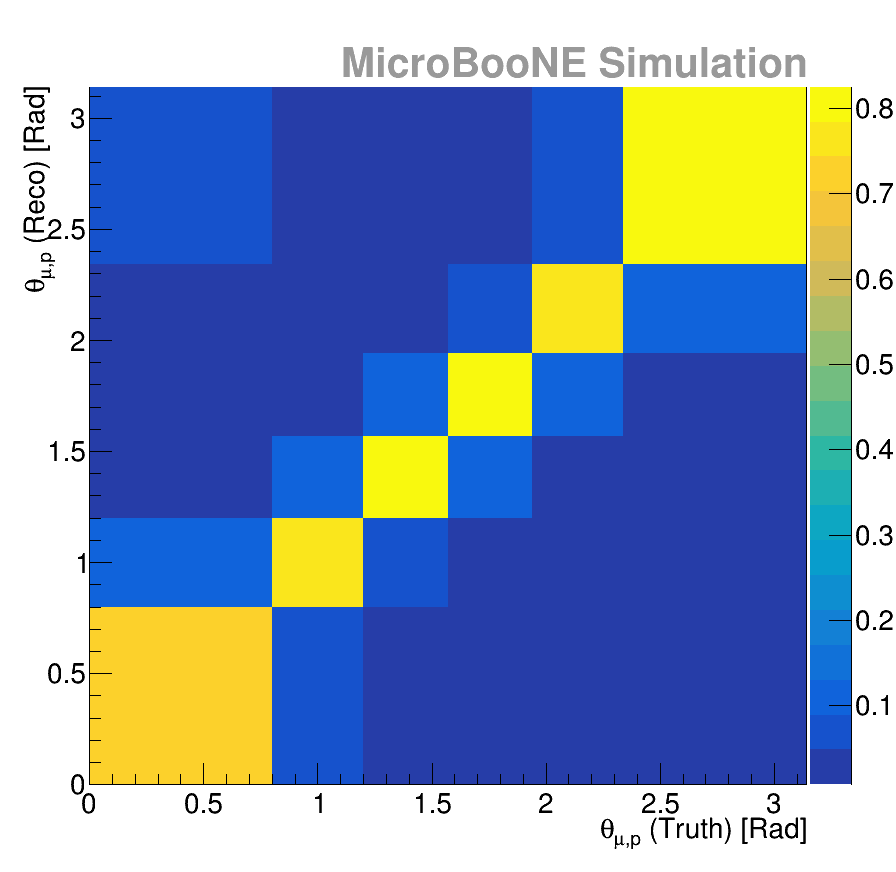}
    \caption{Migration matrix between true and reconstructed bins for reconstructed opening angle ($\theta_{\mu p}$) in simulated CC$0\pi Np$ events.}
    \label{fig:migmat_thetamup}
\end{figure}
\FloatBarrier
\bibliographystyle{apsrev}
\bibliography{references}

\begin{thebibliography}{60}
\expandafter\ifx\csname natexlab\endcsname\relax\def\natexlab#1{#1}\fi
\expandafter\ifx\csname bibnamefont\endcsname\relax
  \def\bibnamefont#1{#1}\fi
\expandafter\ifx\csname bibfnamefont\endcsname\relax
  \def\bibfnamefont#1{#1}\fi
\expandafter\ifx\csname citenamefont\endcsname\relax
  \def\citenamefont#1{#1}\fi
\expandafter\ifx\csname url\endcsname\relax
  \def\url#1{\texttt{#1}}\fi
\expandafter\ifx\csname urlprefix\endcsname\relax\def\urlprefix{URL }\fi
\providecommand{\bibinfo}[2]{#2}
\providecommand{\eprint}[2][]{\url{#2}}

\bibitem[{\citenamefont{Alvarez-Ruso et~al.}(2018)}]{nustec-review}
\bibinfo{author}{\bibfnamefont{L.}~\bibnamefont{Alvarez-Ruso}}
  \bibnamefont{et~al.} (\bibinfo{collaboration}{NuSTEC}),
  \bibinfo{journal}{Prog. Part. Nucl. Phys.} \textbf{\bibinfo{volume}{100}},
  \bibinfo{pages}{1} (\bibinfo{year}{2018}).

\bibitem[{\citenamefont{Abe et~al.}(2018{\natexlab{a}})}]{Abe:2018wpn}
\bibinfo{author}{\bibfnamefont{K.}~\bibnamefont{Abe}} \bibnamefont{et~al.}
  (\bibinfo{collaboration}{T2K Collaboration}), \bibinfo{journal}{Phys. Rev.
  Lett.} \textbf{\bibinfo{volume}{121}}, \bibinfo{pages}{171802}
  (\bibinfo{year}{2018}{\natexlab{a}}).

\bibitem[{\citenamefont{Acero et~al.}(2019)}]{Acero:2019ksn}
\bibinfo{author}{\bibfnamefont{M.~A.} \bibnamefont{Acero}} \bibnamefont{et~al.}
  (\bibinfo{collaboration}{NOvA Collaboration}), \bibinfo{journal}{Phys. Rev.
  Lett.} \textbf{\bibinfo{volume}{123}}, \bibinfo{pages}{151803}
  (\bibinfo{year}{2019}).

\bibitem[{\citenamefont{Antonello et~al.}(2015)}]{Antonello:2015lea}
\bibinfo{author}{\bibfnamefont{M.}~\bibnamefont{Antonello}}
  \bibnamefont{et~al.} (\bibinfo{collaboration}{MicroBooNE, LAr1-ND,
  ICARUS-WA104}) (\bibinfo{year}{2015}), \eprint{arXiv:1503.01520}.

\bibitem[{\citenamefont{Abi et~al.}(2020)}]{Abi:2020evt}
\bibinfo{author}{\bibfnamefont{B.}~\bibnamefont{Abi}} \bibnamefont{et~al.}
  (\bibinfo{collaboration}{DUNE Collaboration}) (\bibinfo{year}{2020}),
  \eprint{arXiv:2002.03005}.

\bibitem[{\citenamefont{Abe et~al.}(2016)}]{HyperK_design_report}
\bibinfo{author}{\bibfnamefont{K.}~\bibnamefont{Abe}} \bibnamefont{et~al.}
  (\bibinfo{collaboration}{Hyper-Kamiokande proto-collaboration}),
  \bibinfo{journal}{KEK preprint}  (\bibinfo{year}{2016}),
  \urlprefix\url{https://lib-extopc.kek.jp/preprints/PDF/2016/1627/1627021.pdf}.

\bibitem[{\citenamefont{Acciarri
  et~al.}(2017{\natexlab{a}})}]{Acciarri:2016smiDET}
\bibinfo{author}{\bibfnamefont{R.}~\bibnamefont{Acciarri}} \bibnamefont{et~al.}
  (\bibinfo{collaboration}{MicroBooNE Collaboration}), \bibinfo{journal}{J. of
  Instrum.} \textbf{\bibinfo{volume}{12}}, \bibinfo{pages}{P02017}
  (\bibinfo{year}{2017}{\natexlab{a}}).

\bibitem[{\citenamefont{Kitagaki et~al.}(1983)}]{Kitagaki:1983px}
\bibinfo{author}{\bibfnamefont{T.}~\bibnamefont{Kitagaki}}
  \bibnamefont{et~al.}, \bibinfo{journal}{Phys. Rev. D}
  \textbf{\bibinfo{volume}{28}}, \bibinfo{pages}{436} (\bibinfo{year}{1983}).

\bibitem[{\citenamefont{Gran et~al.}(2006)}]{Gran:2006jn}
\bibinfo{author}{\bibfnamefont{R.}~\bibnamefont{Gran}} \bibnamefont{et~al.}
  (\bibinfo{collaboration}{K2K Collaboration}), \bibinfo{journal}{Phys. Rev. D}
  \textbf{\bibinfo{volume}{74}}, \bibinfo{pages}{052002}
  (\bibinfo{year}{2006}).

\bibitem[{\citenamefont{Aguilar-Arevalo et~al.}(2010)}]{miniboone-ccqe}
\bibinfo{author}{\bibfnamefont{A.}~\bibnamefont{Aguilar-Arevalo}}
  \bibnamefont{et~al.} (\bibinfo{collaboration}{MiniBooNE Collaboration}),
  \bibinfo{journal}{Phys. Rev. D} \textbf{\bibinfo{volume}{81}},
  \bibinfo{pages}{092005} (\bibinfo{year}{2010}).

\bibitem[{\citenamefont{Martini et~al.}(2011)\citenamefont{Martini, Ericson,
  and Chanfray}}]{Martini:2011wp}
\bibinfo{author}{\bibfnamefont{M.}~\bibnamefont{Martini}},
  \bibinfo{author}{\bibfnamefont{M.}~\bibnamefont{Ericson}}, \bibnamefont{and}
  \bibinfo{author}{\bibfnamefont{G.}~\bibnamefont{Chanfray}},
  \bibinfo{journal}{Phys. Rev. C} \textbf{\bibinfo{volume}{84}},
  \bibinfo{pages}{055502} (\bibinfo{year}{2011}).

\bibitem[{\citenamefont{Nieves et~al.}(2011)\citenamefont{Nieves, Ruiz~Simo,
  and Vicente~Vacas}}]{Nieves:2011ppTHEORY}
\bibinfo{author}{\bibfnamefont{J.}~\bibnamefont{Nieves}},
  \bibinfo{author}{\bibfnamefont{I.}~\bibnamefont{Ruiz~Simo}},
  \bibnamefont{and} \bibinfo{author}{\bibfnamefont{M.~J.}
  \bibnamefont{Vicente~Vacas}}, \bibinfo{journal}{Phys. Rev.}
  \textbf{\bibinfo{volume}{C83}}, \bibinfo{pages}{045501}
  (\bibinfo{year}{2011}).

\bibitem[{\citenamefont{Betancourt et~al.}(2017)}]{Betancourt:2017uso}
\bibinfo{author}{\bibfnamefont{M.}~\bibnamefont{Betancourt}}
  \bibnamefont{et~al.} (\bibinfo{collaboration}{MINERvA Collaboration}),
  \bibinfo{journal}{Phys. Rev. Lett.} \textbf{\bibinfo{volume}{119}},
  \bibinfo{pages}{082001} (\bibinfo{year}{2017}).

\bibitem[{\citenamefont{Ruterbories et~al.}(2019)}]{Ruterbories:2018gub}
\bibinfo{author}{\bibfnamefont{D.}~\bibnamefont{Ruterbories}}
  \bibnamefont{et~al.} (\bibinfo{collaboration}{MINERvA Collaboration}),
  \bibinfo{journal}{Phys. Rev.} \textbf{\bibinfo{volume}{D99}},
  \bibinfo{pages}{012004} (\bibinfo{year}{2019}).

\bibitem[{\citenamefont{Abe et~al.}(2018{\natexlab{b}})}]{Abe:2018pwo}
\bibinfo{author}{\bibfnamefont{K.}~\bibnamefont{Abe}} \bibnamefont{et~al.}
  (\bibinfo{collaboration}{T2K Collaboration}), \bibinfo{journal}{Phys. Rev.}
  \textbf{\bibinfo{volume}{D98}}, \bibinfo{pages}{032003}
  (\bibinfo{year}{2018}{\natexlab{b}}).

\bibitem[{\citenamefont{Abratenko et~al.}(2020{\natexlab{a}})}]{uB_CCQE_2020}
\bibinfo{author}{\bibfnamefont{P.}~\bibnamefont{Abratenko}}
  \bibnamefont{et~al.} (\bibinfo{collaboration}{MicroBooNE Collaboration}),
  \bibinfo{journal}{Phys. Rev. Lett.} \textbf{\bibinfo{volume}{In-Print}}
  (\bibinfo{year}{2020}{\natexlab{a}}), \eprint{arXiv:2006.00108}.

\bibitem[{\citenamefont{Megias et~al.}(2016)\citenamefont{Megias, Amaro,
  Barbaro, Caballero, Donnelly, and Ruiz~Simo}}]{Megias:2016fjk}
\bibinfo{author}{\bibfnamefont{G.}~\bibnamefont{Megias}},
  \bibinfo{author}{\bibfnamefont{J.}~\bibnamefont{Amaro}},
  \bibinfo{author}{\bibfnamefont{M.}~\bibnamefont{Barbaro}},
  \bibinfo{author}{\bibfnamefont{J.}~\bibnamefont{Caballero}},
  \bibinfo{author}{\bibfnamefont{T.}~\bibnamefont{Donnelly}}, \bibnamefont{and}
  \bibinfo{author}{\bibfnamefont{I.}~\bibnamefont{Ruiz~Simo}},
  \bibinfo{journal}{Phys. Rev. D} \textbf{\bibinfo{volume}{94}},
  \bibinfo{pages}{093004} (\bibinfo{year}{2016}).

\bibitem[{\citenamefont{Acciarri et~al.}(2014)}]{Argoneut-2p}
\bibinfo{author}{\bibfnamefont{R.}~\bibnamefont{Acciarri}} \bibnamefont{et~al.}
  (\bibinfo{collaboration}{ArgoNeuT Collaboration}), \bibinfo{journal}{Phys.
  Rev.} \textbf{\bibinfo{volume}{D90}}, \bibinfo{pages}{012008}
  (\bibinfo{year}{2014}).

\bibitem[{\citenamefont{Abratenko et~al.}(2019)}]{Adams:2019iqcINCL}
\bibinfo{author}{\bibfnamefont{P.}~\bibnamefont{Abratenko}}
  \bibnamefont{et~al.} (\bibinfo{collaboration}{MicroBooNE Collaboration}),
  \bibinfo{journal}{Phys. Rev. Lett.} \textbf{\bibinfo{volume}{123}},
  \bibinfo{pages}{131801} (\bibinfo{year}{2019}).

\bibitem[{\citenamefont{Adams et~al.}(2020{\natexlab{a}})}]{uBUVLaser2019}
\bibinfo{author}{\bibfnamefont{C.}~\bibnamefont{Adams}} \bibnamefont{et~al.}
  (\bibinfo{collaboration}{MicroBooNE Collaboration}), \bibinfo{journal}{J. of
  Instrum.} \textbf{\bibinfo{volume}{15}}, \bibinfo{pages}{P07010}
  (\bibinfo{year}{2020}{\natexlab{a}}).

\bibitem[{\citenamefont{Abratenko
  et~al.}(2020{\natexlab{b}})}]{uBcosmicSCE_2020}
\bibinfo{author}{\bibfnamefont{P.}~\bibnamefont{Abratenko}}
  \bibnamefont{et~al.} (\bibinfo{collaboration}{MicroBooNE Collaboration})
  (\bibinfo{year}{2020}{\natexlab{b}}), \eprint{arXiv:2008.09765}.

\bibitem[{\citenamefont{Adams et~al.}(2019)}]{Adams:2019bztCOSMICTAGGER}
\bibinfo{author}{\bibfnamefont{C.}~\bibnamefont{Adams}} \bibnamefont{et~al.}
  (\bibinfo{collaboration}{MicroBooNE Collaboration}), \bibinfo{journal}{J. of
  Instrum.} \textbf{\bibinfo{volume}{14}}, \bibinfo{pages}{P04004}
  (\bibinfo{year}{2019}).

\bibitem[{\citenamefont{Allison et~al.}(2016)}]{Allison:2016lfl}
\bibinfo{author}{\bibfnamefont{J.}~\bibnamefont{Allison}} \bibnamefont{et~al.},
  \bibinfo{journal}{Nucl. Instrum. Meth.} \textbf{\bibinfo{volume}{A835}},
  \bibinfo{pages}{186} (\bibinfo{year}{2016}).

\bibitem[{\citenamefont{Aguilar-Arevalo et~al.}(2009)}]{AAAA_2009_MBflux}
\bibinfo{author}{\bibfnamefont{A.}~\bibnamefont{Aguilar-Arevalo}}
  \bibnamefont{et~al.} (\bibinfo{collaboration}{MiniBooNE Collaboration}),
  \bibinfo{journal}{Phys. Rev. D} \textbf{\bibinfo{volume}{79}},
  \bibinfo{pages}{072002} (\bibinfo{year}{2009}).

\bibitem[{\citenamefont{Andreopoulos et~al.}(2010)}]{Andreopoulos:2009rq}
\bibinfo{author}{\bibfnamefont{C.}~\bibnamefont{Andreopoulos}}
  \bibnamefont{et~al.}, \bibinfo{journal}{Nucl. Instrum. Meth.}
  \textbf{\bibinfo{volume}{A614}}, \bibinfo{pages}{87} (\bibinfo{year}{2010}).

\bibitem[{\citenamefont{{D. H. Wright and M. H.
  Kelsey}}(2015)}]{Wright:2015xia}
\bibinfo{author}{\bibnamefont{{D. H. Wright and M. H. Kelsey}}},
  \bibinfo{journal}{Nucl. Instrum. Meth.} \textbf{\bibinfo{volume}{A804}},
  \bibinfo{pages}{175} (\bibinfo{year}{2015}).

\bibitem[{\citenamefont{{E.L. Snider and G. Petrillo}}(2017)}]{Snider:2017wjd}
\bibinfo{author}{\bibnamefont{{E.L. Snider and G. Petrillo}}},
  \bibinfo{journal}{J. Phys. Conf. Ser.} \textbf{\bibinfo{volume}{898}},
  \bibinfo{pages}{042057} (\bibinfo{year}{2017}).

\bibitem[{\citenamefont{{A. Bodek and J. L. Ritchie}}(1981)}]{Bodek:1980ar}
\bibinfo{author}{\bibnamefont{{A. Bodek and J. L. Ritchie}}},
  \bibinfo{journal}{Phys. Rev.} \textbf{\bibinfo{volume}{D23}},
  \bibinfo{pages}{1070} (\bibinfo{year}{1981}).

\bibitem[{\citenamefont{Llewellyn~Smith}(1972)}]{LlewellynSmith:1971zm}
\bibinfo{author}{\bibfnamefont{C.~H.} \bibnamefont{Llewellyn~Smith}},
  \bibinfo{journal}{Phys. Rept.} \textbf{\bibinfo{volume}{3}},
  \bibinfo{pages}{261} (\bibinfo{year}{1972}).

\bibitem[{\citenamefont{Katori}(2015)}]{Katori:2013eoa}
\bibinfo{author}{\bibfnamefont{T.}~\bibnamefont{Katori}}, \bibinfo{journal}{AIP
  Conf. Proc.} \textbf{\bibinfo{volume}{1663}}, \bibinfo{pages}{030001}
  (\bibinfo{year}{2015}).

\bibitem[{\citenamefont{{D. Rein and L. M. Sehgal}}(1981)}]{Rein:1980wg}
\bibinfo{author}{\bibnamefont{{D. Rein and L. M. Sehgal}}},
  \bibinfo{journal}{Ann. Phys.} \textbf{\bibinfo{volume}{133}},
  \bibinfo{pages}{79} (\bibinfo{year}{1981}).

\bibitem[{\citenamefont{{D.~Rein and L.M.~Sehgal}}(2007)}]{Rein:2007}
\bibinfo{author}{\bibnamefont{{D.~Rein and L.M.~Sehgal}}},
  \bibinfo{journal}{Phys.Lett.B} \textbf{\bibinfo{volume}{657}},
  \bibinfo{pages}{207 } (\bibinfo{year}{2007}).

\bibitem[{\citenamefont{{A. Bodek and U. K. Yang}}(2003)}]{Bodek:2002ps}
\bibinfo{author}{\bibnamefont{{A. Bodek and U. K. Yang}}}, \bibinfo{journal}{J.
  Phys.} \textbf{\bibinfo{volume}{G29}}, \bibinfo{pages}{1899}
  (\bibinfo{year}{2003}).

\bibitem[{\citenamefont{{S. A. Dytman and A. S. Meyer}}(2011)}]{Dytman:2011zz}
\bibinfo{author}{\bibnamefont{{S. A. Dytman and A. S. Meyer}}},
  \bibinfo{journal}{AIP Conf. Proc.} \textbf{\bibinfo{volume}{1405}},
  \bibinfo{pages}{213} (\bibinfo{year}{2011}).

\bibitem[{\citenamefont{{Heck} et~al.}(1998)\citenamefont{{Heck}, {Knapp},
  {Capdevielle}, {Schatz}, and {Thouw}}}]{1998cmcc}
\bibinfo{author}{\bibfnamefont{D.}~\bibnamefont{{Heck}}},
  \bibinfo{author}{\bibfnamefont{J.}~\bibnamefont{{Knapp}}},
  \bibinfo{author}{\bibfnamefont{J.~N.} \bibnamefont{{Capdevielle}}},
  \bibinfo{author}{\bibfnamefont{G.}~\bibnamefont{{Schatz}}}, \bibnamefont{and}
  \bibinfo{author}{\bibfnamefont{T.}~\bibnamefont{{Thouw}}},
  \emph{\bibinfo{title}{{CORSIKA: a Monte Carlo code to simulate extensive air
  showers.}}} (\bibinfo{year}{1998}), \bibinfo{note}{~FZKA-6019}.

\bibitem[{\citenamefont{Adams et~al.}(2018)}]{Adams:2018gbiPROCESS}
\bibinfo{author}{\bibfnamefont{C.}~\bibnamefont{Adams}} \bibnamefont{et~al.}
  (\bibinfo{collaboration}{MicroBooNE Collaboration}), \bibinfo{journal}{J. of
  Instrum.} \textbf{\bibinfo{volume}{13}}, \bibinfo{pages}{P07007}
  (\bibinfo{year}{2018}).

\bibitem[{\citenamefont{Acciarri
  et~al.}(2017{\natexlab{b}})}]{Acciarri:2017sdeNOISE}
\bibinfo{author}{\bibfnamefont{R.}~\bibnamefont{Acciarri}} \bibnamefont{et~al.}
  (\bibinfo{collaboration}{MicroBooNE Collaboration}), \bibinfo{journal}{J. of
  Instrum.} \textbf{\bibinfo{volume}{12}}, \bibinfo{pages}{P08003}
  (\bibinfo{year}{2017}{\natexlab{b}}).

\bibitem[{\citenamefont{Acciarri et~al.}(2018)}]{Acciarri:2017hatPANDORA}
\bibinfo{author}{\bibfnamefont{R.}~\bibnamefont{Acciarri}} \bibnamefont{et~al.}
  (\bibinfo{collaboration}{MicroBooNE Collaboration}), \bibinfo{journal}{Eur.
  Phys. J.} \textbf{\bibinfo{volume}{C78}}, \bibinfo{pages}{82}
  (\bibinfo{year}{2018}).

\bibitem[{\citenamefont{Abratenko et~al.}(2017)}]{Abratenko:2017nkiMCS}
\bibinfo{author}{\bibfnamefont{P.}~\bibnamefont{Abratenko}}
  \bibnamefont{et~al.} (\bibinfo{collaboration}{MicroBooNE Collaboration}),
  \bibinfo{journal}{J. of Instrum.} \textbf{\bibinfo{volume}{12}},
  \bibinfo{pages}{P10010} (\bibinfo{year}{2017}).

\bibitem[{\citenamefont{Adams et~al.}(2020{\natexlab{b}})}]{Adams:2019ssgCAL}
\bibinfo{author}{\bibfnamefont{C.}~\bibnamefont{Adams}} \bibnamefont{et~al.}
  (\bibinfo{collaboration}{MicroBooNE Collaboration}), \bibinfo{journal}{J. of
  Instrum.} \textbf{\bibinfo{volume}{15}}, \bibinfo{pages}{P03022}
  (\bibinfo{year}{2020}{\natexlab{b}}).

\bibitem[{\citenamefont{D'Agostini}(1995)}]{DAgostini:1994fjx}
\bibinfo{author}{\bibfnamefont{G.}~\bibnamefont{D'Agostini}},
  \bibinfo{journal}{Nucl. Instrum. Meth.} \textbf{\bibinfo{volume}{A362}},
  \bibinfo{pages}{487} (\bibinfo{year}{1995}).

\bibitem[{\citenamefont{Koch}(2019)}]{Koch2019}
\bibinfo{author}{\bibfnamefont{L.}~\bibnamefont{Koch}}, \bibinfo{journal}{J. of
  Instrum.} \textbf{\bibinfo{volume}{14}}, \bibinfo{pages}{P09013}
  (\bibinfo{year}{2019}).

\bibitem[{\citenamefont{Li et~al.}(2016)\citenamefont{Li, Tsang, Thorn, Qian,
  Diwan, Joshi, Kettell, Morse, Rao, Stewart et~al.}}]{LI2016160}
\bibinfo{author}{\bibfnamefont{Y.}~\bibnamefont{Li}},
  \bibinfo{author}{\bibfnamefont{T.}~\bibnamefont{Tsang}},
  \bibinfo{author}{\bibfnamefont{C.}~\bibnamefont{Thorn}},
  \bibinfo{author}{\bibfnamefont{X.}~\bibnamefont{Qian}},
  \bibinfo{author}{\bibfnamefont{M.}~\bibnamefont{Diwan}},
  \bibinfo{author}{\bibfnamefont{J.}~\bibnamefont{Joshi}},
  \bibinfo{author}{\bibfnamefont{S.}~\bibnamefont{Kettell}},
  \bibinfo{author}{\bibfnamefont{W.}~\bibnamefont{Morse}},
  \bibinfo{author}{\bibfnamefont{T.}~\bibnamefont{Rao}},
  \bibinfo{author}{\bibfnamefont{J.}~\bibnamefont{Stewart}},
  \bibnamefont{et~al.}, \bibinfo{journal}{Nuclear Instruments and Methods in
  Physics Research Section A: Accelerators, Spectrometers, Detectors and
  Associated Equipment} \textbf{\bibinfo{volume}{816}}, \bibinfo{pages}{160 }
  (\bibinfo{year}{2016}), ISSN \bibinfo{issn}{0168-9002}.

\bibitem[{\citenamefont{Acciarri et~al.}(2013)}]{Acciarri_2013}
\bibinfo{author}{\bibfnamefont{R.}~\bibnamefont{Acciarri}} \bibnamefont{et~al.}
  (\bibinfo{collaboration}{ArgoNeuT Collaboration}), \bibinfo{journal}{J. of
  Instrum.} \textbf{\bibinfo{volume}{8}}, \bibinfo{pages}{P08005}
  (\bibinfo{year}{2013}).

\bibitem[{\citenamefont{Birks}(1951)}]{birks_model}
\bibinfo{author}{\bibfnamefont{J.~B.} \bibnamefont{Birks}},
  \bibinfo{journal}{Proc. Phys. Soc.} \textbf{\bibinfo{volume}{A64}},
  \bibinfo{pages}{874} (\bibinfo{year}{1951}).

\bibitem[{\citenamefont{Amaruso et~al.}(2004)}]{ICARUS_recomb}
\bibinfo{author}{\bibfnamefont{S.}~\bibnamefont{Amaruso}} \bibnamefont{et~al.}
  (\bibinfo{collaboration}{ICARUS Collaboration}), \bibinfo{journal}{Nucl.
  Instrum. Meth.} \textbf{\bibinfo{volume}{A 523}}, \bibinfo{pages}{275}
  (\bibinfo{year}{2004}).

\bibitem[{\citenamefont{Andreopoulos et~al.}(2015)\citenamefont{Andreopoulos,
  Barry, Dytman, Gallagher, Golan, Hatcher, Perdue, and
  Yarba}}]{Andreopoulos:2015wxa}
\bibinfo{author}{\bibfnamefont{C.}~\bibnamefont{Andreopoulos}},
  \bibinfo{author}{\bibfnamefont{C.}~\bibnamefont{Barry}},
  \bibinfo{author}{\bibfnamefont{S.}~\bibnamefont{Dytman}},
  \bibinfo{author}{\bibfnamefont{H.}~\bibnamefont{Gallagher}},
  \bibinfo{author}{\bibfnamefont{T.}~\bibnamefont{Golan}},
  \bibinfo{author}{\bibfnamefont{R.}~\bibnamefont{Hatcher}},
  \bibinfo{author}{\bibfnamefont{G.}~\bibnamefont{Perdue}}, \bibnamefont{and}
  \bibinfo{author}{\bibfnamefont{J.}~\bibnamefont{Yarba}}
  (\bibinfo{year}{2015}), \eprint{arXiv:1510.05494}.

\bibitem[{\citenamefont{Nieves et~al.}(2004)\citenamefont{Nieves, Amaro, and
  Valverde}}]{Nieves:2004wx}
\bibinfo{author}{\bibfnamefont{J.}~\bibnamefont{Nieves}},
  \bibinfo{author}{\bibfnamefont{J.~E.} \bibnamefont{Amaro}}, \bibnamefont{and}
  \bibinfo{author}{\bibfnamefont{M.}~\bibnamefont{Valverde}},
  \bibinfo{journal}{Phys. Rev.} \textbf{\bibinfo{volume}{C70}},
  \bibinfo{pages}{055503} (\bibinfo{year}{2004}), \bibinfo{note}{[Erratum:
  Phys. Rev. C 72,019902(2005)]}.

\bibitem[{\citenamefont{Gran et~al.}(2013)\citenamefont{Gran, Nieves, Sanchez,
  and Vicente~Vacas}}]{Gran:2013kda}
\bibinfo{author}{\bibfnamefont{R.}~\bibnamefont{Gran}},
  \bibinfo{author}{\bibfnamefont{J.}~\bibnamefont{Nieves}},
  \bibinfo{author}{\bibfnamefont{F.}~\bibnamefont{Sanchez}}, \bibnamefont{and}
  \bibinfo{author}{\bibfnamefont{M.~J.} \bibnamefont{Vicente~Vacas}},
  \bibinfo{journal}{Phys. Rev.} \textbf{\bibinfo{volume}{D88}},
  \bibinfo{pages}{113007} (\bibinfo{year}{2013}).

\bibitem[{\citenamefont{Golan et~al.}(2012)\citenamefont{Golan, Juszczak, and
  Sobczyk}}]{Golan:2012wx}
\bibinfo{author}{\bibfnamefont{T.}~\bibnamefont{Golan}},
  \bibinfo{author}{\bibfnamefont{C.}~\bibnamefont{Juszczak}}, \bibnamefont{and}
  \bibinfo{author}{\bibfnamefont{J.~T.} \bibnamefont{Sobczyk}},
  \bibinfo{journal}{Phys. Rev.} \textbf{\bibinfo{volume}{C86}},
  \bibinfo{pages}{015505} (\bibinfo{year}{2012}).

\bibitem[{\citenamefont{Hayato}(2009)}]{Hayato:2009zz}
\bibinfo{author}{\bibfnamefont{Y.}~\bibnamefont{Hayato}},
  \bibinfo{journal}{Acta Phys. Polon. B} \textbf{\bibinfo{volume}{40}},
  \bibinfo{pages}{2477} (\bibinfo{year}{2009}).

\bibitem[{\citenamefont{Buss et~al.}(2012)\citenamefont{Buss, Gaitanos,
  Gallmeister, van Hees, Kaskulov, Lalakulich, Larionov, Leitner, Weil, and
  Mosel}}]{Buss:2011mx}
\bibinfo{author}{\bibfnamefont{O.}~\bibnamefont{Buss}},
  \bibinfo{author}{\bibfnamefont{T.}~\bibnamefont{Gaitanos}},
  \bibinfo{author}{\bibfnamefont{K.}~\bibnamefont{Gallmeister}},
  \bibinfo{author}{\bibfnamefont{H.}~\bibnamefont{van Hees}},
  \bibinfo{author}{\bibfnamefont{M.}~\bibnamefont{Kaskulov}},
  \bibinfo{author}{\bibfnamefont{O.}~\bibnamefont{Lalakulich}},
  \bibinfo{author}{\bibfnamefont{A.~B.} \bibnamefont{Larionov}},
  \bibinfo{author}{\bibfnamefont{T.}~\bibnamefont{Leitner}},
  \bibinfo{author}{\bibfnamefont{J.}~\bibnamefont{Weil}}, \bibnamefont{and}
  \bibinfo{author}{\bibfnamefont{U.}~\bibnamefont{Mosel}},
  \bibinfo{journal}{Phys. Rept.} \textbf{\bibinfo{volume}{512}},
  \bibinfo{pages}{1} (\bibinfo{year}{2012}).

\bibitem[{\citenamefont{Kuzmin et~al.}(2004)\citenamefont{Kuzmin, Lyubushkin,
  and Naumov}}]{Kuzmin:2003ji}
\bibinfo{author}{\bibfnamefont{K.~S.} \bibnamefont{Kuzmin}},
  \bibinfo{author}{\bibfnamefont{V.~V.} \bibnamefont{Lyubushkin}},
  \bibnamefont{and} \bibinfo{author}{\bibfnamefont{V.~A.}
  \bibnamefont{Naumov}}, \bibinfo{journal}{Phys. Part. Nucl.}
  \textbf{\bibinfo{volume}{35}}, \bibinfo{pages}{S133} (\bibinfo{year}{2004}).

\bibitem[{\citenamefont{{Ch. Berger and L. M.
  Sehgal}}(2007)}]{Berger:2007rqRES}
\bibinfo{author}{\bibnamefont{{Ch. Berger and L. M. Sehgal}}},
  \bibinfo{journal}{Phys. Rev.} \textbf{\bibinfo{volume}{D76}},
  \bibinfo{pages}{113004} (\bibinfo{year}{2007}).

\bibitem[{\citenamefont{{Ch. Berger and L. M.
  Sehgal}}(2009)}]{Berger:2008xsCOH}
\bibinfo{author}{\bibnamefont{{Ch. Berger and L. M. Sehgal}}},
  \bibinfo{journal}{Phys. Rev.} \textbf{\bibinfo{volume}{D79}},
  \bibinfo{pages}{053003} (\bibinfo{year}{2009}).

\bibitem[{\citenamefont{Graczyk and Sobczyk}(2008)}]{Graczyk:2007xk}
\bibinfo{author}{\bibfnamefont{K.~M.} \bibnamefont{Graczyk}} \bibnamefont{and}
  \bibinfo{author}{\bibfnamefont{J.~T.} \bibnamefont{Sobczyk}},
  \bibinfo{journal}{Phys. Rev. D} \textbf{\bibinfo{volume}{77}},
  \bibinfo{pages}{053003} (\bibinfo{year}{2008}), \eprint{0709.4634}.

\bibitem[{\citenamefont{{U. K. Yang and A. Bodek}}(2000)}]{Yang:1999xg}
\bibinfo{author}{\bibnamefont{{U. K. Yang and A. Bodek}}},
  \bibinfo{journal}{Eur. Phys. J.} \textbf{\bibinfo{volume}{C13}},
  \bibinfo{pages}{241} (\bibinfo{year}{2000}).

\bibitem[{\citenamefont{Salcedo et~al.}(1988)\citenamefont{Salcedo, Oset,
  Vicente-Vacas, and Garcia-Recio}}]{Salcedo:1987md}
\bibinfo{author}{\bibfnamefont{L.~L.} \bibnamefont{Salcedo}},
  \bibinfo{author}{\bibfnamefont{E.}~\bibnamefont{Oset}},
  \bibinfo{author}{\bibfnamefont{M.~J.} \bibnamefont{Vicente-Vacas}},
  \bibnamefont{and}
  \bibinfo{author}{\bibfnamefont{C.}~\bibnamefont{Garcia-Recio}},
  \bibinfo{journal}{Nucl. Phys.} \textbf{\bibinfo{volume}{A484}},
  \bibinfo{pages}{557} (\bibinfo{year}{1988}).

\bibitem[{\citenamefont{{V. R. Pandharipande and Steven C.
  Pieper}}(1992)}]{Pandharipande:1992zz}
\bibinfo{author}{\bibnamefont{{V. R. Pandharipande and Steven C. Pieper}}},
  \bibinfo{journal}{Phys. Rev.} \textbf{\bibinfo{volume}{C45}},
  \bibinfo{pages}{791} (\bibinfo{year}{1992}).

\bibitem[{\citenamefont{{U. Mosel and K. Gallmeister}}(2018)}]{Mosel:2017ssx}
\bibinfo{author}{\bibnamefont{{U. Mosel and K. Gallmeister}}},
  \bibinfo{journal}{Phys. Rev.} \textbf{\bibinfo{volume}{C97}},
  \bibinfo{pages}{045501} (\bibinfo{year}{2018}).

\end{thebibliography}

\end{document}


\section{Supplemental material}

\FloatBarrier\subsection{Tables for $\cos(\theta_{\mu}^{reco})$} 
\begin{table}[ht] 
\begin{tabular}{|c|c|c|c|c|} \hline 
Bin & Min. & Max. & Cross Section [$10^{-38}$~cm$^{2}/$nucleus] & Uncertainty [$10^{-38}$~cm$^{2}/$nucleus] \\ \hline 
1 & -1.0 & -0.82 & 2.21 & 1.34 \\ \hline 
2 & -0.82 & -0.66 & 2.59 & 1.19 \\ \hline 
3 & -0.66 & -0.39 & 3.04 & 1.14 \\ \hline 
4 & -0.39 & -0.16 & 2.79 & 1.34 \\ \hline 
5 & -0.16 & 0.05 & 3.05 & 1.61 \\ \hline 
6 & 0.05 & 0.25 & 5.65 & 2.89 \\ \hline 
7 & 0.25 & 0.43 & 6.14 & 2.84 \\ \hline 
8 & 0.43 & 0.59 & 9.99 & 4.25 \\ \hline 
9 & 0.59 & 0.73 & 14.04 & 4.41 \\ \hline 
10 & 0.73 & 0.83 & 20.31 & 5.67 \\ \hline 
11 & 0.83 & 0.91 & 25.62 & 6.16 \\ \hline 
12 & 0.91 & 1.0 & 21.04 & 4.08 \\ \hline 
\end{tabular} 
\caption{Table of cross section values for $\cos(\theta_{\mu}^{reco})$} 
\end{table}

\begin{table}[ht] 
\Rotatebox{90}{\begin{tabular}{|c|c|c|c|c|c|c|c|c|c|c|c|c|} \hline 
\ & 1 & 2 & 3 & 4 & 5 & 6 & 7 & 8 & 9 & 10 & 11 & 12 \\ \hline 
1 & 1.7934 & 0.9270 & -0.0605 & -0.4256 & 0.0173 & -0.8999 & -0.5651 & -1.0836 & -0.3254 & -0.2682 & -0.4757 & 0.7041 \\ \hline 
2 & 0.9270 & 1.4106 & 0.3556 & 0.4003 & 0.7832 & 0.6165 & 0.9385 & 1.1049 & 1.9390 & 2.1075 & 2.4235 & 0.8802 \\ \hline 
3 & -0.0605 & 0.3556 & 1.3019 & 1.1164 & 0.7086 & 1.9313 & 1.9166 & 3.0322 & 2.8167 & 4.4216 & 4.6857 & 1.4270 \\ \hline 
4 & -0.4256 & 0.4003 & 1.1164 & 1.7856 & 1.4433 & 3.1686 & 3.1661 & 4.7591 & 4.5361 & 5.8172 & 6.2097 & 1.3219 \\ \hline 
5 & 0.0173 & 0.7832 & 0.7086 & 1.4433 & 2.5863 & 3.7252 & 3.8775 & 5.7256 & 6.2840 & 6.9444 & 6.9237 & 1.4021 \\ \hline 
6 & -0.8999 & 0.6165 & 1.9313 & 3.1686 & 3.7252 & 8.3284 & 7.7514 & 11.6528 & 11.2337 & 13.8667 & 13.3612 & 1.5554 \\ \hline 
7 & -0.5651 & 0.9385 & 1.9166 & 3.1661 & 3.8775 & 7.7514 & 8.0405 & 11.5437 & 11.5158 & 14.0011 & 13.4258 & 2.2315 \\ \hline 
8 & -1.0836 & 1.1049 & 3.0322 & 4.7591 & 5.7256 & 11.6528 & 11.5437 & 18.0482 & 17.4522 & 21.4629 & 21.0347 & 4.0317 \\ \hline 
9 & -0.3254 & 1.9390 & 2.8167 & 4.5361 & 6.2840 & 11.2337 & 11.5158 & 17.4522 & 19.4160 & 22.0725 & 22.1412 & 6.0539 \\ \hline 
10 & -0.2682 & 2.1075 & 4.4216 & 5.8172 & 6.9444 & 13.8667 & 14.0011 & 21.4629 & 22.0725 & 32.0930 & 26.6263 & 6.6603 \\ \hline 
11 & -0.4757 & 2.4235 & 4.6857 & 6.2097 & 6.9237 & 13.3612 & 13.4258 & 21.0347 & 22.1412 & 26.6263 & 37.9812 & 14.2555 \\ \hline 
12 & 0.7041 & 0.8802 & 1.4270 & 1.3219 & 1.4021 & 1.5554 & 2.2315 & 4.0317 & 6.0539 & 6.6603 & 14.2555 & 16.6867 \\ \hline 
\end{tabular} 
}\caption{Table of covariances for $\cos(\theta_{\mu}^{reco})$. 
Units are in $10^{-38}$~cm$^{-2}/nucleon$, as in the cross section results.}
\end{table}

\begin{table}[ht] 
\Rotatebox{90}{\begin{tabular}{|c|c|c|c|c|c|c|c|c|c|c|c|c|} \hline 
\backslashbox{True bin}{Reconstructed bin} & 1 & 2 & 3 & 4 & 5 & 6 & 7 & 8 & 9 & 10 & 11 & 12 \\ \hline 
1 & 0.8234 & 0.1170 & 0.0048 & 0.0000 & 0.0000 & 0.0000 & 0.0000 & 0.0000 & 0.0000 & 0.0048 & 0.0286 & 0.0215 \\ \hline 
2 & 0.0986 & 0.7053 & 0.1565 & 0.0000 & 0.0000 & 0.0000 & 0.0030 & 0.0000 & 0.0106 & 0.0167 & 0.0061 & 0.0030 \\ \hline 
3 & 0.0038 & 0.0981 & 0.7237 & 0.1283 & 0.0019 & 0.0000 & 0.0058 & 0.0179 & 0.0103 & 0.0019 & 0.0026 & 0.0058 \\ \hline 
4 & 0.0010 & 0.0024 & 0.1420 & 0.6633 & 0.1632 & 0.0077 & 0.0082 & 0.0029 & 0.0000 & 0.0034 & 0.0005 & 0.0053 \\ \hline 
5 & 0.0004 & 0.0000 & 0.0004 & 0.1499 & 0.6568 & 0.1785 & 0.0069 & 0.0019 & 0.0000 & 0.0008 & 0.0000 & 0.0046 \\ \hline 
6 & 0.0002 & 0.0005 & 0.0005 & 0.0030 & 0.1332 & 0.6806 & 0.1752 & 0.0039 & 0.0007 & 0.0000 & 0.0005 & 0.0018 \\ \hline 
7 & 0.0007 & 0.0000 & 0.0009 & 0.0003 & 0.0010 & 0.1242 & 0.6730 & 0.1931 & 0.0046 & 0.0007 & 0.0007 & 0.0009 \\ \hline 
8 & 0.0000 & 0.0000 & 0.0014 & 0.0000 & 0.0001 & 0.0016 & 0.1150 & 0.6915 & 0.1863 & 0.0029 & 0.0004 & 0.0008 \\ \hline 
9 & 0.0003 & 0.0001 & 0.0003 & 0.0000 & 0.0001 & 0.0004 & 0.0018 & 0.1067 & 0.7126 & 0.1730 & 0.0040 & 0.0007 \\ \hline 
10 & 0.0001 & 0.0002 & 0.0003 & 0.0000 & 0.0001 & 0.0002 & 0.0001 & 0.0016 & 0.1232 & 0.7075 & 0.1628 & 0.0041 \\ \hline 
11 & 0.0003 & 0.0002 & 0.0006 & 0.0000 & 0.0002 & 0.0000 & 0.0005 & 0.0002 & 0.0030 & 0.1071 & 0.7413 & 0.1466 \\ \hline 
12 & 0.0001 & 0.0000 & 0.0000 & 0.0000 & 0.0001 & 0.0001 & 0.0002 & 0.0000 & 0.0004 & 0.0018 & 0.0992 & 0.8982 \\ \hline 
\end{tabular} 
}\caption{Table of smearing matrix for $\cos(\theta_{\mu}^{reco})$. 
Each row represents a true bin, and each column represents a reconstructed bin.  Each value is the fraction of events in the true bin which are reconstructed in the respective reconstructed bin. 
}\end{table}

\FloatBarrier\subsection{Tables for $\cos(\theta_{p}^{reco})$} 
\begin{table}[ht] 
\begin{tabular}{|c|c|c|c|c|} \hline 
Bin & Min. & Max. & Cross Section [$10^{-38}$~cm$^{2}/$nucleus] & Uncertainty [$10^{-38}$~cm$^{2}/$nucleus] \\ \hline 
1 & -1.0 & -0.5 & 1.48 & 0.54 \\ \hline 
2 & -0.5 & 0.0 & 1.92 & 0.79 \\ \hline 
3 & 0.0 & 0.27 & 3.62 & 1.36 \\ \hline 
4 & 0.27 & 0.45 & 9.96 & 3.15 \\ \hline 
5 & 0.45 & 0.62 & 14.54 & 4.31 \\ \hline 
6 & 0.62 & 0.76 & 16.78 & 5.27 \\ \hline 
7 & 0.76 & 0.86 & 20.52 & 7.10 \\ \hline 
8 & 0.86 & 0.94 & 22.96 & 6.40 \\ \hline 
9 & 0.94 & 1.0 & 20.85 & 4.85 \\ \hline 
\end{tabular} 
\caption{Table of cross section values for $\cos(\theta_{p}^{reco})$} 
\end{table}

\begin{table}[ht] 
\Rotatebox{90}{\begin{tabular}{|c|c|c|c|c|c|c|c|c|c|} \hline 
\ & 1 & 2 & 3 & 4 & 5 & 6 & 7 & 8 & 9 \\ \hline 
1 & 0.2969 & 0.3030 & 0.4939 & 1.1582 & 1.3929 & 2.0472 & 2.1497 & 1.9569 & 1.1126 \\ \hline 
2 & 0.3030 & 0.6277 & 0.5008 & 1.2688 & 1.5086 & 2.3315 & 1.8130 & 1.9718 & 1.9636 \\ \hline 
3 & 0.4939 & 0.5008 & 1.8496 & 3.6296 & 4.7684 & 5.3382 & 8.1096 & 6.9332 & 1.7090 \\ \hline 
4 & 1.1582 & 1.2688 & 3.6296 & 9.9288 & 12.2306 & 15.1181 & 19.8528 & 17.4264 & 3.5951 \\ \hline 
5 & 1.3929 & 1.5086 & 4.7684 & 12.2306 & 18.6101 & 19.3953 & 28.4204 & 25.4440 & 4.5663 \\ \hline 
6 & 2.0472 & 2.3315 & 5.3382 & 15.1181 & 19.3953 & 27.7419 & 30.6366 & 27.0616 & 7.0598 \\ \hline 
7 & 2.1497 & 1.8130 & 8.1096 & 19.8528 & 28.4204 & 30.6366 & 50.4159 & 41.9922 & 7.8924 \\ \hline 
8 & 1.9569 & 1.9718 & 6.9332 & 17.4264 & 25.4440 & 27.0616 & 41.9922 & 40.9950 & 9.9232 \\ \hline 
9 & 1.1126 & 1.9636 & 1.7090 & 3.5951 & 4.5663 & 7.0598 & 7.8924 & 9.9232 & 23.4981 \\ \hline 
\end{tabular} 
}\caption{Table of covariances for $\cos(\theta_{p}^{reco})$. 
Units are in $10^{-38}$~cm$^{-2}/nucleon$, as in the cross section results.}
\end{table}

\begin{table}[ht] 
\Rotatebox{90}{\begin{tabular}{|c|c|c|c|c|c|c|c|c|c|} \hline 
\backslashbox{True bin}{Reconstructed bin} & 1 & 2 & 3 & 4 & 5 & 6 & 7 & 8 & 9 \\ \hline 
1 & 0.8306 & 0.0670 & 0.0108 & 0.0147 & 0.0188 & 0.0159 & 0.0217 & 0.0154 & 0.0051 \\ \hline 
2 & 0.1060 & 0.7694 & 0.0331 & 0.0139 & 0.0119 & 0.0216 & 0.0185 & 0.0123 & 0.0133 \\ \hline 
3 & 0.0238 & 0.0548 & 0.6666 & 0.1671 & 0.0290 & 0.0279 & 0.0151 & 0.0099 & 0.0058 \\ \hline 
4 & 0.0104 & 0.0129 & 0.0771 & 0.7190 & 0.1527 & 0.0115 & 0.0081 & 0.0052 & 0.0031 \\ \hline 
5 & 0.0102 & 0.0117 & 0.0158 & 0.0947 & 0.7345 & 0.1172 & 0.0082 & 0.0059 & 0.0018 \\ \hline 
6 & 0.0113 & 0.0120 & 0.0102 & 0.0164 & 0.1106 & 0.7222 & 0.1002 & 0.0137 & 0.0033 \\ \hline 
7 & 0.0185 & 0.0153 & 0.0096 & 0.0105 & 0.0216 & 0.1094 & 0.6815 & 0.1223 & 0.0115 \\ \hline 
8 & 0.0295 & 0.0173 & 0.0072 & 0.0096 & 0.0139 & 0.0207 & 0.0969 & 0.7174 & 0.0873 \\ \hline 
9 & 0.0396 & 0.0176 & 0.0091 & 0.0060 & 0.0106 & 0.0179 & 0.0273 & 0.1222 & 0.7498 \\ \hline 
\end{tabular} 
}\caption{Table of smearing matrix for $\cos(\theta_{p}^{reco})$. 
Each row represents a true bin, and each column represents a reconstructed bin.  Each value is the fraction of events in the true bin which are reconstructed in the respective reconstructed bin. 
}\end{table}

\FloatBarrier\subsection{Tables for $p_{\mu}^{reco}$} 
\begin{table}[ht] 
\begin{tabular}{|c|c|c|c|c|} \hline 
Bin & Min. & Max. & Cross Section [$10^{-38}$~cm$^{2}/$nucleus] & Uncertainty [$10^{-38}$~cm$^{2}/$nucleus] \\ \hline 
1 & 0.1 & 0.18 & 4.96 & 5.29 \\ \hline 
2 & 0.18 & 0.3 & 15.33 & 4.83 \\ \hline 
3 & 0.3 & 0.48 & 21.93 & 6.60 \\ \hline 
4 & 0.48 & 0.75 & 15.66 & 4.64 \\ \hline 
5 & 0.75 & 1.14 & 7.16 & 1.65 \\ \hline 
6 & 1.14 & 2.5 & 0.72 & 0.20 \\ \hline 
\end{tabular} 
\caption{Table of cross section values for $p_{\mu}^{reco}$} 
\end{table}

\begin{table}[ht] 
\begin{tabular}{|c|c|c|c|c|c|c|} \hline 
\ & 1 & 2 & 3 & 4 & 5 & 6 \\ \hline 
1 & 27.9365 & 17.2163 & 24.3234 & 16.7276 & 5.5649 & 0.6573 \\ \hline 
2 & 17.2163 & 23.3313 & 27.0439 & 19.2995 & 6.4874 & 0.7270 \\ \hline 
3 & 24.3234 & 27.0439 & 43.5974 & 29.4177 & 9.8565 & 1.1398 \\ \hline 
4 & 16.7276 & 19.2995 & 29.4177 & 21.5012 & 7.0441 & 0.7499 \\ \hline 
5 & 5.5649 & 6.4874 & 9.8565 & 7.0441 & 2.7370 & 0.2742 \\ \hline 
6 & 0.6573 & 0.7270 & 1.1398 & 0.7499 & 0.2742 & 0.0411 \\ \hline 
\end{tabular} 
\caption{Table of covariances for $p_{\mu}^{reco}$. 
Units are in $10^{-38}$~cm$^{-2}/nucleon$, as in the cross section results.}
\end{table}

\begin{table}[ht] 
\begin{tabular}{|c|c|c|c|c|c|c|} \hline 
\backslashbox{True bin}{Reconstructed bin} & 1 & 2 & 3 & 4 & 5 & 6 \\ \hline 
1 & 0.9012 & 0.0839 & 0.0112 & 0.0000 & 0.0000 & 0.0037 \\ \hline 
2 & 0.0262 & 0.8959 & 0.0655 & 0.0091 & 0.0024 & 0.0009 \\ \hline 
3 & 0.0012 & 0.0884 & 0.7750 & 0.1175 & 0.0144 & 0.0035 \\ \hline 
4 & 0.0005 & 0.0132 & 0.1511 & 0.6772 & 0.1415 & 0.0165 \\ \hline 
5 & 0.0002 & 0.0084 & 0.0461 & 0.2104 & 0.5954 & 0.1394 \\ \hline 
6 & 0.0000 & 0.0046 & 0.0422 & 0.0893 & 0.2560 & 0.6079 \\ \hline 
\end{tabular} 
\caption{Table of smearing matrix for $p_{\mu}^{reco}$. 
Each row represents a true bin, and each column represents a reconstructed bin.  Each value is the fraction of events in the true bin which are reconstructed in the respective reconstructed bin. 
}\end{table}

\FloatBarrier\subsection{Tables for $p_{p}^{reco}$} 
\begin{table}[ht] 
\begin{tabular}{|c|c|c|c|c|} \hline 
Bin & Min. & Max. & Cross Section [$10^{-38}$~cm$^{2}/$nucleus] & Uncertainty [$10^{-38}$~cm$^{2}/$nucleus] \\ \hline 
1 & 0.3 & 0.41 & 27.97 & 6.76 \\ \hline 
2 & 0.41 & 0.495 & 29.04 & 6.35 \\ \hline 
3 & 0.495 & 0.56 & 29.10 & 7.75 \\ \hline 
4 & 0.56 & 0.62 & 26.42 & 8.16 \\ \hline 
5 & 0.62 & 0.68 & 23.27 & 7.45 \\ \hline 
6 & 0.68 & 0.74 & 20.18 & 6.75 \\ \hline 
7 & 0.74 & 0.8 & 15.26 & 5.34 \\ \hline 
8 & 0.8 & 0.87 & 10.98 & 4.69 \\ \hline 
9 & 0.87 & 0.93 & 7.24 & 4.13 \\ \hline 
10 & 0.93 & 1.2 & 2.18 & 1.19 \\ \hline 
\end{tabular} 
\caption{Table of cross section values for $p_{p}^{reco}$} 
\end{table}

\begin{table}[ht] 
\Rotatebox{90}{\begin{tabular}{|c|c|c|c|c|c|c|c|c|c|c|} \hline 
\ & 1 & 2 & 3 & 4 & 5 & 6 & 7 & 8 & 9 & 10 \\ \hline 
1 & 45.7017 & 35.3128 & 38.9875 & 42.1566 & 40.2467 & 33.8402 & 26.6454 & 21.5815 & 13.3534 & 4.8874 \\ \hline 
2 & 35.3128 & 40.3708 & 43.1251 & 43.4089 & 41.8683 & 35.6314 & 27.8538 & 23.5982 & 16.4211 & 5.1985 \\ \hline 
3 & 38.9875 & 43.1251 & 60.0924 & 58.9237 & 51.4485 & 46.1571 & 34.3899 & 30.1461 & 25.5225 & 7.3045 \\ \hline 
4 & 42.1566 & 43.4089 & 58.9237 & 66.6254 & 54.3791 & 49.9635 & 37.2171 & 32.8809 & 28.1581 & 7.9232 \\ \hline 
5 & 40.2467 & 41.8683 & 51.4485 & 54.3791 & 55.5041 & 47.1985 & 36.1147 & 30.9086 & 21.6284 & 6.3370 \\ \hline 
6 & 33.8402 & 35.6314 & 46.1571 & 49.9635 & 47.1985 & 45.5509 & 32.7279 & 28.3037 & 20.7805 & 5.6280 \\ \hline 
7 & 26.6454 & 27.8538 & 34.3899 & 37.2171 & 36.1147 & 32.7279 & 28.5185 & 21.8546 & 15.4636 & 4.4421 \\ \hline 
8 & 21.5815 & 23.5982 & 30.1461 & 32.8809 & 30.9086 & 28.3037 & 21.8546 & 21.9778 & 13.5438 & 3.9134 \\ \hline 
9 & 13.3534 & 16.4211 & 25.5225 & 28.1581 & 21.6284 & 20.7805 & 15.4636 & 13.5438 & 17.0655 & 4.1015 \\ \hline 
10 & 4.8874 & 5.1985 & 7.3045 & 7.9232 & 6.3370 & 5.6280 & 4.4421 & 3.9134 & 4.1015 & 1.4113 \\ \hline 
\end{tabular} 
}\caption{Table of covariances for $p_{p}^{reco}$. 
Units are in $10^{-38}$~cm$^{-2}/nucleon$, as in the cross section results.}
\end{table}

\begin{table}[ht] 
\Rotatebox{90}{\begin{tabular}{|c|c|c|c|c|c|c|c|c|c|c|} \hline 
\backslashbox{True bin}{Reconstructed bin} & 1 & 2 & 3 & 4 & 5 & 6 & 7 & 8 & 9 & 10 \\ \hline 
1 & 0.9412 & 0.0437 & 0.0078 & 0.0036 & 0.0018 & 0.0003 & 0.0009 & 0.0002 & 0.0000 & 0.0004 \\ \hline 
2 & 0.2459 & 0.7210 & 0.0256 & 0.0036 & 0.0015 & 0.0013 & 0.0005 & 0.0003 & 0.0000 & 0.0005 \\ \hline 
3 & 0.0467 & 0.2046 & 0.7120 & 0.0326 & 0.0017 & 0.0013 & 0.0005 & 0.0003 & 0.0004 & 0.0000 \\ \hline 
4 & 0.0229 & 0.0656 & 0.1795 & 0.6993 & 0.0301 & 0.0009 & 0.0009 & 0.0004 & 0.0004 & 0.0000 \\ \hline 
5 & 0.0190 & 0.0390 & 0.0675 & 0.1755 & 0.6698 & 0.0280 & 0.0007 & 0.0003 & 0.0000 & 0.0001 \\ \hline 
6 & 0.0107 & 0.0247 & 0.0473 & 0.0723 & 0.1898 & 0.6348 & 0.0191 & 0.0008 & 0.0000 & 0.0006 \\ \hline 
7 & 0.0107 & 0.0181 & 0.0269 & 0.0522 & 0.0844 & 0.2040 & 0.5900 & 0.0128 & 0.0002 & 0.0007 \\ \hline 
8 & 0.0090 & 0.0150 & 0.0194 & 0.0350 & 0.0581 & 0.0962 & 0.1987 & 0.5654 & 0.0028 & 0.0004 \\ \hline 
9 & 0.0083 & 0.0128 & 0.0207 & 0.0191 & 0.0414 & 0.0756 & 0.1053 & 0.2560 & 0.4599 & 0.0008 \\ \hline 
10 & 0.0139 & 0.0148 & 0.0238 & 0.0305 & 0.0364 & 0.0550 & 0.0718 & 0.1079 & 0.1720 & 0.4737 \\ \hline 
\end{tabular} 
}\caption{Table of smearing matrix for $p_{p}^{reco}$. 
Each row represents a true bin, and each column represents a reconstructed bin.  Each value is the fraction of events in the true bin which are reconstructed in the respective reconstructed bin. 
}\end{table}

\FloatBarrier\subsection{Tables for $\theta_{\mu-p}^{reco}$} 
\begin{table}[ht] 
\begin{tabular}{|c|c|c|c|c|} \hline 
Bin & Min. & Max. & Cross Section [$10^{-38}$~cm$^{2}/$nucleus] & Uncertainty [$10^{-38}$~cm$^{2}/$nucleus] \\ \hline 
1 & 0.0 & 0.8 & 1.15 & 0.42 \\ \hline 
2 & 0.8 & 1.2 & 3.96 & 0.96 \\ \hline 
3 & 1.2 & 1.57 & 8.88 & 2.53 \\ \hline 
4 & 1.57 & 1.94 & 11.00 & 3.61 \\ \hline 
5 & 1.94 & 2.34 & 7.64 & 2.60 \\ \hline 
6 & 2.34 & 3.14 & 2.09 & 0.80 \\ \hline 
\end{tabular} 
\caption{Table of cross section values for $\theta_{\mu-p}^{reco}$} 
\end{table}

\begin{table}[ht] 
\begin{tabular}{|c|c|c|c|c|c|c|} \hline 
\ & 1 & 2 & 3 & 4 & 5 & 6 \\ \hline 
1 & 0.1740 & 0.2579 & 0.6364 & 0.6839 & 0.3653 & 0.0985 \\ \hline 
2 & 0.2579 & 0.9197 & 2.1438 & 2.7486 & 1.8152 & 0.4344 \\ \hline 
3 & 0.6364 & 2.1438 & 6.4252 & 8.2692 & 5.0467 & 0.9837 \\ \hline 
4 & 0.6839 & 2.7486 & 8.2692 & 13.0130 & 8.5595 & 1.8212 \\ \hline 
5 & 0.3653 & 1.8152 & 5.0467 & 8.5595 & 6.7773 & 1.6190 \\ \hline 
6 & 0.0985 & 0.4344 & 0.9837 & 1.8212 & 1.6190 & 0.6399 \\ \hline 
\end{tabular} 
\caption{Table of covariances for $\theta_{\mu-p}^{reco}$. 
Units are in $10^{-38}$~cm$^{-2}/nucleon$, as in the cross section results.}
\end{table}

\begin{table}[ht] 
\begin{tabular}{|c|c|c|c|c|c|c|} \hline 
\backslashbox{True bin}{Reconstructed bin} & 1 & 2 & 3 & 4 & 5 & 6 \\ \hline 
1 & 0.7362 & 0.1018 & 0.0269 & 0.0281 & 0.0330 & 0.0740 \\ \hline 
2 & 0.0619 & 0.7518 & 0.1227 & 0.0230 & 0.0185 & 0.0221 \\ \hline 
3 & 0.0089 & 0.0651 & 0.8030 & 0.0992 & 0.0120 & 0.0118 \\ \hline 
4 & 0.0073 & 0.0131 & 0.1072 & 0.7898 & 0.0671 & 0.0155 \\ \hline 
5 & 0.0133 & 0.0134 & 0.0169 & 0.1104 & 0.7685 & 0.0775 \\ \hline 
6 & 0.0215 & 0.0181 & 0.0204 & 0.0215 & 0.0933 & 0.8252 \\ \hline 
\end{tabular} 
\caption{Table of smearing matrix for $\theta_{\mu-p}^{reco}$. 
Each row represents a true bin, and each column represents a reconstructed bin.  Each value is the fraction of events in the true bin which are reconstructed in the respective reconstructed bin. 
}\end{table}